\documentclass[final,3p,times]{elsyearticle}

\usepackage{graphicx}

\usepackage{amssymb}

\usepackage[english,francais]{babel}

\def\og{``}
\def\fg{"}

\usepackage[utf8]{inputenc}
\usepackage[T1]{fontenc}
\usepackage{amsmath}
\usepackage{stmaryrd}
\usepackage{color}
\usepackage{hyperref}
\hypersetup{breaklinks=true}
\hypersetup{colorlinks=true}
\hypersetup{urlcolor=blue}
\hypersetup{citecolor=black}
\hypersetup{linkcolor=black}
\usepackage{cleveref}
\newcommand{\bea}{\begin{eqnarray}}
\newcommand{\eea}{\end{eqnarray}}
\newcommand{\be}{\begin{equation}}
\newcommand{\ee}{\end{equation}}
\newcommand{\rr}{\mathbf{r}}
\newcommand{\kk}{\mathbf{k}}

\newcommand{\QQ}{\mathbf{Q}}

\newcommand{\FF}{\mathbf{F}}

\newcommand{\qq}{\mathbf{q}}
\newcommand{\JJ}{\mathbf{J}}

\newcommand{\vv}{\mathbf{v}}
\newcommand{\nn}{\mathbf{n}}

\newcommand{\mD}{\mathcal{D}}
\newcommand{\mA}{\mathcal{A}}
\newcommand{\zero}{\mathbf{0}}

\newcommand{\veps}{\varepsilon}

\newcommand{\ii}{\textrm{i}}
\newcommand{\eee}{\textrm{e}}
\newcommand{\dd}{\mathrm{d}}
\newcommand{\etab}{{\boldmath$\eta$}}
\newcommand{\kapb}{{\boldmath$\kappa$}}

\DeclareMathOperator\ch{ch}
\DeclareMathOperator\sh{sh}
\DeclareMathOperator\re{Re}
\DeclareMathOperator\im{Im}
\usepackage{float}

\begin{document}

\begin{frontmatter}


\selectlanguage{english}
\title{Random walk of a massive quasiparticle in the phonon gas of an ultralow temperature superfluid}


\author{Yvan Castin}

\address{Laboratoire Kastler Brossel, ENS-Universit\'e PSL, CNRS, Universit\'e de la Sorbonne and Coll\`ege de France, 24 rue Lhomond, 75231 Paris, France}



\begin{abstract}
We consider in dimension 3 a homogeneous superfluid at very low temperature $T$ having two types of excitations, (i) gapless acoustic phonons with a linear dispersion relation at low wave number, and (ii) gapped $\gamma$ quasiparticles with a quadratic (massive) dispersion relation in the vicinity of its extrema. Recent works [Nicolis and Penco (2018), Castin, Sinatra and Kurkjian (2017, 2019)], extending the historical study by Landau and Khalatnikov on the phonon-roton interaction in liquid helium 4, have explicitly determined the scattering amplitude of a thermal phonon on a $\gamma$ quasiparticle at rest to leading order in temperature. We generalize this calculation to the case of a $\gamma$ quasiparticle of arbitrary subsonic group velocity, with a rigorous construction of the $S$ matrix between exact asymptotic states, taking into account the unceasing phonon-phonon and phonon-$\gamma$ interaction, which dresses the incoming and emerging phonon and $\gamma$ quasiparticle by virtual phonons; this sheds new light on the Feynman diagrams of phonon-$\gamma$ scattering. In the whole domain of the parameter space (wave number $k$, interaction strength, etc.) where the $\gamma$ quasiparticle is energetically stable with respect to the emission of phonons of arbitrary wavevector, we can therefore characterize the erratic motion it performs in the superfluid due to its unceasing collisions with thermal phonons, through (a) the mean force $F(k)$ and (b) longitudinal and transverse momentum diffusion coefficients $D_\sslash(k)$ and $D_\perp(k)$ coming into play in a Fokker-Planck equation, then, at long times when the quasiparticle has thermalized, (c) the spatial diffusion coefficient $\mathcal{D}^{\rm spa}$, independent of $k$.  At the location $k_0$ of an extremum of the dispersion relation, where the group velocity of the quasiparticle vanishes, $F(k)$ varies linearly with velocity with an isotropic viscous friction coefficient $\alpha$ that we calculate; if $k_0=0$, the momentum diffusion is also isotropic and $F(k_0)=0$; if $k_0>0$, it is not ($D_\sslash(k_0)\neq D_\perp(k_0)$), and $F(k_0)$ is non-zero but subleading with respect to $\alpha$ by one order in temperature. The velocity time correlation function, whose integral gives $\mathcal{D}^{\rm spa}$, also distinguishes between these two cases ($k_0$ is now the location of the minimum): if $k_0=0$, it decreases exponentially, with the expected viscous damping rate of the mean velocity; if $k_0>0$, it is bimodal and has a second component, with an amplitude lower by a factor $\propto T$, but with a lower damping rate in the same ratio (it is the thermalization rate of the velocity direction); this balances that. We also characterize analytically the behavior of the force and of the momentum diffusion in the vicinity of any sonic edge of the stability domain where the quasiparticle speed tends to the speed of sound in the superfluid. The general expressions given in this work are supposedly exact to leading order in temperature (order $T^8$ for $F(k)$, order $T^9$ for $D_\sslash(k)$, $D_\perp(k)$ and $F(k_0)$, order $T^{-7}$ for $\mathcal{D}^{\rm spa}$). They however require an exact knowledge of the dispersion relation of the $\gamma$ quasiparticle and of the equation of state of the superfluid at zero temperature. We therefore illustrate them in the BCS approximation, after calculating the stability domain, for a fermionic $\gamma$ quasiparticle (an unpaired fermion) in a superfluid of unpolarized spin 1/2 fermions, a system that can be realised with cold atoms in flat bottom traps; this domain also exhibits an interesting, unobserved first order subsonic instability line where the quasi-particle is destabilized by emission of phonons of finite wave vectors, in addition to the expected sonic instability line resulting from Landau's criterion. By the way, we refute the thesis of Lerch, Bartosch and Kopietz (2008), stating that there would be no fermionic quasiparticle in such a superfluid.
\vskip 0.25\baselineskip

\end{abstract} 
\end{frontmatter}


\selectlanguage{english}

\section{Position of the problem and model system considered} 

Certain spatially homogeneous three-dimensional superfluids present at arbitrarily low temperature two types of excitations. The first type corresponds to a branch of acoustic excitation, of angular eigenfrequency $ \omega_q $ tending linearly to zero with the wave number $ q $, 
\be
\label{eq:001}
\omega_q\underset{q\to 0}= c q \left[1 + \frac{\gamma_\phi}{8} \left(\frac{\hbar q}{m c}\right)^2 + O(q^4 \ln q)\right]
\ee
where $ c $ is the speed of sound at zero temperature, $ m $ the mass of a particle of the superfluid and $ \gamma_\phi $ a dimensionless curvature; this branch is always present in a superfluid subjected to short-range interactions, and the associated quanta are bosonic quasiparticles, the phonons, denoted by $\phi$. We recall the exact hydrodynamic relationship $ mc^2 = \rho \dd \mu/\dd \rho $ where $ \mu $ is the chemical potential and $ \rho $ the density in the ground state. The second type of excitation is not guaranteed: it corresponds to quasiparticles, called here $ \gamma $ for brevity, presenting an energy gap $ \Delta_*> 0 $ and a behavior of massive particle, i.e.\ with a parabolic dispersion relation around the minimum: 
\be
\label{eq:002}
\epsilon_k \underset{k\to k_0}{=} \Delta_* + \frac{\hbar^2 (k-k_0)^2}{2 m_*} 
+ \frac{\hbar^2(k-k_0)^3b}{3m_*} + O(k-k_0)^4
\ee
The effective mass $ m_* $ is positive, but the position $ k_0 $ of the minimum in the space of wave numbers can be positive or zero depending on the case; the coefficient $ b $ has the dimension of a length and is {\sl a priori} 
nonzero only if $ k_0> 0 $. The excitations of the two types are coupled together, in the sense that a $\gamma$ quasiparticle can absorb or emit phonons, for example. 

At zero temperature, a $\gamma$ quasiparticle of wave number $ k $ fairly close to $ k_0 $, therefore of energy fairly close to $ \Delta_* $, has a sufficiently low group velocity, in particular lower than the speed of sound, and cannot emit phonons without violating the conservation of energy-momentum, according to a fairly classic argument due to Landau, at least if there is also conservation of the total number of quasi-particles $ \gamma $, as is the case for an impurity, i.e. an atom of another species in the superfluid, or of the parity of this total number, as is the case for an elementary fermionic excitation of the superfluid (a known counterexample to the purely kinematic argument of Landau is that of the biroton in helium 4, whose dispersion relation is of the form (\ref{eq:002}) with $k_0 = 0 $ and which can decay into phonons \cite{Greytak, Pita1973}).  The quasiparticle is then stable and moves forward in the fluid in a ballistic manner, without damping. 

At a nonzero but arbitrarily low temperature $ T $, the acoustic excitation branch is thermally populated, so that a phonon gas at equilibrium coexists with the superfluid. The $\gamma$ quasiparticle, which was stable at zero temperature, now undergoes random collisions with the phonons and performs a random walk in momentum space $ \hbar \kk $. In the quasi-classical limit where the width $ \Delta k $ of the wave number distribution of $ \gamma $ is large enough, so that the coherence length $ 1/\Delta k $ of $ \gamma $ is much lower than the typical wavelength of thermal phonons, 
\be
\label{eq:003}
\frac{1}{\Delta k} \ll \frac{\hbar c}{k_B T}
\ee
we can characterize this random walk by a mean force $ \FF (\kk) $ and a momentum diffusion matrix $ \underline{\underline{D}} (\kk) $ inserted in a Fokker-Planck equation. The $\gamma$ quasiparticle also performs a random walk in position space, a Brownian motion, which is characterized at long times by a spatial diffusion coefficient $ \mD^{\rm spa} $. The objective of this work is to calculate the force and the diffusion to leading order in temperature, for any value of the wave number $ k $ where the $\gamma$ quasiparticle is stable at zero temperature. For this, it is necessary to determine the scattering amplitude of a phonon on the $\gamma$ quasiparticle to leading order in the wave number $ q $ of the phonon for any subsonic speed of $ \gamma $, which, to our knowledge, has not been done in the literature; references \cite{Penco, PRLphigam} are for example limited to low speeds, as evidenced by the rotonic action (30) of reference \cite{Penco} limited to the first two terms of our expansion (\ref{eq:002}). 

Several systems have the two types of excitations required. In the case of a bosonic system, one immediately thinks of superfluid liquid helium 4, the single excitation branch of which includes both a linear acoustic departure at zero wave number and a quadratic relative minimum at nonzero wavenumber $k_0$; the massive $\gamma$ quasiparticle associated with the minimum is a roton according to the terminology used. Although there is no conservation law of their number, the rotons of wave number $ k $ close to $ k_0 $ are stable with respect to the emission of phonons of arbitrary wave vectors \cite{Penco}. In the case of a fermionic system, come to mind the cold atom gases with two spin states $ \uparrow $ and $ \downarrow $: in the non-polarized gas, that is to say with exactly the same number of fermions in each spin state, at sufficiently low temperature, the fermions form bound pairs $ \uparrow \downarrow $ under the effect of attractive interactions between atoms of opposite spins, these pairs form a condensate and a superfluid with an acoustic excitation branch. A pair-breaking excitation would create two fragments, each one being a fermionic $\gamma$ quasiparticle of gap $ \Delta_* $ (the half-binding energy of the pair) and of parabolic dispersion relation around the minimum. To have a single $\gamma$ quasiparticle present, it suffices to add to the unpolarized gas a single $ \uparrow $ or $ \downarrow $ fermion, which will remain unpaired and form the desired quasiparticle. \footnote{At nonzero temperature, there is a nonzero fraction of thermally broken bound pairs, but it is a $ O (T^{\nu} \exp (-\Delta_*/k_B T)) $ ($ \nu = 1/2 $ if $ k_0> 0 $, $ \nu = 3/2 $ if $ k_0 = 0 $) which we neglect in the following.} The spatially homogeneous case is achievable in a flat-bottom potential box \cite{Hadzibabic, Zwierleina, Zwierleinb, Zwierleinc}.

To fix the ideas, we will make {\sl explicit} calculations of the stability domain of $ \gamma $ at $ T = 0 $, then of the force and of the diffusion which it undergoes at $ T> 0 $, only in the case of a gas of fermionic cold atoms, using for lack of a better BCS theory, for any interaction strength since the scattering length $ a $ between $ \uparrow $ and $ \downarrow $ is adjustable at will by Feshbach resonance \cite{Thomas2002, Salomon2003, Grimm2004b, Ketterle2004, Salomon2010, Zwierlein2012}. It then seems to us that the random walk of $ \gamma $ in the momentum or position space is accessible experimentally, since we can separate the unpaired fermionic atom from the paired fermions to manipulate and image it, by transforming those in strongly bound dimer molecules by fast Feshbach ramp towards the BEC (Bose-Einstein condensate) limit $ a \to 0^+$ \cite{KetterleVarenna}.  In order to make the discussion less abstract, let us outline a possible experimental protocol:
\begin{enumerate}
\item in a flat bottom trap, we prepare at very low temperature a very weakly polarized gas of cold fermionic atoms, with a little more fermions in the spin state $ \uparrow $ than in the state $ \downarrow $; the $ N_\uparrow-N_\downarrow $ unpaired fermions give rise to as many $\gamma$ quasi-particles in the interacting gas. This idea was implemented at MIT \cite{KetterleGap}. At this point, the quasi-particles $ \gamma $ have a momentum distribution $ \propto \exp[- \hbar^2 (k-k_0)^2/2 m_* k_B T]$ and a uniform position distribution.
\item by a slow Feshbach ramp on the magnetic field, we modify the scattering length $ a $ between atoms adiabatically, without putting the gas out of thermal equilibrium. We can thus change at will the position $ k_0 = 0 $ or $ k_0> 0 $ of the minimum of $ \epsilon_k $, reversibly. Whenever we want to act on the $ \gamma $ quasi-particles to manipulate their distribution in momentum or in position space by an electromagnetic field (laser, radiofrequency) {\sl without affecting the bound pairs}, or simply to image these distributions \cite{KetterleVarenna}, we perform a fast Feshbach ramp (only the internal state of the bound pairs follows, the gas does not have time to thermalize) towards a very weak and positive scattering length $ a_{\rm action} $; at this value, the bound pairs exist as strongly bound dimers of electromagnetic field resonance frequency very different from that of unpaired atoms, which allows the desired selective action, as was done at the ENS \cite{Christophemolec}; if necessary, a rapid Feshbach ramp is performed to return to the initial scattering length.
\item to observe the diffusion in position space (section \ref{sec:carac}, equation (\ref{eq:143})), it is necessary initially to spatially filter the quasi-particles $\gamma$, for example by sending a pushing or internal-state depopulating field masked by an opaque disc, which eliminates the quasi-particles outside a cylinder (as in point 2 so as not to disturb the bound pairs); one filters in the same way according to an orthogonal direction, to leave intact a ball of $ \gamma $ quasi-particles in position space, of which one can then measure the spread over time (as in point 2).
\item to access the mean force and the momentum diffusion (section \ref{sec:carac}, equation (\ref{eq:103})) at wave vector $\kk_{\rm target} $, we prepare the $\gamma$ quasi-particles with a narrow momentum distribution out of equilibrium around $\kk =\kk_{\rm target} $, then we measure as a function of time the mean of $\kk$ and its variances and covariances. The narrow distribution results, for example, from a Raman transfer from $\kk\simeq \zero $ to $\kk\simeq \kk_{\rm target} $ of the quasiparticles in the intermediate phase $a=a_{\rm action} $ of point 2, where we have managed to have a distribution centered on the zero momentum by prior adiabatic passage in a regime $ k_0 = 0 $.
\end{enumerate}
This would allow, if not to measure, at least to constrain the scattering amplitude $ \mA $ between phonons and $\gamma$ quasiparticle, on which $ \FF $, $ \underline{\underline{D}} $ and $ \mathcal{D}^{\rm spa} $ depend. There is a strong motivation for this: the exact expression of this amplitude to leading order in $ q $ is not yet unanimous, references \cite{Penco} and \cite{PRLphigam} remaining in disagreement, even if we take into account as in the erratum \cite{PRLerr} the interaction between phonons omitted in \cite{PRLphigam}, and the experiments have to our knowledge not yet settled the problem \cite{Fok}. We will take advantage of this to make more convincing and more solid the computation of $ \mA $ from the effective low-energy Hamiltonian of reference \cite{PRLphigam}.  
\section{Stability domain in the momentum space of the $\gamma$ quasiparticle at $ T = 0 $} 
\label{sec:stab} 

Let us consider a $\gamma$ quasiparticle of initial wave vector $ \kk $ in the superfluid at zero temperature, therefore in the initial absence of phonons or other excitations, and study the stability of this quasiparticle with respect to the emission of excitations in the superfluid. In this part, for illustrative purposes, we assume that the number of $\gamma$ quasiparticles is conserved, except at the end of the section where it is conserved modulo $ 2 $, and we use mean field dispersion relations for the $\gamma$ quasi-particle and the phonons (in this case, BCS theory and Anderson's RPA), with real values.

\paragraph{Emission of phonons} First, the $\gamma$ quasiparticle can a priori emit any number $ n \geq 1 $ of phonons of wave vectors $ \qq_i $, $ 1 \leq i \leq n $, recoiling to conserve momentum. The corresponding energy change is equal to 
\be
\label{eq:004}
\Delta E = \epsilon_{\kk-\sum_{i=1}^{n} \qq_i} + \left(\sum_{i=1}^{n} \hbar \omega_{\qq_i}\right) -\epsilon_\kk
\ee
where we note indifferently $ \omega_\qq $ or $ \omega_q $ the angular frequency dispersion relation of phonons and $ \epsilon_\kk $ or $ \epsilon_k $ the energy dispersion relation of the $\gamma$ quasiparticle. When the $ \qq_i $ span the existence domain $ D^* $ of the acoustic branch, and that $ n $ spans $ \mathbb{N}^* $, the energy $ \epsilon_\kk+\Delta E $ of the final state spans an energy continuum. There are thus two possible cases: (i) $ \Delta E $ is always positive, the energy $ \epsilon_\kk $ is at the lower edge of the continuum, \footnote{To see it, it is enough to make the $ \qq_i $ tend to zero in the expression of $ \Delta E $.} the initial state $ | \gamma: \kk \rangle $ of the quasiparticle remains a discrete state and $ \gamma $ is {\it stable} at the considered wave vector $ \kk $; (ii) $ \Delta E $ is not always positive, the initial energy $ \epsilon_\kk $ of the quasiparticle is inside the energy continuum, the resonant emission of phonons (with $ \Delta E = 0 $) is possible, the initial state $ | \gamma: \kk \rangle $ dilutes in the continuum and gives rise to a complex energy resonance, and the $\gamma$ quasiparticle is {\it unstable} at the considered wave vector $ \kk $. 

\begin{figure} [t] 
\centerline{\includegraphics [width = 0.23 \textwidth, clip =]{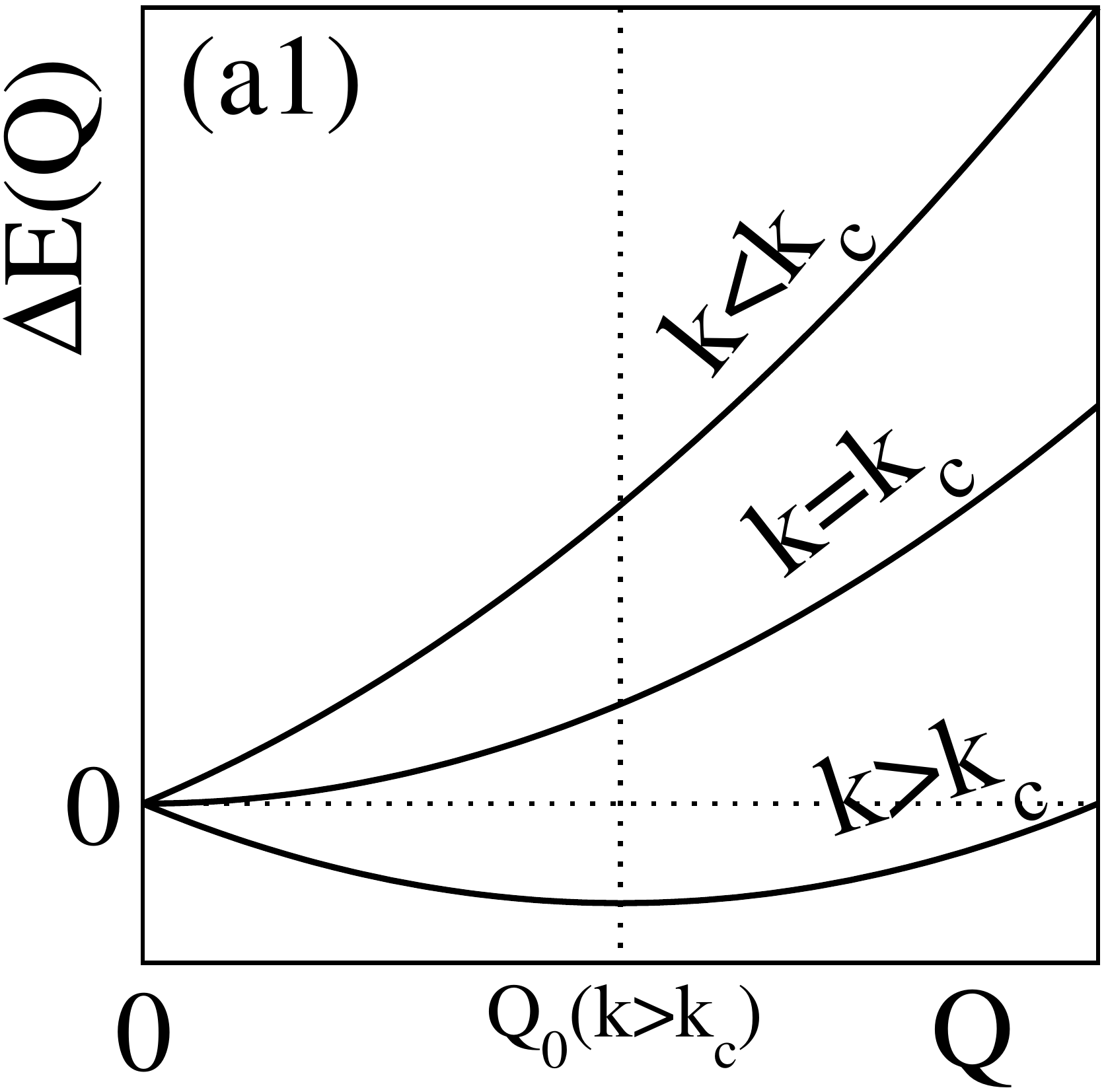} \hspace{3mm} \includegraphics [width = 0.23 \textwidth, clip =]{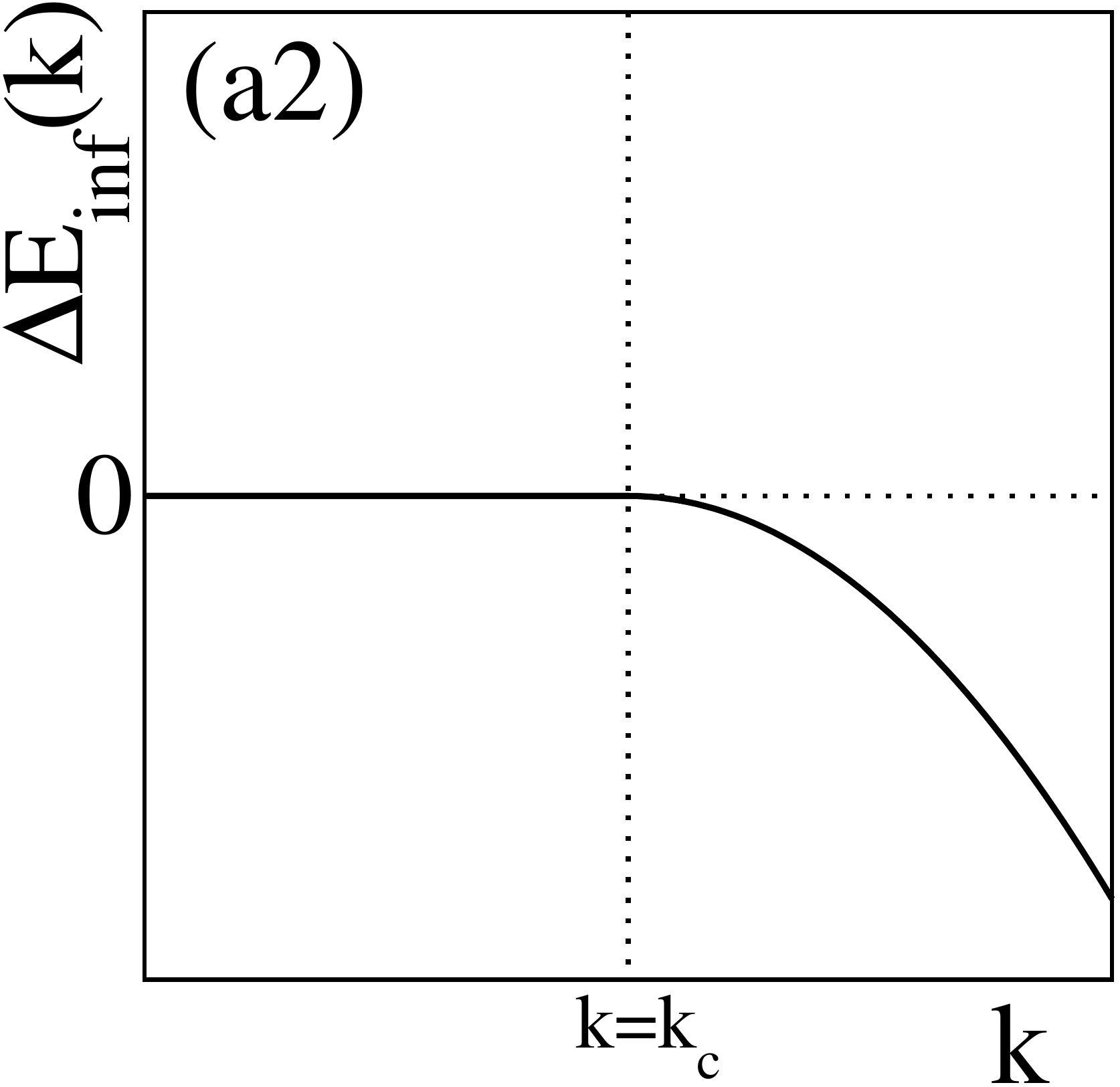} \hspace{3mm} \includegraphics [width = 0.23 \textwidth, clip =]{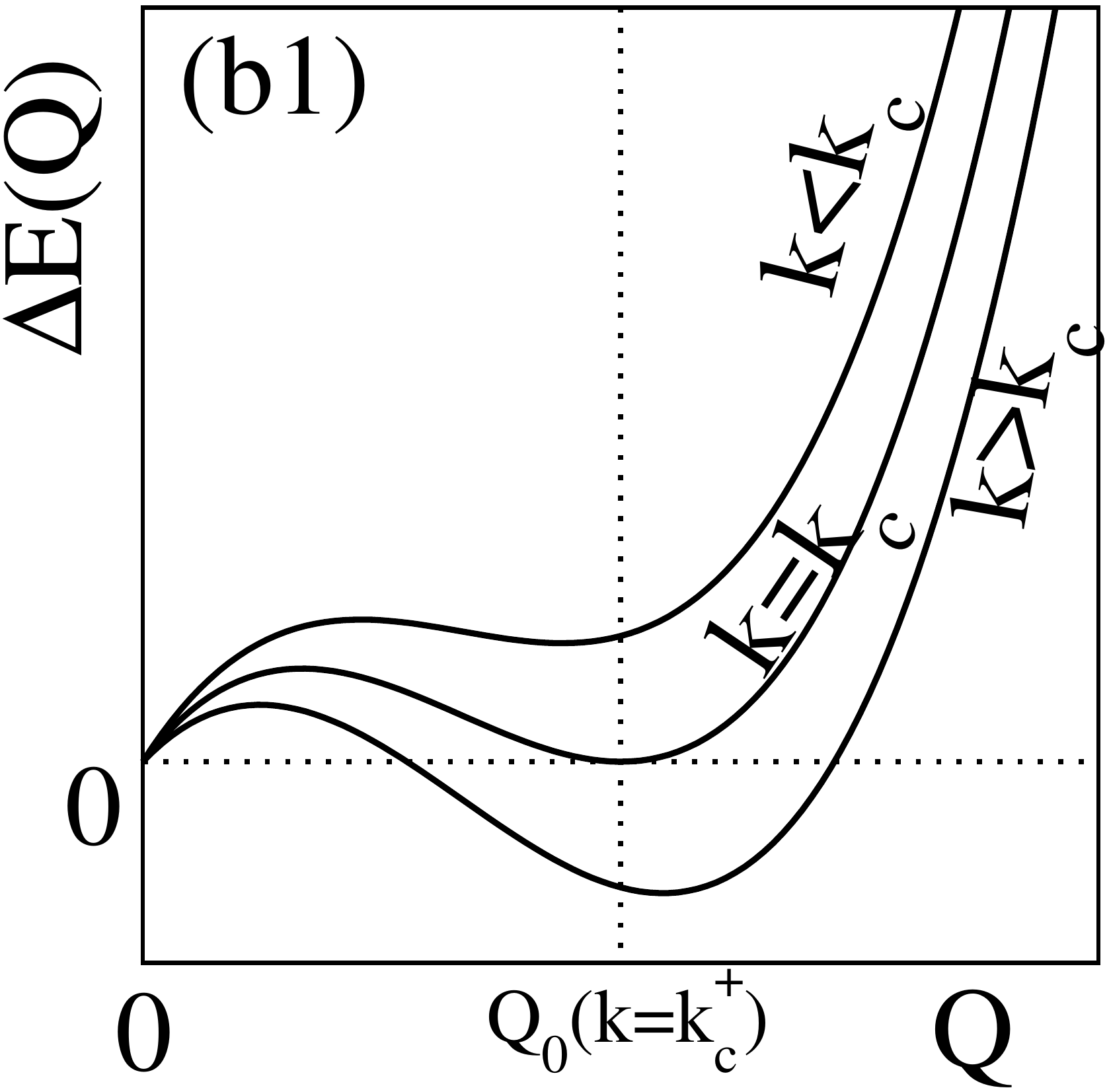} \hspace{3mm} \includegraphics [width = 0.23 \textwidth, clip =]{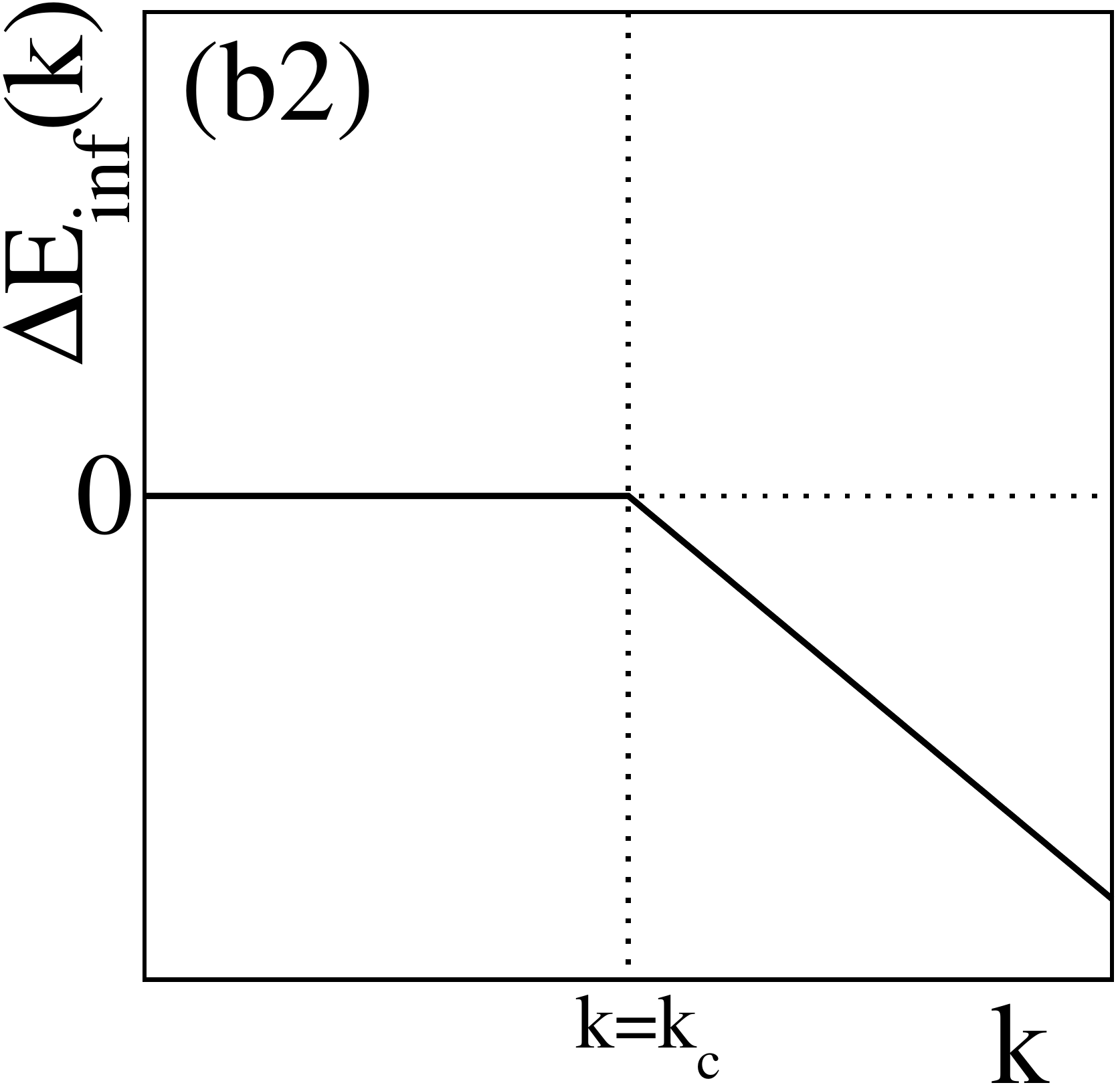}} 
\caption{The different scenarios of destabilization of a massive $\gamma$ quasiparticle of wave number $ k $ by emission of phonons in a superfluid when the total number of $ \gamma $ is conserved and the dispersion relations are real (like those of mean field theories). (a) The group velocity $v_k$ of the quasiparticle changes from subsonic to supersonic when $ k $ crosses $ k_c $ (for example from left to right); the change of energy $ \Delta E (\QQ) $ at the emission of a fixed total phonon momentum $ \hbar \QQ $ (see equation (\ref{eq:006})) is minimal in $ \QQ = \zero $ for $ k <k_c $ and in $ \QQ = \QQ_0 (\kk)\neq \zero $ for $ k> k_c $, where the modulus $ Q_0 (k) $ tends linearly to zero when $ k \to k_c^+$, as in (a1); the absolute minimum $ \Delta E_{\rm inf} (k) $ of the energy change deviates from zero quadratically near $ k_c $ and the destabilization is second order, as in (a2). (b) The group velocity $v_k$ remains subsonic when $ k $ crosses $ k_c $, but the minimum of $ \Delta E (\QQ) $ in $ \QQ = \QQ_0 (\kk)\neq\zero $, which was only relative for $ k <k_c $, becomes absolute (and negative) for $ k> k_c $, as in (b1); $ \Delta E_{\rm inf} (k) $ deviates from zero linearly and the destabilization is first order, as in (b2).  In (a1) and (b1), the figure is drawn in $\QQ$  space along the direction of $ \kk $ or $ - \kk $, depending on whether the group velocity $ v_ {| \kk- \QQ_0 (\kk) |} $ of $ \gamma $ after emission is $> 0 $ or $ <0 $ (this is the direction of the minimizer $\QQ_0(\kk) $ if $ Q_0 (k)> 0 $, see  footnote \ref{note:coli}, and the one chosen in equation (\ref{eq:008}) otherwise).} 
\label{fig:destab} 
\end{figure}

To decide between the two cases, we must determine the lower edge $ \epsilon_\kk+\Delta E_{\rm inf} (\kk) $ of the continuum, minimizing $ \Delta E $ on the number and the wave vectors of the emitted phonons. This minimization can be carried out in two stages: with a fixed total wave vector of emitted phonons, $ \QQ = \sum_{i = 1}^n \qq_i $, the energy is minimized to obtain an effective acoustic dispersion relation \footnote{Since the existence domain $ D^* $ does not contain the zero wave vector, we must in principle take the infimum on the $ \qq_i $. We reduce ourselves to taking a minimum by adding the non-physical zero vector to $ D^* $, $ D = D^* \cup \{\zero \} $, and by extending $ \omega_\qq $ by continuity, $ \omega_{\qq = \mathbf{0}} = 0 $.} 
\be
\label{eq:005}
\hbar\omega_{\rm eff}(\QQ)  = \min_{n\in \mathbb{N}^*}\ \min_{\{(\qq_i)_{1\leq i\leq n} \in D^n | \sum_{i=1}^{n} \qq_i=\QQ\}}\ \sum_{i=1}^n \hbar \omega_{\qq_i}
\ee
then we minimize on $ \QQ $: 
\be
\label{eq:006}
\Delta E_{\rm inf}(\kk) = \min_\QQ \Delta E(\QQ) \quad\mbox{with}\quad \Delta E(\QQ) \equiv \epsilon_{\kk-\QQ} -\epsilon_\kk + \hbar\omega_{\rm eff}(\QQ)
\ee
If $ \Delta E_{\rm inf} (\kk) <0 $, the $\gamma$ quasiparticle is unstable at the wave vector $ \kk $, otherwise it is stable. It is instructive to carry out a local study at low emitted total wave number $ Q $: from the behavior (\ref{eq:001}) and the simplifying remarks which follow, we derive $ \omega_{\rm eff} (\QQ) \underset{Q \to 0}{=} c Q+O (Q^3) $ which leads to the expansion 
\be
\label{eq:008}
\Delta E(\QQ) \underset{Q\to 0}{=} \hbar c Q (1-|v_k/c|) +\frac{1}{2} Q^2 \frac{\dd^2\epsilon_k}{\dd k^2} + O(Q^3)
\ee
where the direction of $ \QQ $, chosen to minimize the energy, is that of $ \kk $ if the group velocity $ v_k = \dd \epsilon_k/\hbar \dd k $ of the $\gamma$ quasiparticle is positive, and that of $-\kk $ otherwise. This provides a first, well known scenario of instability: if the $\gamma$ quasiparticle is supersonic ($ | v_k |> c $), it can slow down by emitting phonons of arbitrarily low wave number. If the transition from an interval of $ k $ where $ \gamma $ is stable to an interval of $ k $ where $ \gamma $ is unstable takes place according to this scenario, that is to say by crossing the speed of sound in the critical number $ k_c $, $ \Delta E_{\rm inf} (k) $ varies quadratically in the neighborhood of $ k_c $ on the unstable side, according to the law deduced from equation (\ref{eq:008}) \footnote{The simplified notation $ \Delta E_{\rm inf} (k) $ takes advantage of the rotational invariance of $ \Delta E_{\rm inf} (\kk) $. Very close to $ k_c $, on the supersonic side, the quadratic approximation (\ref{eq:008}) reaches its minimum in $ Q_0 (k) = (| v_k |/c-1) \hbar c/(\dd^2 \epsilon_k/\dd k^2) $, and this minimum is equal to $ \Delta E_{\rm inf} (k) =-[\hbar c (1-| v_k |/c)]^2/(2 \dd^2 \epsilon_k/\dd k^2) $. It remains to linearize the group velocity in the vicinity of $ k = k_c $, $ \hbar v_k \simeq \pm \hbar c+\frac{\dd^2 \epsilon_k}{\dd k^2} |_{k = k_c} (k-k_c) $, to obtain (\ref{eq:010}).}: 
\be
\label{eq:010}
\Delta E_{\rm inf}(k) \underset{k\to k_c}{=} -\frac{1}{2} \frac{\dd^2\epsilon_k}{\dd k^2}\Big|_{k=k_c}(k-k_c)^2+O(k-k_c)^3
\ee
For the internal consistency of this scenario, the second derivative of $ \epsilon_k $ must be positive in $ k = k_c $: we have $ | v_{k_c} | = c $ and $ \Delta E (k = k_c, \QQ) $ must reach its minimum in $ Q = 0 $; otherwise, the transition to $ \Delta E_{\rm inf} (k) <0 $ would have taken place before the sonic threshold. We are therefore dealing here with a second order destabilization, see figure \ref{fig:destab}a. A second scenario of instability is that the group velocity remains subsonic but that $ \Delta E (\QQ) $ admits an absolute negative minimum in $\QQ_0(\kk)\neq\zero$: the $\gamma$ quasiparticle reduces its energy by emitting phonons of necessarily not infinitesimal total momentum. When $ k $ passes from the stable zone to the unstable zone, the position of the minimum of $ \Delta E (\QQ) $ then jumps discontinuously from zero to the nonzero value $ \QQ_0 (\kk) $ in $ k = k_c $, $ \Delta E_{\rm inf} (k) $ varies linearly in the vicinity of $ k_c $ on the unstable side \footnote{Assuming $\QQ_0(\kk)$ collinear to $\kk$ (see footnote \ref{note:coli}), we take the derivative of  $ \Delta E_{\rm inf} (k) = \Delta E (k, Q_0 (k)) $ with respect to $ k $ (with obvious notations) and we use the fact that $ \partial_Q \Delta E = 0 $ in $ Q = Q_0 (k) $ to obtain $ \dd \Delta E_{\rm int} (k)/\dd k = \partial_k \Delta E (k, Q_0 (k)) $, which {\it a priori} has a nonzero limit in $ k = k_c $ (if the group velocities of $ \gamma $ differ before and after the phononic emission).} and the destabilization is first order, see figure \ref{fig:destab}b. \footnote{\label{note:coli} By isotropy, only the first contribution to $ \Delta E (\QQ) $ in equation (\ref{eq:006}) depends on the cosine $ u $ of the angle between $ \kk $ and $ \QQ $. For $ k> 0 $ and $ Q = Q_0 (k)> 0 $ fixed and $ u $ spanning $ [- 1,1] $, two cases arise. (i) $ \Delta E (\QQ) $ is minimal at the edge $ u = 1 $ or $ u = -1 $ ($\QQ_0(\kk)$ and $\kk$ are collinear): its derivative with respect to $ u $ is not necessarily zero at the minimum, but must be negative or positive respectively; one has $\dd\epsilon_{\kk-\QQ}/\dd u = (-\hbar k Q /|\kk-\QQ|) v_{|\kk-\QQ|} $ hence the sign of the final group velocity of the quasi-particle stated in the caption of figure \ref{fig:destab}. (ii) $ \Delta E (\QQ) $ is minimal in $u_0 \in]-1,1[$ ($ \kk $ and $ \QQ_0 (\kk) $ are not collinear); then $ \dd \epsilon_{\kk- \QQ} / \dd u $ is zero at the minimum and we are in the particular case $ v_ {| \kk- \QQ_0 (\kk)|} = \zero $; by expressing further the vanishing of the first differential $ \Delta E (\QQ) $ with respect to $ \QQ $ in $ \QQ_0 (\kk) $, we find that the phononic group velocity must also be zero , $ \dd\omega_{\rm eff} (\QQ_0 (\kk)) / \dd Q = 0 $; the positivity of the second differential requires that $ \epsilon_K $ and $ \omega_{\rm eff} (\QQ) $ have a minimum in $K=|\kk-\QQ_0(\kk)|$ and $\QQ=\QQ_0(\kk)$.}

Let us explicitly determine the stability map of a $\gamma$ quasiparticle in a pair-condensed gas of fermions, using the approximate BCS theory. The dispersion relation of the $\gamma$ quasiparticle is then given by 
\be
\label{eq:011}
\epsilon_k=\left[\left(\frac{\hbar^2 k^2}{2m}-\mu\right)^2+\Delta^2\right]^{1/2}
\ee
where $ m $ is the mass of a fermion, $ \mu $ is the chemical potential of the gas and $ \Delta> 0 $ its order parameter. If $ \mu> 0 $, the minimum of $ \epsilon_k $ is reached in $ k_0 = (2m \mu)^{1/2}/\hbar> 0 $, is equal to $ \Delta_* = \Delta $ and leads to an effective mass $ m_* = m \Delta/2 \mu $. If on the other hand $ \mu <0 $, the minimum of $ \epsilon_k $ is reached in $ k_0 = 0 $, is equal to $ \Delta_* = (\Delta^2+\mu^2)^{1/2} $ and leads to the effective mass $ m_* = m \Delta_*/| \mu | $. The phonon dispersion relation is deduced from Anderson's RPA or, which amounts to the same thing, from time-dependent BCS theory linearized around the stationary solution. It has been studied in detail in references \cite{CKS, concavite, vcrit}. Let's just say that it has a rotationally invariant existence domain $ D $ of the connected compact form $ q \leq q_{\rm sup} $ for a scattering length $ a <0 $ (that is $ 0 <\Delta/\mu <1.162 $ according to BCS theory), of the two-connected-component form $ q \leq q_{\rm sup} $ and $ q \geq q_{\rm inf}>q_{\rm sup} $ for $ 1.162 <\Delta/\mu <1.729 $, and given by $ \mathbb{R}^3 $ for $ \Delta/\mu> 1.729 $ or $ \mu <0 $. The calculation of $ \hbar \omega_{\rm eff} (\QQ) $ by numerical energy minimization on the number of phonons $ n $ and their wave vectors $ \qq_i $ at fixed total wave vector $\QQ$, as in equation (\ref{eq:005}), is facilitated by the following remarks: \footnote{Here are brief justifications, knowing that $ \omega_q $ is an increasing positive function of $ q $. (a): if $ D = B (\zero, q_{\rm sup}) $ and $ P $ projects orthogonally on $ \QQ $, then (i) $ D $ is stable by the action of $ P $, (ii) the substitution $ \qq_i \to P \qq_i $ does not change the total wave vector and does not increase the energy; we can therefore limit ourselves to $ \qq_i $ collinear with $ \QQ $. If $ \qq_i $ and $ \qq_j $ are collinear but with opposite directions, with $ q_i \geq q_j $, we lower the energy with fixed total wave vector (without leaving $ D $) by the substitution $ (\qq_i, \qq_j) \to (\qq_i+\qq_j, \zero) $; we can therefore limit ourselves to $ \qq_i $ collinear with $ \QQ $ and in the same direction. (b), (c): if $ q_i <q_j $ are in a concavity (convexity) interval of $ \omega_q $, and we set $ Q_{ij} = q_i+q_j $, the function $ q \mapsto \omega_q+\omega_{Q_{ij}-q} $ is of positive (negative) derivative in $ q = q_i <Q_{ij}/2 $, because the derivative of $ \omega_q $ is decreasing (increasing), so we lower the energy by reducing (increasing) $ q_i $ at fixed $ Q_{ij} $.} 
\begin{itemize} 
\item [(a)] if $ D $ is connected, we can impose without losing anything on the energy that all $ \qq_i $ are collinear with $ \QQ $ and of the same direction, as we do in the continuation of these remarks. It remains to minimize on $ n $ and the wave numbers $ q_i $. 
\item [(b)] if $ \omega_q $ is concave over the interval $ [q_a, q_b] $, and two test wave numbers $ q_i $ and $ q_j $ are in this interval, we lower the energy at fixed $ Q $ by symmetrically moving them apart from their mean value $ (q_i+q_j)/2 $ until one of the two wave numbers reaches $ q_a $ or $ q_b $. If there were $ s \geq 2 $ test numbers in the interval, we are thus reduced to a configuration with $ s_a $ wave numbers in $ q_a $, $ s_b $ in $ q_b $ and zero or one in the interval. 
\item [(c)] if $ \omega_q $ is convex over the interval $ [q_a, q_b] $, and two test wave numbers $ q_i $ and $ q_j $ are in this interval, we lower the energy at fixed $ Q $ by making them converge symmetrically towards their average value $ (q_i+q_j)/2 $. If there are more than two test numbers in the interval, they are all coalesced to their average value. 
\item [(d)] if in addition $ q_a = 0 $, the energy is lowered at fixed $ Q $ by replacing the coalesced test wave number $ Q_{\rm coa} $ by a divergent number $ S $ of phonons of identical infinitesimal wave numbers $ Q_{\rm coa}/S $, which allows us to linearize the dispersion relation in $ q = 0 $ and leads to the coalesced energy $ \hbar c Q_{\rm coa} $. 
\end{itemize} 
Let us give a first example of reduction of the problem in the case $ 1.221 <\Delta/\mu <1.710 $, by limiting ourselves to the lower connected component of the existence domain $ D $, the ball of center $ \zero $ with radius $ q_{\rm sup} $, where the acoustic branch presents two inflection points $ q_{a} $ and $ q_{b} $ \cite{concavite}: $ \omega_q $ is convex over the interval $ [ 0, q_a] $, concave on $ [q_a, q_b] $ and again convex on $ [q_b, q_{\rm sup}] $. We can therefore parametrize the energy of the phonons in this ball as follows, 
\be
\label{eq:012}
\hbar\omega_{\rm eff}(Q)= \hbar c Q_1 + \hbar\omega_{q_2}+ n_3 \hbar\omega_{q_3} \quad\mbox{or}\quad \hbar c Q_1 + n_3 \hbar\omega_{q_3}
\ee
with the constraint $ Q = Q_1+q_2+n_3 q_3 $ or $ Q = Q_1+n_3 q_3 $, $ Q_1 $ any positive number, $ q_2 \in ] q_a, q_b [$, $ n_3 \in \mathbb{N} $ and $ q_3 \in [q_b, q_{\rm sup}] $. We can simplify further by noting that there can be a phonon $ q_2 $ only if $ Q_1 = 0 $. \footnote{If $ Q_1> 0 $ in the presence of a phonon $ q_2 $, let's perform the variation $ (Q_1, q_2) \to (Q_1+\eta, q_2-\eta) $ where $ \eta $ is infinitesimal of any sign. The corresponding energy change in (\ref{eq:012}) is $ \delta E = \hbar [c-v (q_2)] \eta+(1/2) \hbar [\dd v (q_2)/\dd q] \eta^2+O (\eta^3) $, where $ v (q) = \dd \omega_q/\dd q $ is the group velocity of the phonons. $ \delta E $ must always be positive since we deviate from the minimum energy. The coefficient of $ \eta $ must therefore be zero, which is not excluded {\sl a priori}; that of $ \eta^2 $ must be positive, which is prohibited by the strict concavity of the acoustic branch in $ q_2 $. If $ Q_1 = 0 $, $ \eta $ is necessarily positive and this reasoning only imposes $v(q_2)\leq c$.} It remains to numerically minimize $ \hbar c Q_1+n_3 \hbar \omega_{q_3} $ or $ \hbar \omega_{q_2}+n_3 \hbar \omega_{q_3} $ with respect to the remaining independent parameters. Let us give two other examples: in the case $ \Delta> 1.729 \mu $ or $ \mu <0 $, the acoustic branch exists and is convex on all $ \mathbb{R}^+$, so that $ \hbar \omega_{\rm eff} (Q) = \hbar c Q $; in the case $ 0 <\Delta/\mu <0.869 $, the acoustic branch is concave over its entire existence domain $ q \in] 0, q_{\rm sup}] $, so that $ \hbar \omega_{\rm eff} (Q) = \hbar \omega_{Qn q_{\rm sup}}+n \hbar \omega_{q_{\rm sup}} $ where $ n $ is the integer part of $ Q/q_{\rm sup} $.  In practice, even in the RPA approximation, there is no known analytical expression of $ \omega_q $. However, for the stability study of $ \gamma $, we generally need to know the acoustic branch over its entire domain of existence, not just at low $ q $; we therefore reuse the numerical results on $ \omega_q $ obtained in reference \cite{vcrit}.

\begin{figure} [t] 
\centerline{\includegraphics [width = 0.3 \textwidth, clip =]{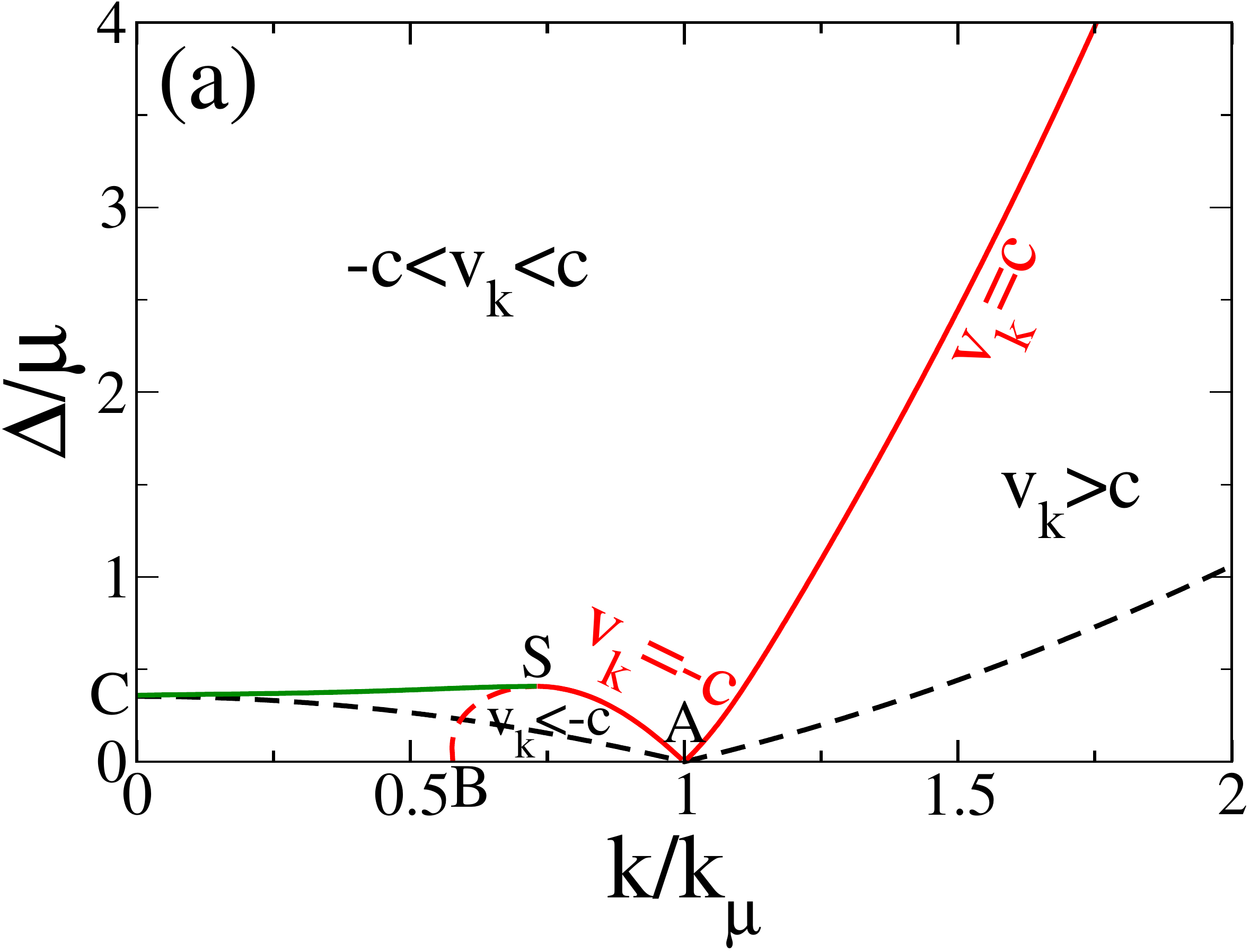} \hspace{3mm} \includegraphics [width = 0.3 \textwidth, clip =]{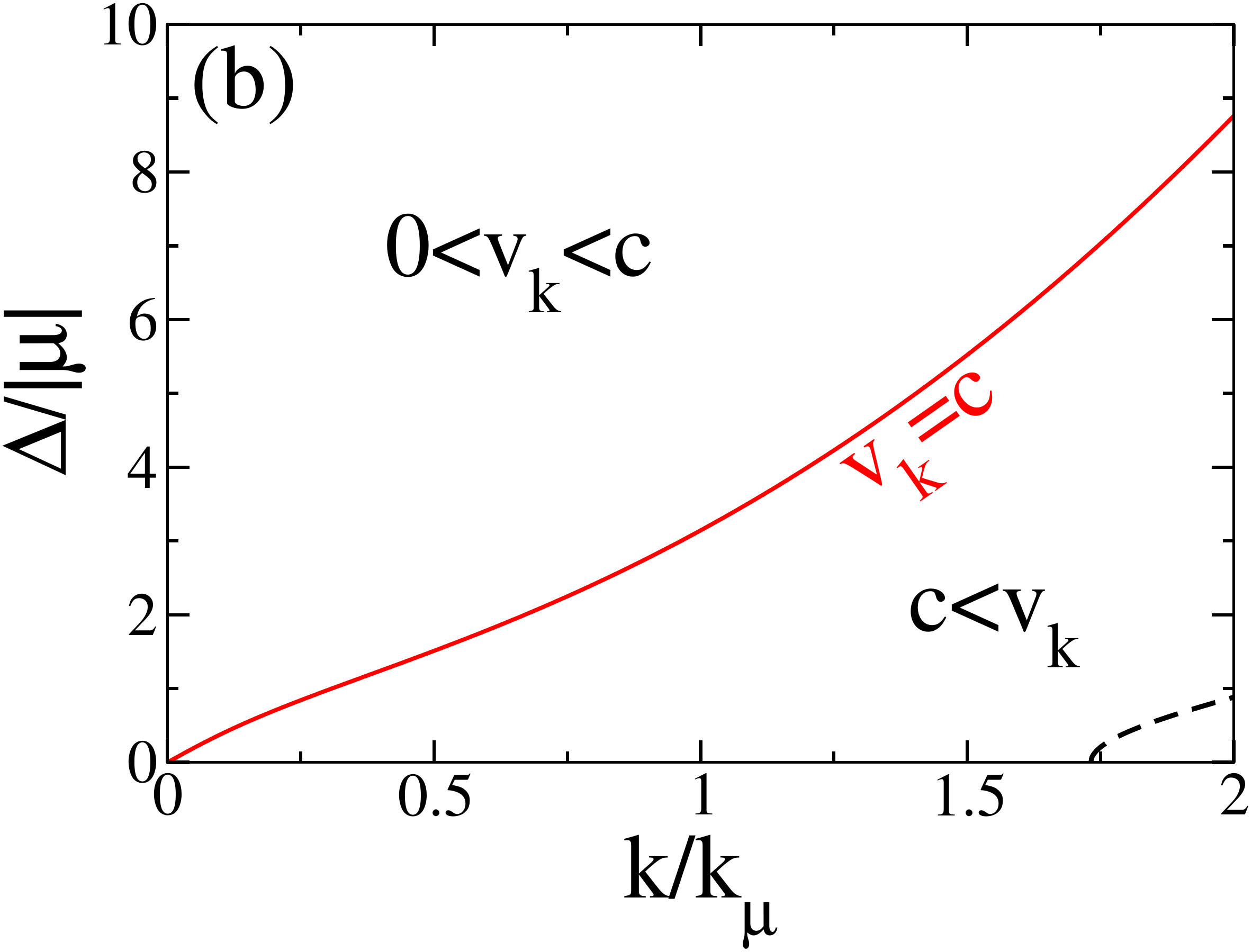} \hspace{3mm} \includegraphics [width = 0.3 \textwidth, clip =]{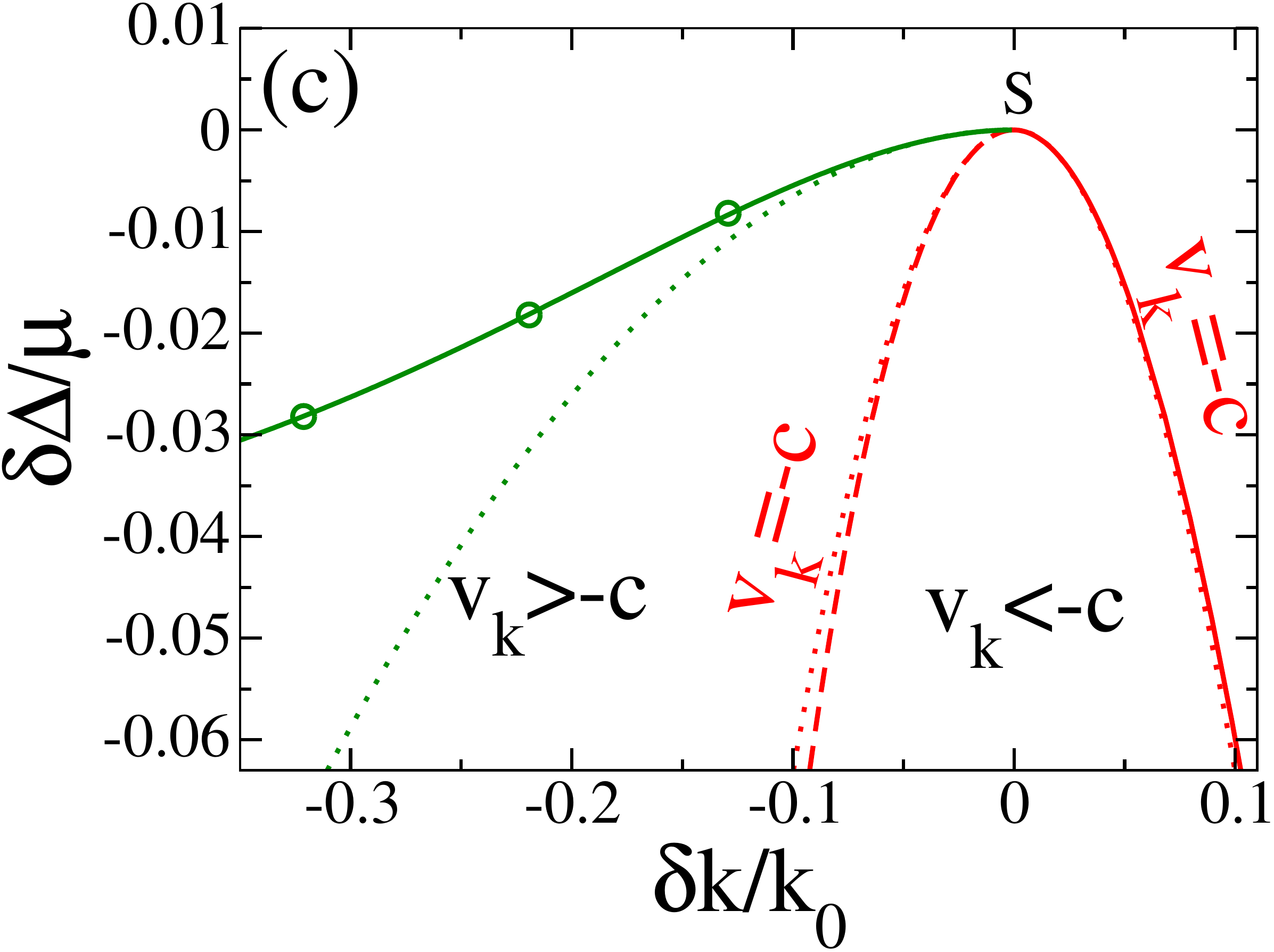}} 
\caption{Stability map of a fermionic $\gamma$ quasiparticle in a non-polarized pair-condensed gas of fermions at zero temperature, in the plane (wave number $ k $ of $ \gamma $, interaction strength), predicted by BCS theory. (a) Case of a chemical potential $ \mu> 0 $. (b) Case of a chemical potential $ \mu <0 $. (c) Enlargement of the case (a) around the critical point S, after refocusing on S ($ \delta k = k-k_{\rm S} $ and $ \delta \Delta = \Delta-\Delta_{\rm S} $ with $ k_{\rm S} \simeq 0.731 k_\mu $ and $ \Delta_{\rm S} \simeq 0.408 \mu $); in solid lines, the results of (a) [the green line is actually an interpolation of the points (circles) actually calculated, which correspond to the values of $\Delta/\mu$ of reference \cite{vcrit} for which the dispersion relation $ q \mapsto \omega_\qq $ of the RPA has been determined numerically], in dotted lines the quadratic departures (\ref{eq:016}) with $ A \simeq 1.121 \, k_\mu^{-1}, B \simeq 14.178 \, \mu k_\mu^{-3}, C \simeq 5.929 \, \mu k_\mu^{-3} $. Red: sonic lines $ v_k = \pm c $ of second order destabilization (destabilization by emission of phonons of infinitesimal wave numbers), where $ v_k $ is the group velocity of $ \gamma $ and $ c $ the speed of sound. In green: first order destabilization line (destabilization by emission of phonons with finite total wave number). In black: lines of destabilization by breaking a bound pair $ \uparrow \downarrow $ [we take $ (n = 0, s = 1) $ in equation (\ref{eq:020}); for $ \mu> 0 $ and $ k <3k_0 $, these are the lower bound cancellation lines in equation (\ref{eq:021})]. Dashed lines: already mentioned destabilization lines that are placed below other destabilization lines and therefore do not change the stability map. $ k $ is in units of $ k_\mu = (2m | \mu |)^{1/2}/\hbar $ (for $ \mu> 0 $, it is also the wave number $ k_0 $ minimizing the energy $ \epsilon_k $ of the $\gamma$ quasiparticle; for $ \mu <0 $, $ k_0 = 0 $). The strength of the interactions is identified by $ \Delta/| \mu | $, where $ \Delta> 0 $ is the order parameter of the fermion gas.} 
\label{fig:map} 
\end{figure} 

The stability map obtained in the plane (wave number, interaction strength) in the case $ \mu> 0 $ is shown in figure \ref{fig:map}a. The stability domain is limited lower right of $ k = k_0 $ by the asymptotically parabolic positive sonic destabilization line $ v_k = c $, \footnote{\label{note:equiv} For $ \mu> 0 $, we find $ \Delta/\mu \sim y (k/k_0)^2 $ if $ k/k_0 \to+\infty $ at fixed $ v_k/c = 1 $. Here $ y \simeq 1.828 $ is the positive solution of $ y (1+y^2) = \frac{2 \Delta}{mc^2} \Big |_{\mu = 0} = 3 (Y+1/Y) $ with $ Y = \pi^2/[2 \Gamma^4 (3/4)] $.}, to the left of $ k = k_0 $ first by the first order destabilization line CS then by the negative sonic destabilization line SA (on which $ v_k =-c $). The ascending part BS of this sonic line (dashed line in the figure), is masked by the instability at finite $ Q $ and therefore has no physical meaning. We note that the dispersion relation (\ref{eq:011}) has a parabolic maximum in $ k = 0 $ (here $ \mu> 0 $): the corresponding massive quasiparticle is sometimes called {\it maxon}. BCS theory therefore predicts that the maxon is stable for fairly strong interactions, $ \Delta/\mu> (\Delta/\mu)_{\rm C} \simeq $ 0.35. In the opposite limit of a weak interaction, the stability domain is reduced to the narrow interval centered on $ k_0 $ and of half-width $ m \Delta/\sqrt{2} \hbar^2 k_0 $. The case $ \mu <0 $, shown in figure \ref{fig:map}b, is poorer: the stability domain is simply limited on the right by the positive sonic line, given by $ \hbar k/m = c $ in the BEC limit (i.e.\ with a slope $ 4 $ at the origin in figure \ref{fig:map}b in BCS theory) and by $ \Delta = y \hbar^2 k^2/2m $ in the limit $ \mu = 0^-$ (i.e.\  an asymptotic parabolic law in figure \ref{fig:map}b), with the same coefficient $ y $ as in footnote \ref{note:equiv}. 

A study of the destabilization lines in the vicinity of the summit point S of figure \ref{fig:map}a can be performed analytically. The dispersion relation (\ref{eq:011}) is conveniently considered as a function $ \epsilon (k, \Delta) $ of $ k $ and $ \Delta $ at fixed scattering length $ a $; similarly, the acoustic branch is seen as a function $ \omega (q, \Delta) $. On the sonic line SA (SB), the second derivative $ \partial_k^2 \epsilon $ is positive (negative), as shown in the discussion below equation (\ref{eq:010}), and vanishes by continuity at the summit point. The coordinates $ (k_{\rm S}, \Delta_{\rm S}) $ of S in the plane $ (k, \Delta) $ are therefore deduced from the system 
\be
\label{eq:014}
\partial_k \epsilon_k|_{\rm S} = - \hbar c|_{\rm S} \quad \mbox{and}\quad \partial_k^2 \epsilon_k|_{\rm S} =0
\ee
Near the critical point S, we find that $ \delta k \equiv k-k_{\rm S} $ and the minimizer $ Q_0 (k) $ are infinitesimals of the first order, while $ \delta \Delta \equiv \Delta-\Delta_{\rm S} $ is an infinitesimal of the second order. It then suffices to expand $ \Delta E (\QQ) $ to order three, using the fact that $ \QQ $, infinitesimal of the first order, is antiparallel to $ \kk $ as it is said below equation (\ref{eq:008}), and that $ \omega_{\rm eff} (\QQ) = \omega (Q, \Delta) $ in the concave case ($ \Delta <0.869 \mu $) at low $ Q $ ($ Q <q_{\rm sup} $), 
\be
\label{eq:015}
\Delta E(\QQ) \underset{Q\to 0}{=} Q \left[A\,\delta \Delta + \frac{1}{2} B (\delta k)^2\right] + \frac{1}{2} Q^2 B\,\delta k + \frac{1}{3} Q^3 C + O(Q^4)
\quad\mbox{with}\quad \left\{ \begin{array}{l}
A= \partial_k\partial_\Delta \epsilon|_{\rm S} + \hbar\partial_\Delta c|_{\rm S} >0 \\
B= \partial_k^3 \epsilon|_{\rm S} >0 \\
C= \frac{1}{2}B+ \frac{3}{8} \hbar c \gamma_\phi \left(\frac{\hbar}{m c}\right)^2|_{\rm S} >0
\end{array}
\right.
\ee
where the absence of infinitely small terms of the first or second order results from the system (\ref{eq:014}), coefficient $ \gamma_\phi $ is that of equation (\ref{eq:001}) and the signs are predicted by BCS theory. One then obtains the quadratic departure of lines SC and BSA of point S as in figure \ref{fig:map}c: \footnote{Let us write the right-hand side of (\ref{eq:015}) in the form $ \Delta E = c_1 Q+c_2 Q^2/2+c_3 Q^3/3 $. On the sonic destabilization line, we simply have $ c_1 = 0 $. On the first-order destabilization line, there is $ Q_c $ such that $ \Delta E (Q_c) = \dd \Delta E (Q_c)/\dd Q = 0 $, as figure \ref{fig:destab}b1 suggests, i.e.\  the discriminant of the polynomial of degree 3 in $ Q $ vanishes, $ (c_1 c_2/2)^2-(4/3) c_3 c_1^3 = 0 $; we find $ Q_c =-3 c_2/4 c_3 $, which must be $> 0 $, which imposes $ \delta k <0 $.} 
\be
\label{eq:016}
\delta \Delta \underset{\delta k\to 0^-}{\stackrel{\rm SC}{\sim}} \frac{(3B-8C)B}{16 A C}(\delta k)^2\ ,\quad \delta \Delta \underset{\delta k\to 0}{\stackrel{\rm BSA}{\sim}} -\frac{B}{2A} (\delta k)^2
\ee
Although interesting, all these results on the destabilization of $ \gamma $ by coupling to phonons are approximate, let us remember. They are based on the dispersion relation (\ref{eq:011}) of the quasi-particle, resulting from the zero-order BCS theory, which precisely ignores the coupling of $ \gamma $ to the phonons; however, the effect of this coupling on $ \epsilon_k $ is {\sl a priori} far from being negligible in the vicinity of an instability of $ \gamma $, in particular first-order subsonic, as shown in reference \cite{Pita1959}. In this context, an experimental verification in a gas of cold fermionic atoms, to our knowledge never made, would be welcome.

\begin{figure}[t]
\centerline{\includegraphics[width=0.20\textwidth,clip=]{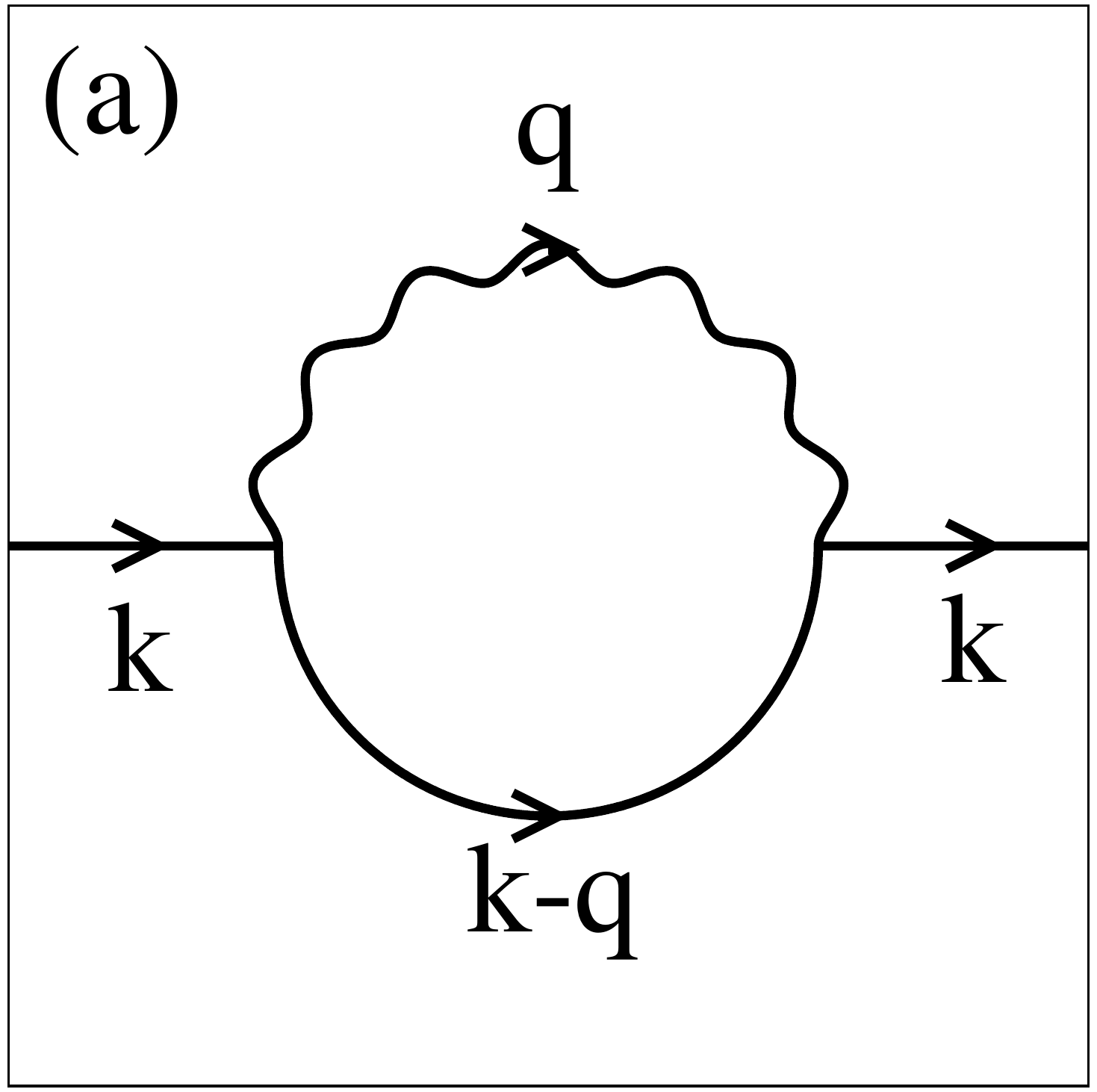}\hspace{3mm}\includegraphics[width=0.20\textwidth,clip=]{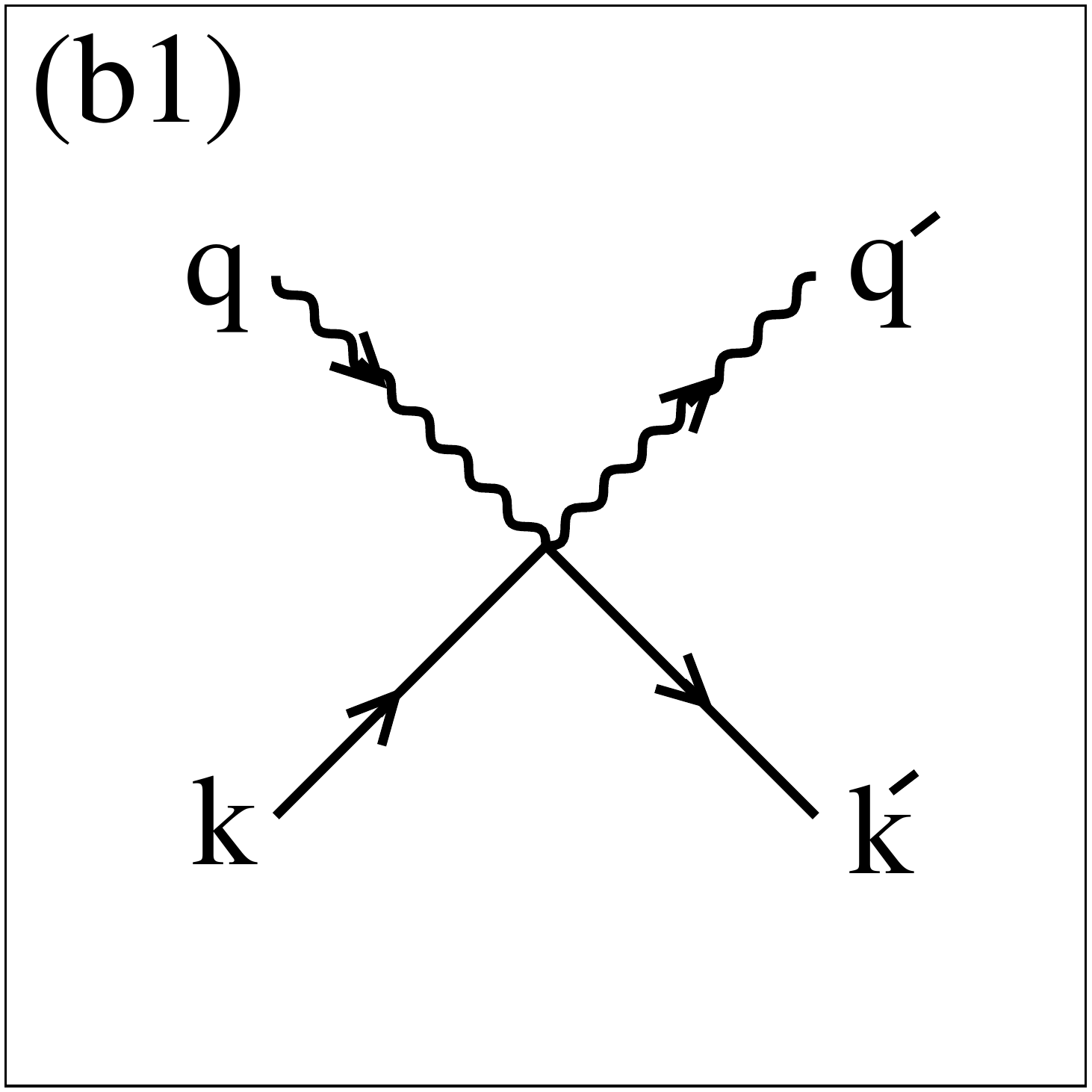}\hspace{3mm}\includegraphics[width=0.20\textwidth,clip=]{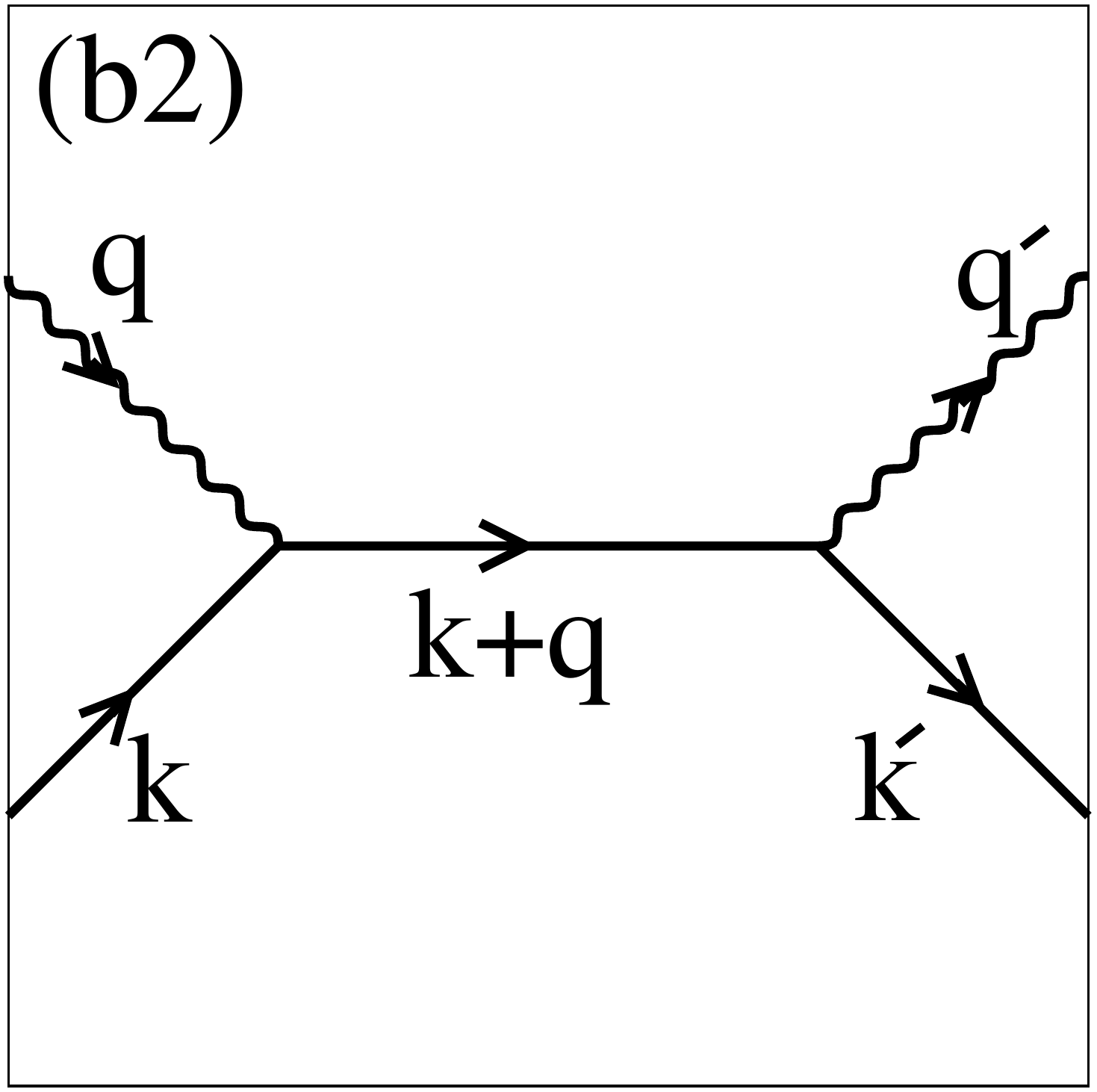}\hspace{3mm}\includegraphics[width=0.20\textwidth,clip=]{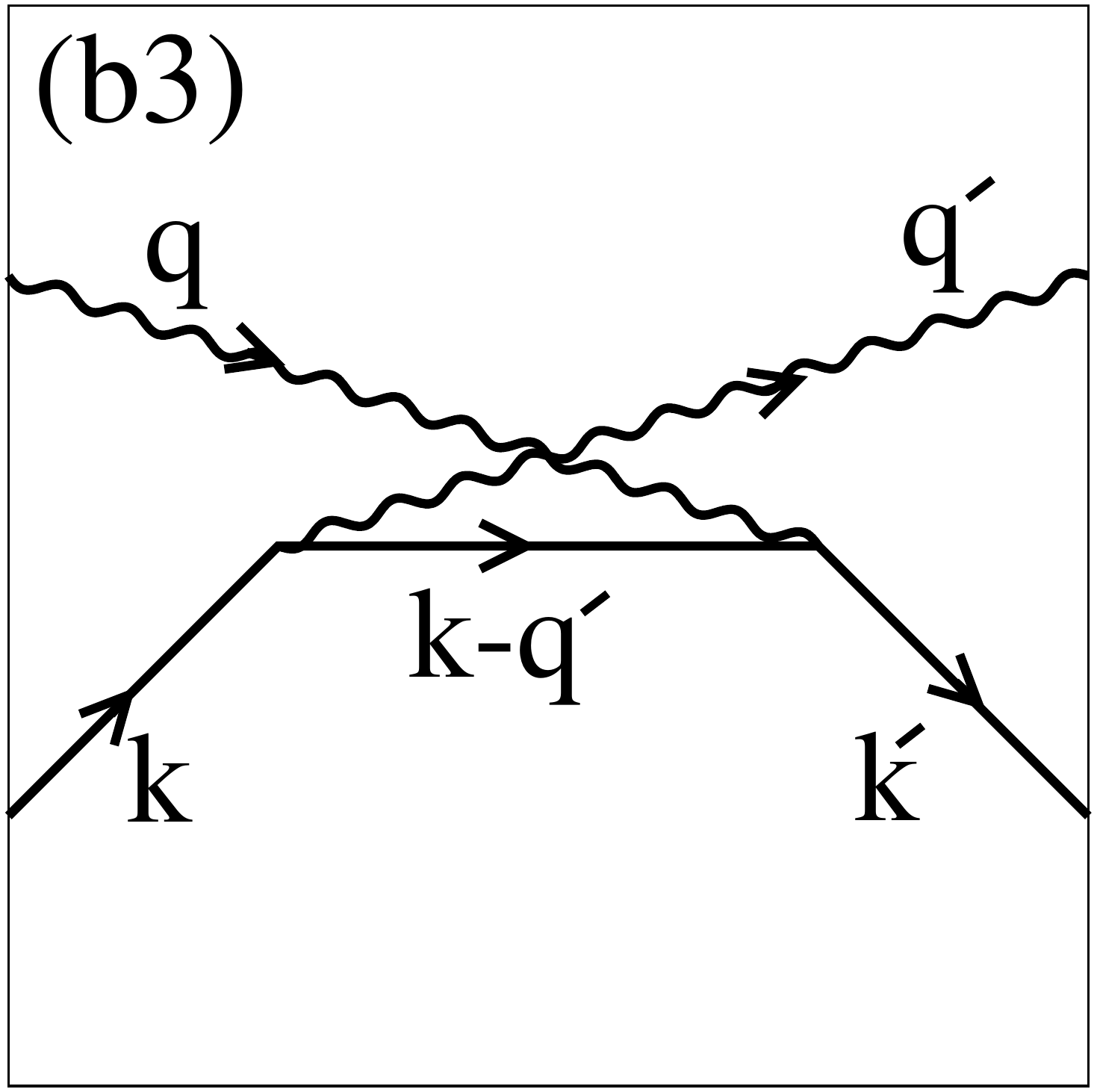}}
\centerline{\includegraphics[width=0.20\textwidth,clip=]{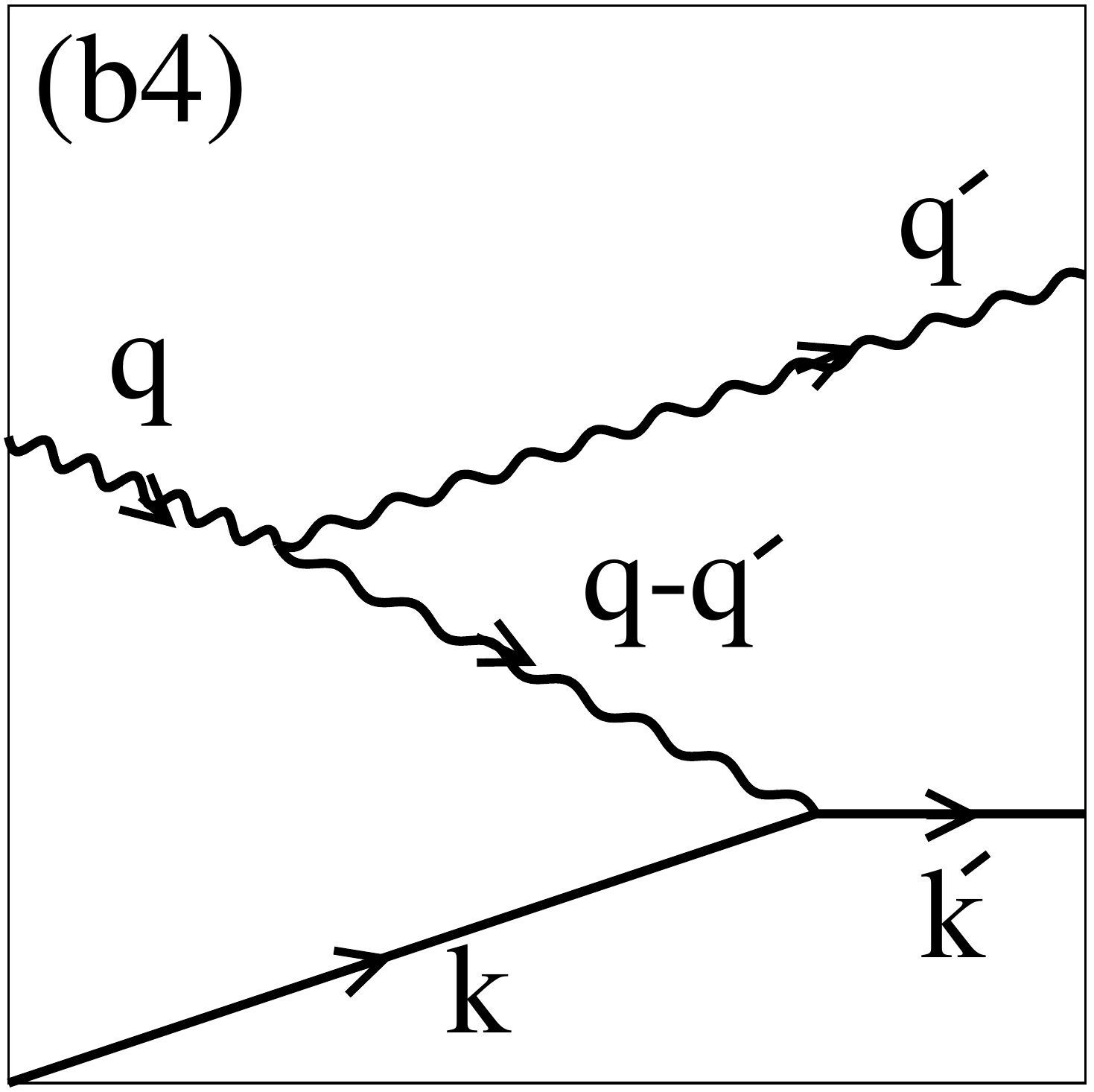}\hspace{3mm}\includegraphics[width=0.20\textwidth,clip=]{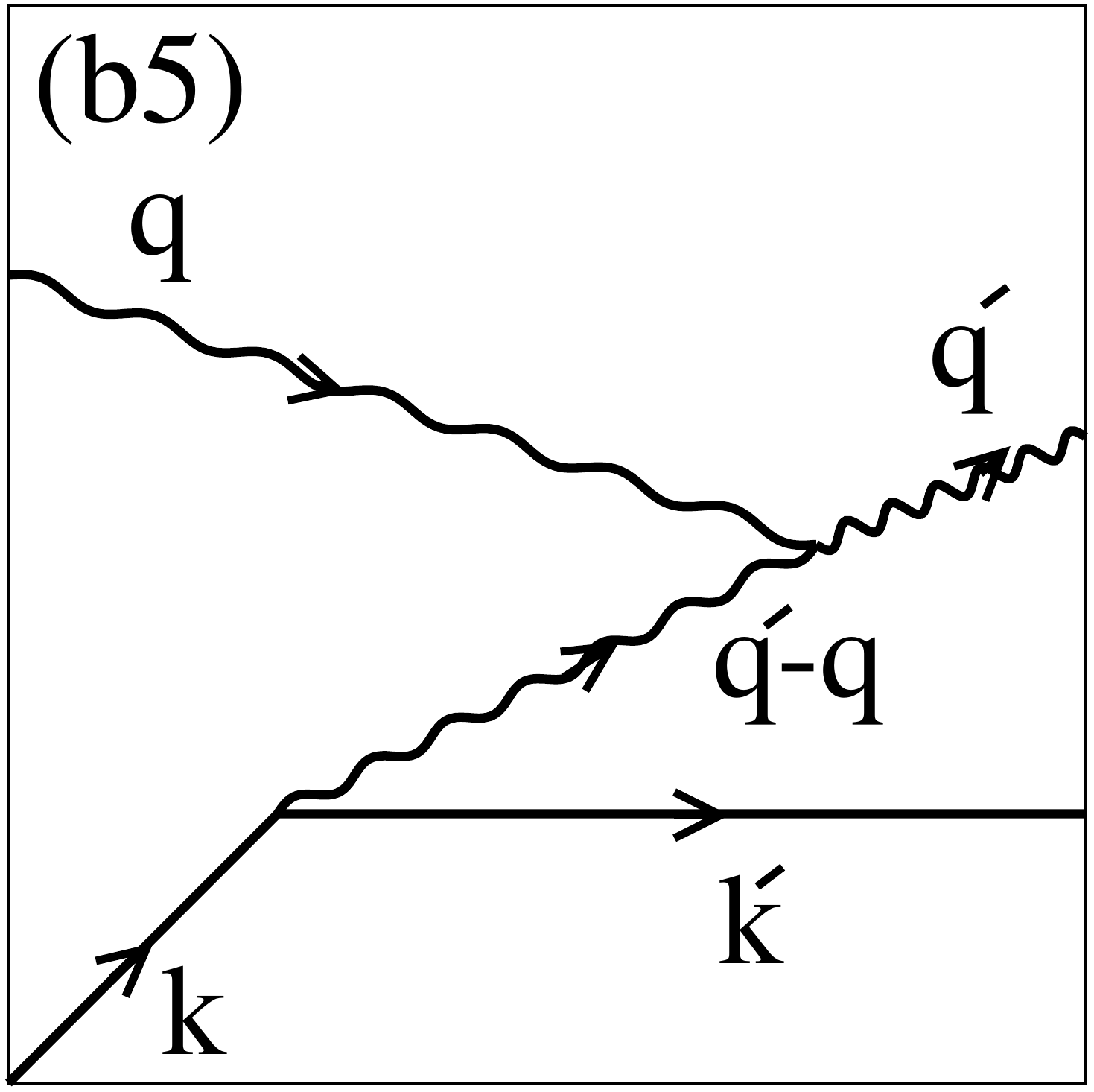}\hspace{3mm}\includegraphics[width=0.20\textwidth,clip=]{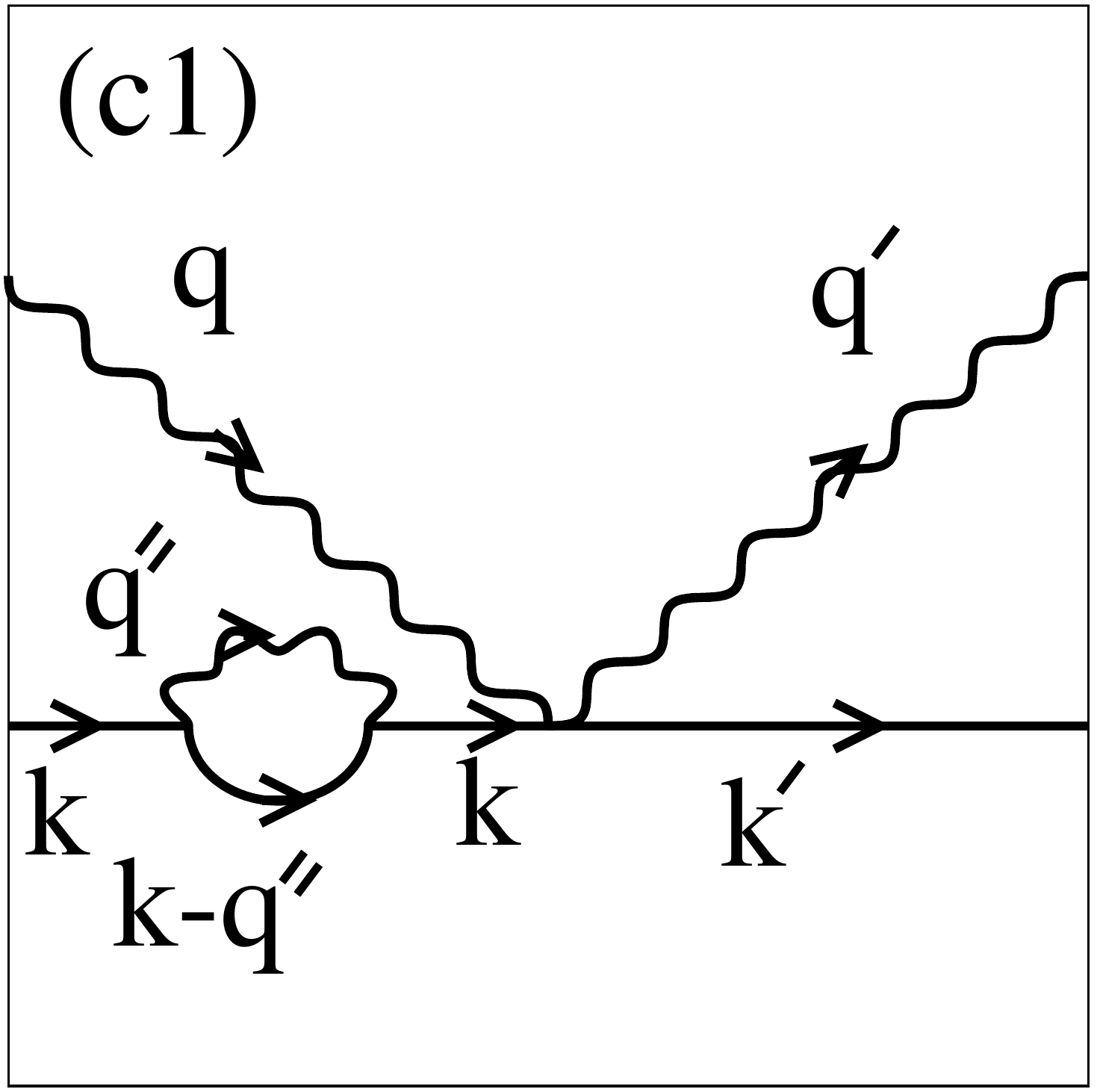}\hspace{3mm}\includegraphics[width=0.20\textwidth,clip=]{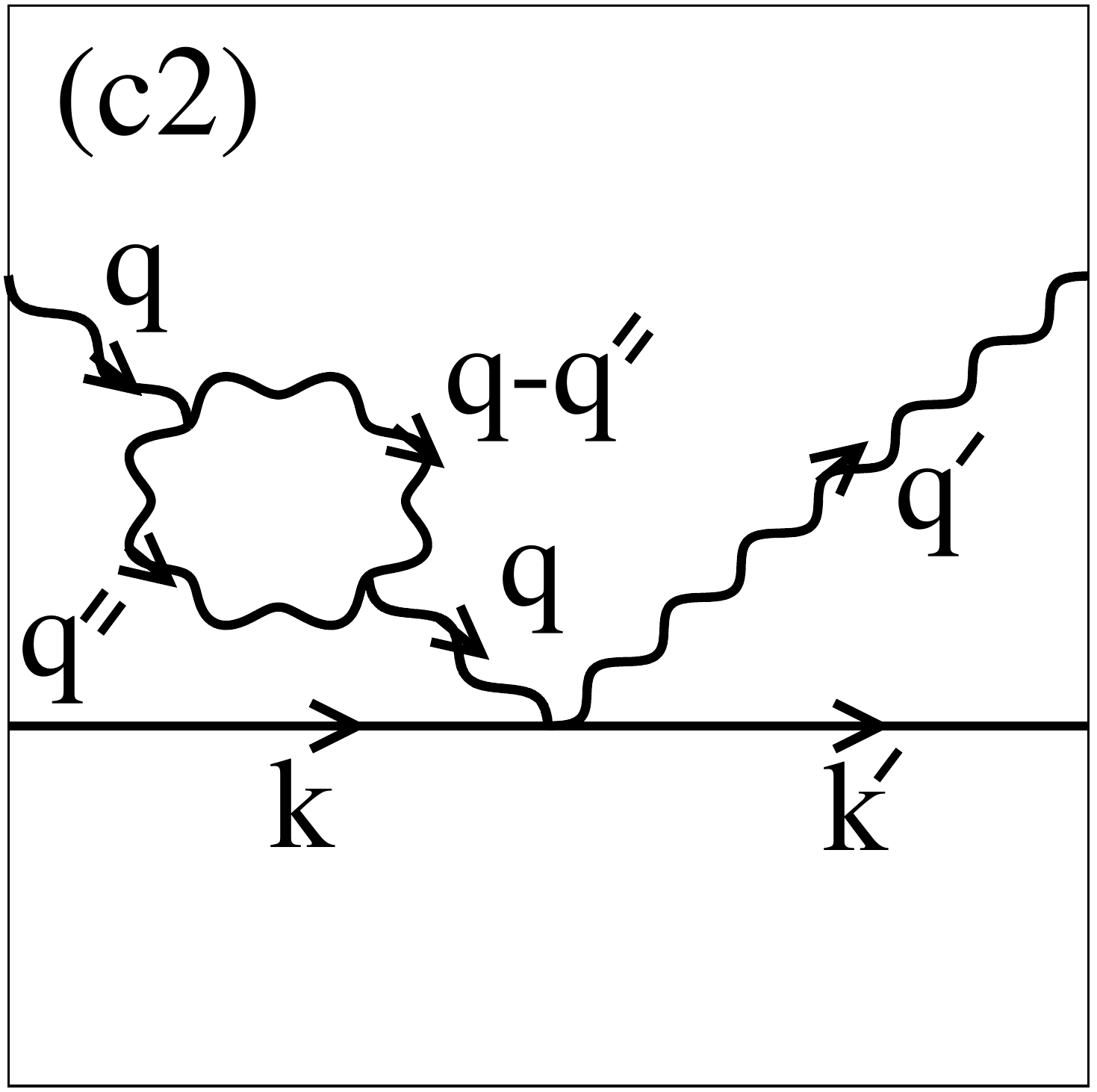}}
\caption{Useful diagrams resulting from the interaction between $\gamma$ quasiparticle (straight line or curve) and phonons (wavy line). (a) One-loop contribution to the propagator of the $\gamma$ quasiparticle, see equation (\ref{eq:018}). (b) Contributions to the scattering amplitude $ | \phi: \qq, \gamma: \kk \rangle \to | \phi: \qq', \gamma: \kk' \rangle $ of a phonon on the $\gamma$ quasiparticle to leading order in temperature ($ q, q '= O (T) \to 0 $, $ \kk $ fixed); the numbering of (b1) to (b5) corresponds, in this order, to the terms $ \mathcal{T}_1 $ to $ \mathcal{T}_5 $ of equation (\ref{eq:038}). (c) Examples of infinite diagrams of order three in the interaction in the perturbative series of the scattering amplitude (\ref{eq:038}) between bare states; as an explicit calculation shows, they do not appear in the scattering amplitude (\ref{eq:041}) between exact asymptotic states.} 
\label{fig:diag} 
\end{figure}

\paragraph{Non-zero spectral weight} For  $ \gamma $ to be a truly stable quasiparticle at wave vector $ \kk $, it is not enough that its energy $ \epsilon_\kk $ is at the lower edge of the energy continuum to which it is coupled by phonon emission. It must also be a quasiparticle, that is to say it has a nonzero spectral weight. Mathematically, this means that its retarded propagator, considered as a function of the complex energy $ z $, has a nonzero residue $ Z $ in the eigenenergy $ z = \epsilon_\kk $. To leading order in the $ \gamma $-phonon coupling $ \hat{V}_{\phi \gamma} $, the propagator is given by the one-loop diagram of figure \ref{fig:diag}a: 
\be
\label{eq:018}
\langle \gamma:\kk| \frac{1}{z-\hat{H}} |\gamma:\kk\rangle \stackrel{\rm one\, loop}{=} \frac{1}{z-\epsilon^{(0)}_\kk-\int_{q<\Lambda} \frac{\dd^3 q}{(2\pi)^3} \frac{|\langle\phi:\qq,\gamma:\kk-\qq|\mathcal{V}_{\phi\gamma}|\gamma:\kk\rangle|^2}{z-(\hbar\omega_\qq^{(0)}+\epsilon_{\kk-\qq}^{(0)})}}
\ee
The complete Hamiltonian $ \hat{H} $ and the interaction operator $ \hat{V}_{\phi \gamma} $ are given by equation (\ref{eq:033}) of the following section and their volume matrix elements in \ref{app:rigor}, in an low-energy effective theory in principle exact in the limit of low phononic wave numbers \cite{LandauKhalatnikov}. $ \epsilon^{(0)}_\kk $ and $ \hbar \omega_\qq^{(0)} $ are the bare excitation energies and $ \Lambda $ is an ultraviolet cutoff on the phonon wave number. By taking the derivative of the denominator in (\ref{eq:018}) with respect to $ z $, and by replacing to leading order the bare quantities by their effective value, we obtain the residue 
\be
\label{eq:019}
Z_\kk \stackrel{\rm one\, loop}{=} \frac{1}{1+\int_{q<\Lambda} \frac{\dd^3 q}{(2\pi)^3} \frac{|\langle\phi:\qq,\gamma:\kk-\qq|\mathcal{V}_{\phi\gamma}|\gamma:\kk\rangle|^2}{[\epsilon_\kk-(\hbar\omega_\qq+\epsilon_{\kk-\qq})]^2}}
\ee
In this expression, in the denominator of the integrand, the energy difference cannot vanish for $ \qq \neq \zero $, by supposed stability of the $\gamma$ quasiparticle; it tends linearly to zero when $ q \to 0 $. In the numerator of the integrand, the matrix element of $ \hat{V}_{\phi \gamma} $ tends to zero as $ q^{1/2} $, see equations (\ref{eq:802b}) and (\ref{eq:803}) of \ref{app:rigor}. It is a robust property: it simply expresses the fact that $ \gamma $ couples directly to the density quantum fluctuations in the superfluid, which are of volume amplitude $ (\hbar \rho q/2 mc)^{1/2} $ in the phonon mode of wave vector $ \qq $ ($ \rho $ is the average density), as predicted by quantum hydrodynamics \cite{LandauKhalatnikov}. The integral in (\ref{eq:019}) is therefore convergent. Thus, at the order considered, in a superfluid gas of fermions in dimension 3, the effective theory of Landau and Khalatnikov excludes any suppression of the spectral weight of the fermionic quasi-particle by infrared singularity, whatever the value of the scattering length $ a $ or the interaction strength \footnote{For example, even in the BEC limit $ k_{\rm F} a \to 0^+$ and for $ k \to 0 $, equation (\ref{eq:019}) predicts a finite correction $ 1-Z_\kk \approx mc 
\Lambda^2/\hbar \rho $.}. 
This directly contradicts and, it seems to us, refutes the conclusions of reference \cite{Lerch2008}, which is based on a one-loop calculation, but in a microscopic model \footnote{Even if the authors do not not explicitly say so, their approach predicts a divergence $ \propto q^{-1/2} $ of the matrix element of the $\phi-\gamma$ coupling , as shown by the explicit integration of their equations (9) and (10) on the Matsubara frequency, i.e.\  on the energy component of their $ P $ quadrivector. The integral in our equation (\ref{eq:019}) would effectively present, in this case, a logarithmic infrared divergence.}. 

To be complete, let us briefly give the predictions of the perturbative result (\ref{eq:019}) on the behavior of the residue in the vicinity of the instability threshold $ k = k_c $ of the $\gamma$ quasi-particle (on the stable side). In the case of sonic instability ($ |v_{k_c}|/c = 1 $), we must now expand the energy difference at the denominator of the integrand around $ \qq = \zero $ one step further, to second order in $ q $; \footnote{We take in the denominator $\epsilon_{\kk-\qq}\! + \! \hbar \omega_\qq \! - \! \epsilon_\kk \simeq \hbar c q \{1 \! - \! u e_k \! + \! (q / 2k) [(1 \! - \! u^2) e_k \! + \! u^2 e_{kk}] \} $, anticipating the notations (\ref{eq:047}). If $ e_ {k} \to \pm 1 $, the integral over $ u = \cos\widehat {(\kk, \qq)} $ is dominated by $ u \simeq \pm 1 $ and we can approximate $ u^2 $ by $ 1 $, $ 1 \! - \! u^2 $ by $ 2 (1 \! \mp \! u) $, and the factor $ (u \! + \! e_\rho)^2 $ to the numerator by $ (1 \! \pm \! e_\rho)^2$.} no infrared divergence appears in the integral even for $ k = k_c $, the spectral weight is not suppressed but only exhibits, as a function of $ k $, a $ (1- | v_k | / c) \ln (1- | v_k | / c) $ singularity. In the case of subsonic instability (destabilization of the first order), supposing it to be due, to remain within the framework of approximation (\ref {eq:019}), to the emission of a single phonon of wave vector $ \QQ_0 (\kk) \neq \zero $ in the linear part of the acoustic branch (case $ \omega_{\rm eff} (\QQ_0) = \omega_{\QQ_0} \simeq c Q_0 $), the integral is now dominated by wave vectors $ \qq $ close to $ \QQ_0 (\kk) \neq \zero $, where the energy difference vanishes at the threshold instability (see figure \ref{fig:destab}b1); by expanding in the denominator the energy difference to second order in $ \qq- \QQ_0 (\kk) $ around $ \QQ_0 (\kk) $, \footnote{In the general case, $\QQ_0(\kk)$ and $\kk$ are collinear, see footnote \ref{note:coli}. One then takes $\epsilon_{\kk-\qq}\!+\!\hbar\omega_\qq\!-\!\epsilon_\kk\simeq C_k+A_k[1-(\hat{\delta\qq}\cdot\hat{\QQ}_0(\kk))^2](\delta q)^2+ B_k (\hat{\delta\qq}\cdot\hat{\QQ}_0(\kk))^2 (\delta q)^2$ with $\delta\qq\!=\!\qq\!-\!\QQ_0(\kk)$, $\hat{\QQ}=\QQ/Q$ the direction of vector $\QQ$, $A_k=\frac{1}{2K} \frac{\dd\epsilon_k}{\dd k}|_{k=K}+\frac{1}{2 Q_0(k)} \hbar \frac{\dd\omega_q}{\dd q}|_{q=Q_0(k)}$, $B_k=\frac{1}{2} \frac{\dd^2\epsilon_k}{\dd k^2}|_{k=K}+\frac{1}{2}\hbar\frac{\dd^2\omega_q}{\dd q^2}|_{q=Q_0(k)}$, $C_k=\Delta E(\QQ_0(\kk))$ and $K=|\kk-\QQ_0(\kk)|$. Whereas $A_k$ and $B_k$ have a finite limit when $k\to k_c$, $C_k$ tends linearly to zero.} and by approximating the numerator of the integrand by its value $ \propto Q_0(k)$ in $ \qq = \QQ_0 (\kk) $, we find this time that there is a suppression of the spectral weight of the quasi-particle at the instability threshold: when $ k \to k_c $, the value of the integral diverges as $ | \Delta E (\QQ_0 (\kk)) |^{-1/2} $ and the residue $ Z_\kk $ tends to zero as $ | \Delta E (\QQ_0 (\kk))|^{1/2} $. These predictions, in this form, are in agreement with the more advanced, non-perturbative study of reference \cite{Pita1959} on the Green function of an elementary excitation near its instability threshold in a boson gas (even if this reference emphasizes the dispersion relation of the excitation, not its spectral weight). \footnote {Looking more closely in the subsonic case, $\Delta E (\QQ_0 (\kk)) $ approaches zero like $|k-k_c|$ for a dispersion relation $\epsilon_k$ preexisting to phonon coupling (like that of mean field (\ref{eq:011})) and as $ (k-k_c)^2 $ for the true dispersion relation of reference \cite{Pita1959} (which takes into account phonon coupling in a self-consistent way).}

\paragraph{Emission of broken pairs} Our previous stability discussion only takes into account the emission of phonons. In a gas of paired fermions, it neglects the fact that the initial fermionic $\gamma$ quasiparticle can, by collision with bound pairs, break one or more, say a number $ s $, if it has sufficient energy. In this case, the final state contains $ n $ phonons and $ 2s+1 $ fermionic quasi-particles, including the initial quasiparticle which recoiled, and expression (\ref{eq:004}) of the change of energy must be generalized as follows: 
\be
\label{eq:020}
\Delta E = \epsilon_{\kk-\sum_{j=1}^{2s}\kk_j-\sum_{i=1}^{n} \qq_i} + \left(\sum_{j=1}^{2s} \epsilon_{\kk_j}\right) + \left(\sum_{i=1}^{n} \hbar \omega_{\qq_i}\right) -\epsilon_\kk
\ee
Let us show however that the emission of broken pairs does not change the BCS stability map of figure \ref{fig:map}. Suppose indeed that $ s \geq 1 $. As $ \epsilon_{\kk} \geq \Delta_* $ and $ \hbar \omega_{\qq} \geq 0 $ for all wave vectors, we have the lower bound 
\be
\label{eq:021}
\Delta E^{(s\neq 0)}\geq 3 \Delta_* - \epsilon_\kk
\ee
So there can be instability by broken pair emission only if $ \epsilon_\kk> 3 \Delta_* $. We verify however that, within the framework of BCS theory, the zone $ \epsilon_\kk> 3 \Delta_* $ is strictly included in the unstable zone of phonon emission \footnote{For $ \mu> 0 $, we verify first in figure \ref{fig:map}a that the line $ \epsilon_\kk = 3 \Delta $, that is to say $ \Delta = | \hbar^2 k^2/2m-\mu |/\sqrt{8} $, is below the lines of instability CSA and $ v_k = c $ [the only dubious case is the limit $ k/k_\mu \to 0 $, where the dashed line seems to join the green line in $ \Delta/\mu = 1/\sqrt{8} $; the chain of inequalities $ \Delta E_{\rm inf}^{s = 0} (k) \leq \Delta+\hbar \omega_{| k-k_0 |}-\epsilon_k \leq 3 \Delta-\epsilon_k \leq \Delta E^{(s \neq 0)} $ allows to conclude, the first inequality coming from the choice $ n = 1 $ and $ \qq_1 = (1-k_0/k) \kk $ in equation (\ref{eq:004}), the second from $ \hbar \omega_\qq \leq 2 \Delta $ over the whole existence domain of the acoustic branch \cite{CKS, vcrit} and the third from equation (\ref{eq:021})]. Then we check it outside the frame of the figure, using in particular the equivalent given in footnote \ref{note:equiv}. For $ \mu <0 $, we verify numerically that $ v_k> c $ on the line $ \epsilon_\kk = 3 (\Delta^2+\mu^2)^{1/2} $, which is obvious in the BEC limit $ \mu \to-\infty $, and in the limit $ \mu \to 0^-$ taking into account the same footnote \ref{note:equiv}.}. To be complete, we calculate and show in black dashed lines in figure \ref{fig:map} the line of destabilization by emission of a broken pair ($ s = 1 $ and $ n = 0 $ in equation (\ref{eq:020})); it reduces in figure \ref{fig:map}a ($ \mu> 0 $) to $ \epsilon_\kk = 3 \Delta $ because $ k $ is $ <3 k_0 $, and in figure \ref{fig:map}b ($ \mu <0 $) to $ \epsilon_\kk = 3 \epsilon_{\kk/3} $ because $ k \mapsto \epsilon_\kk $ is increasing convex.

\section{Scattering amplitude of the $\gamma$ quasiparticle on a low energy phonon} 
\label{sec:ampdiff} 

In the problem which interests us, the $\gamma$ quasiparticle, with an initial wave vector $ \kk $ ensuring its stability at zero temperature within the meaning of section \ref{sec:stab}, is immersed in the phonon gas of the superfluid at nonzero but very low temperature $ T $, in particular $ k_B T \ll mc^2, \Delta_* $. The $\gamma$ quasiparticle cannot absorb phonons while conserving the energy-momentum, since its group velocity $ v_k $ is subsonic. To see it, it suffices to expand the energy variation after absorption of $ n $ phonons of wave vectors $ \qq_i $ to first order in $ q_i = O (k_B T/\hbar c) $: 
\be
\label{eq:031}
\Delta E_{\rm abs} = \epsilon_{\kk+\sum_{i=1}^{n} \qq_i} -\left( \epsilon_\kk + \sum_{i=1}^{n} \hbar \omega_{\qq_i}\right) \sim 
-\sum_{i=1}^{n} \hbar c q_i \left(1-\frac{v_k}{c} \hat{\kk}\cdot\hat{\qq}_i\right) < 0
\ee
where we have introduced the directions of the wave vectors $ \hat{\kk} = \kk/k $ and $ \hat{\qq}_i = \qq_i/q_i $. On the other hand, nothing prevents the $ \gamma$ quasiparticle {\sl from scattering} phonons, that is to say from absorbing and reemitting a certain nonzero number. At low temperature, the dominant process is the scattering of a phonon, 
\be
\label{eq:032}
|\gamma:\kk,\phi:\qq\rangle\to |\gamma:\kk',\phi:\qq'\rangle \quad\mbox{with}\quad \kk'=\kk+\qq-\qq' \ \mbox{and}\ \qq'\neq\qq
\ee
whose probability amplitude on the energy shell must now be calculated in the limit $ q, q '\to 0 $. \footnote{On may wonder why $ \gamma $ could not scatter an incident phonon of infinitesimal wave vector $ \qq $ in a finite wave vector mode $ \qq '$. However, if this were the case, $ \qq '$ would have a nonzero limit $ \qq_0' $ when $ \qq \to \zero $ and the process $ | \gamma: \kk \rangle \to | \gamma: \kk-\qq_0 ', \phi: \qq_0' \rangle $ would conserve energy, in contradiction with the assumption of stability of the $\gamma$ quasiparticle.} 

\paragraph{The Hamiltonian} For that, we start from the effective low energy Hamiltonian $\hat{H}$ obtained by quantum hydrodynamics for the phononic part and by a local-density approximation, valid in the quasi-classical limit (\ref{eq:003}), for the coupling between phonons and $\gamma$ quasiparticle, in the quantization volume with periodic boundary conditions $ [0, L]^3 $ whose size will be made to tend to infinity \cite{Penco,PRLphigam,LandauKhalatnikov,Annalen}: 
\be
\label{eq:033}
\hat{H}=\hat{H}_0+\hat{V} \quad\mbox{with}\quad \left\{\begin{array}{l} \displaystyle \hat{H}_0 = E_0 +\sum_{\qq\neq\zero}^{q<\Lambda} \hbar \omega_\qq^{(0)} \hat{b}_\qq^\dagger \hat{b}_\qq +\sum_\kk \epsilon_\kk^{(0)} \hat{\gamma}_\kk^\dagger \hat{\gamma}_\kk \\
\displaystyle \hat{V} = \hat{V}_{\phi\phi} + \hat{V}_{\phi\gamma} \quad\mbox{and}\quad \hat{V}_{\phi\gamma} = \hat{H}_3^{\phi\gamma} +\hat{H}_4^{\phi\gamma}
\end{array}
\right.
\ee
Let us content ourselves here with qualitatively describing the different contributions, since their explicit expressions are given elsewhere (in particular in \ref{app:rigor}): 
\begin{itemize} 
\item The interaction-free Hamiltonian $ \hat{H}_0 $ is quadratic in the creation and annihilation operators $ \hat{b}_\qq^\dagger $ and $ \hat{b}_\qq $ of a phonon of wave vector $ \qq $, $ \hat{\gamma}_\kk^\dagger $ and $ \hat{\gamma}_\kk $ of a $\gamma$ quasiparticle of wave vector $ \kk $, operators obeying the usual bosonic commutation relations for phonons, and fermionic anticommutation relations for the $\gamma$ quasiparticle of a superfluid of fermions, for example $ [\hat{b}_\qq, \hat{b}_{\qq '}^\dagger] = \delta_{\qq, \qq'} $. It involves the bare eigenenergies $ \hbar \omega_\qq^{(0)} $ and $ \epsilon_\kk^{(0)} $ of quasiparticles, which will be shifted by the effect of interactions to give true or effective eigenenergies $ \hbar \omega_\qq $ and $ \epsilon_\kk $. It comprises a cutoff $ \Lambda $ on the wave number of the phonons preventing an ultraviolet divergence of these energy shifts; this is inevitable in an effective low energy theory, which cannot describe the effect of interactions for large wave numbers. The simplest cutoff choice here is $ \Lambda = A k_B T/\hbar c $, with a constant $ A \gg 1 $; once the $\phi-\gamma$ scattering amplitude calculated to leading order in temperature, one can make $ A $ tend to $+\infty $ without triggering any divergence, as we will see. 
\item The interaction Hamiltonian $ \hat{V} $ consists of the interaction operator between phonons, denoted $ \hat{V}_{\phi \phi} $, and the interaction operator between phonons and $\gamma$ quasiparticle, noted $ \hat{V}_{\phi \gamma} $. We omit here the interaction operator $ \hat{V}_{\gamma \gamma} $ between $\gamma$ quasiparticles, since there is only one in the system \footnote{In a microscopic theory of fermion gas, it would be different, the effective $\phi-\gamma$ interaction appearing as a sub-diagram of a $\gamma-\gamma$ interaction not conserving the total number of $\gamma$ quasiparticles (except modulo 2) \cite{Zwerger}.}. 
\item The interaction between phonons results to leading order from 3-body processes, of the type $ \hat{b}^\dagger \hat{b}^\dagger \hat{b} $ (Beliaev-like decay of a phonon into two phonons) or $ \hat{b}^\dagger \hat{b} \hat{b} $ (Landau recombination of two phonons in one), which can be resonant (conserve energy-momentum) if the acoustic branch is initially convex, or of the type $ \hat{b}^\dagger \hat{b}^\dagger \hat{b}^\dagger $ or $ \hat{b} \hat{b} \hat{b} $, never resonant. Paradoxically, the interactions $ \hat{b}^\dagger \hat{b} \hat{b} $ and $ \hat{b}^\dagger \hat{b}^\dagger \hat{b} $ between phonons contribute to the $\phi-\gamma$ scattering amplitude to leading order \cite{Penco}; their unfortunate omission in reference \cite{PRLphigam} has been corrected \cite{PRLerr}. To sub-leading orders, $ \hat{V}_{\phi \phi} $ has 4-body \cite{LandauKhalatnikov, Annalen}, 5-body \cite{Khalat5,Khalat5bis}, etc. processes, as quantum hydrodynamics allows in principle to describe, \footnote{\label{note:grad} In order to be consistent, account should then be taken of so-called \og gradient corrections \fg \, in the lower-order terms, in the sense of references \cite{SonWingate, Salasnich}, as it is done in section V.D of reference \cite{Annalen}.} but which do not play a role in our problem. 
\item The interaction of the $\gamma$ quasiparticle with the phonons consists to leading order in an absorption $ \hat{b} \hat{\gamma}^\dagger \hat{\gamma} $ or emission $ \hat{b}^\dagger \hat{\gamma}^\dagger \hat{\gamma} $ process of a phonon; these terms being cubic in the creation and annihilation operators, we store them in $ \hat{H}_3^{\phi \gamma} $. To sub-leading order, it involves the direct scattering of a phonon $ \hat{b}^\dagger \hat{b} \hat{\gamma}^\dagger \hat{\gamma} $, the direct absorption of 2 phonons (double absorption) $ \hat{b} \hat{b} \hat{\gamma}^\dagger \hat{\gamma} $ and the reverse process of double emission $ \hat{b}^\dagger \hat{b}^\dagger \hat{\gamma}^\dagger \hat{\gamma} $, quartic contributions all stored in $ \hat{H}_4^{\phi \gamma} $. It will not be useful here to go beyond this, which the approach of reference \cite{PRLphigam} as such would not allow us to do anyway (see our footnote \ref{note:grad}). 
\item The matrix elements of $ \hat{V}_{\phi \phi} $ in quantum hydrodynamics only depend on the equation of state of the gas at zero temperature, i.e.\ chemical potential $ \mu (\rho) $ considered as a function of the density $ \rho $ in the ground state, and its derivatives with respect to $ \rho $ at fixed interatomic scattering length $ a $.\footnote{As long as we have not got rid of the $ \Lambda $ cut-off, we must take the bare equation of state in the matrix elements \cite{CRASbrou}.} The matrix elements of $ \hat{V}_{\phi \gamma} $ deduced from local density approximation depend on the dispersion relation of the $\gamma$ quasiparticle and its first and second derivatives with respect to $ \rho $ at fixed wave vector $ \kk $ and scattering length $ a $. 
\end{itemize} 

\paragraph{The usual $ S $ matrix} Let us calculate the $\phi-\gamma$ scattering amplitude as the transition probability amplitude between the initial state $ | \mathrm{i} \rangle = | \gamma: \kk, \phi: \qq \rangle $ and the final state $ | \mathrm{f} \rangle = | \gamma: \kk ', \phi: \qq' \rangle $ as in equation (\ref{eq:032}), by the method of the $S$ matrix (section B${}_{\rm III}$.1 of reference \cite{CCTbordeaux}), that is in the limit of an infinite evolution time. As the asymptotic states are taken as eigenstates of $ \hat{H}_0 $, the transition is only allowed if it conserves the corresponding energy, i.e.\  the sum of the bare energies: 
\be
\label{eq:035}
E_{\rm i}^{(0)}\equiv \epsilon_\kk^{(0)}+\hbar\omega_\qq^{(0)} = E_{\rm f}^{(0)}\equiv\epsilon_{\kk'}^{(0)}+\hbar\omega_{\qq'}^{(0)}
\ee
The transition amplitude is then given by the matrix element of the operator $ \hat{T} (z) $ between $ | \mathrm{i} \rangle $ and $ | \mathrm{f} \rangle $ on the energy shell, i.e.\ for $ z = E_{\rm i}^{(0)}+\ii \eta $, $ \eta \to 0^+$: 
\be
\label{eq:036}
A_{\rm fi}^{(0)} = \langle \mathrm{f}| \hat{V} + \hat{V} \frac{1}{E_{\rm i}^{(0)}+\ii \eta-\hat{H}} \hat{V} |\mathrm{i}\rangle
\ee
To leading order at low temperature ($ q, q '= O (T) \to 0 $ at fixed $ k $), a general analysis of the perturbative series in $ \hat{V} $ of result (\ref{eq:036}), exposed in \ref{app:rigor} and which we will come back to, suggests that we can limit ourselves to order two in $ \hat{V} $, that is replace $ \hat{H} $ by $ \hat{H}_0 $ in the denominator of equation (\ref{eq:036}). We then eliminate the remaining sub-leading contributions, as explained in \ref{app:rigor}, to obtain: 
\be
\label{eq:038}
A^{(0)}_{\rm fi} \underset{T\to 0}{\sim} \mathcal{T}_1 + \mathcal{T}_2 +\mathcal{T}_3 + \mathcal{T}_4 + \mathcal{T}_5
\ee
with 
\bea
\label{eq:338}
\mathcal{T}_1 &=& L^{-3}\langle\phi:\qq',\gamma:\kk'|\mathcal{H}_4^{\phi\gamma}|\phi:\qq,\gamma:\kk\rangle \\
\mathcal{T}_2 &=& L^{-3}\frac{\langle \phi:\qq', \gamma:\kk'|\mathcal{H}_3^{\phi\gamma}|\gamma:\kk+\qq\rangle\langle\gamma:\kk+\qq|\mathcal{H}_3^{\phi\gamma}|\phi:\qq,\gamma:\kk\rangle}{\hbar\omega_\qq^{(0)}+\epsilon_\kk^{(0)}-\epsilon_{\kk+\qq}^{(0)}} \\
\mathcal{T}_3 &=& L^{-3}\frac{\langle\gamma:\kk'|\mathcal{H}_3^{\phi\gamma}|\phi:\qq,\gamma:\kk-\qq'\rangle\langle\phi:\qq',\gamma:\kk-\qq'|\mathcal{H}_3^{\phi\gamma}|\gamma:\kk\rangle}{\epsilon^{(0)}_\kk-\hbar\omega^{(0)}_{\qq'}-\epsilon^{(0)}_{\kk-\qq'}} \\
\label{eq:341}
\mathcal{T}_4 &=& L^{-3}\frac{\langle\gamma:\kk'|\mathcal{H}_3^{\phi\gamma}|\phi:\qq-\qq',\gamma:\kk\rangle\langle\phi:\qq',\phi:\qq-\qq'|\mathcal{V}_{\phi\phi}|\phi:\qq\rangle}{\hbar\omega^{(0)}_\qq-\hbar\omega^{(0)}_{\qq-\qq'}-\hbar\omega^{(0)}_{\qq'}}\\
\label{eq:342}
\mathcal{T}_5 &=& L^{-3}\frac{\langle\phi:\qq'|\mathcal{V}_{\phi\phi}|\phi:\qq'-\qq,\phi:\qq\rangle \langle\phi:\qq'-\qq,\gamma:\kk'|\mathcal{H}_3^{\phi\gamma}|\gamma:\kk\rangle}{\epsilon^{(0)}_\kk-\epsilon^{(0)}_{\kk'}-\hbar\omega^{(0)}_{\qq'-\qq}}
\eea
The successive terms in the right-hand side of (\ref{eq:038}) are represented by diagrams (b1) to (b5) of figure \ref{fig:diag}, and the volume matrix elements in the numerator are given by equations (\ref{eq:802a}), (\ref{eq:802b}) and (\ref{eq:802c}) of \ref{app:rigor}, sometimes up to a Hermitian conjugation; in the last term, we used the fact that $ \kk-\kk '= \qq'-\qq $. Bare eigenenergies differ from effective energies by $ O (T^3) $ terms as shown by ordinary perturbation theory\footnote{The energy shift $ \delta \epsilon_\kk $ of a single $\gamma$ quasiparticle or $ \hbar \delta \omega_\qq $ of a single phonon is nonzero to order 2 in $ \hat{V} $. To this order, it is therefore deduced from a one-loop diagram as in figure \ref{fig:diag}a. From equation (\ref{eq:018}), we get $ \delta \epsilon_\kk \simeq \int_{q <\Lambda} \frac{\dd^3 q}{(2 \pi)^3} \frac{| \langle \phi: \qq, \gamma: \kk-\qq | \mathcal{V}_{\phi \gamma} | \gamma: \kk \rangle |^2}{\epsilon_\kk^{(0)}+\ii \eta-(\hbar \omega_\qq^{(0)}+\epsilon_{\kk-\qq}^{(0)})} $; the numerator and the denominator of the integrand are of order $ T $, like the cutoff $ \Lambda $, so $ \delta \epsilon_\kk $ is of order $ T^3 $. Similarly, $ \hbar \delta \omega_\qq \simeq \int_{q <\Lambda} \frac{\dd^3 q '}{(2 \pi)^3} \frac{| \langle \phi: \qq ', \phi: \qq-\qq' | \mathcal{V}_{\phi \phi} | \phi: \qq \rangle |^2/2}{\hbar \omega_\qq^{(0)}+\ii \eta-(\hbar \omega_{\qq '}^{(0)}+\hbar \omega_{\qq-\qq'}^{(0)})} $; the numerator is $ \approx qq '| \qq-\qq' | $ and the denominator is expanded using equation (\ref{eq:001}) keeping the curvature term near the zero angle between $ \qq '$ and $ \qq $, which gives $ \re \hbar \omega_\qq \approx T^5 \ln T $ and, if $ \gamma_\phi> 0 $, an nonzero imaginary part $ \im \hbar \omega_\qq^{(0)} \approx T^5 $.}; however, to leading order for $ A_{\rm fi} $, which is of order one in $ T $, it suffices to expand the numerators and the denominators of the second and third terms of (\ref{eq:038}) up to the sub-leading relative order $ T $ i.e.\ up to order $ T^2 $, the rest can be written directly to leading order \footnote{Indeed, the contributions of the second and third terms exactly cancel to their leading order $ T^0 $, while those of the other three terms are immediately of order $ T $.}. One can thus replace the bare energies by the effective energies in the energy denominators and the matrix elements of (\ref{eq:038}), as well as in the conservation of energy (\ref{eq:035}), which gives exactly expression (3) of reference \cite{PRLerr}. 

Our $ S $ matrix calculation is however not fully convincing. The general analysis of \ref{app:rigor} mentioned above ignores the existence, at orders in $ \hat{V} $ greater than or equal to three, of infinite (and not divergent) diagrams. In these diagrams, one of the energy denominators, giving the difference between $ E_{\rm i}^{(0)} = E_{\rm f}^{(0)} $ and the energy of the intermediate state, is exactly equal to zero, and not on a zero measure set \footnote{If the energy denominator were canceled inside an integral on the wave vector of an internal phonon, the infinitesimal purely imaginary shift $+\ii \eta $ of $ E_{\rm i}^{(0)} $ would give a finite integral in the theory of distributions.}. This phenomenon occurs each time the intermediate state returns to the initial state $ | \mathrm{i} \rangle $ or passes in advance to the final state $ | \mathrm{f} \rangle $. Examples are given in figure \ref{fig:diag}c, to order three in $ \hat{V} $. In addition, as we will see, in the limit of a zero group velocity $ v_k \to 0 $, our scattering amplitude (\ref{eq:038}) is not even in agreement with that of reference \cite{Penco}, which encourages us to be more rigorous. 

\paragraph{Exact asymptotic states} The catastrophic appearance of infinite terms in the perturbative series of amplitude (\ref{eq:036}) is a phenomenon known in quantum field theory and is not surprising. Indeed, the expression of the $S$ matrix at the origin of relations (\ref{eq:035}, \ref{eq:036}) comes from ordinary quantum mechanics, where the total number of particles is a conserved quantity, as in the collision of two atoms. 

Here, on the other hand, the Hamiltonian $ \hat{H} $ conserves the number of $\gamma$ quasiparticles, but not the number of phonons. The $\gamma$ quasiparticle in fact never ceases to interact with the phononic field, even at the moments infinitely anterior to or infinitely subsequent to its collision with the incident phonon, by emitting and reabsorbing {\sl virtual} or {\sl captive} phonons, which the conservation of energy-momentum prevents from escaping to infinity. The correct asymptotic states of the $\gamma$ quasiparticle to consider in scattering theory are therefore its true stationary states $ || \gamma: \kk \rangle $ and $ || \gamma: \kk '\rangle $ of eigenenergy $ \epsilon_\kk $ and $ \epsilon_{\kk '} $, dressed by captive phonons, rather than the bare non stationary states $ | \gamma: \kk \rangle = \hat{\gamma}^\dagger_\kk | \mathrm{vacuum} \rangle $ and $ | \gamma: \kk '\rangle = \hat{\gamma}^\dagger_{\kk'} | \mathrm{vacuum} \rangle $ \footnote{\label{note:defgamhab} To give a precise definition of $ || \gamma: \kk \rangle $, consider the subspace $ \mathcal{E}_\kk $ generated by repeated action of the interaction Hamiltonian $ \hat{V} $ on the vector $ | \gamma: \kk \rangle $, that is to say on a bare $\gamma$ quasiparticle in the presence of the phonon vacuum. $ \mathcal{E}_\kk $ is thus the superposition of states with arbitrary number of phonons and arbitrary wave vectors, in the presence of a bare $ \gamma $ quasiparticle having recoiled. It is stable under the action of the complete Hamiltonian $ \hat{H} $. Under the condition of acoustic stability stated below equation (\ref{eq:006}), we expect that $ \hat{H} $ has in $ \mathcal{E}_\kk $ only one discrete energy $ \epsilon_\kk $, corresponding to the ground state $ || \gamma: \kk \rangle $, and located at the lower edge of a continuum of eigenenergies. $ \epsilon_\kk $ is the pole of the exact propagator in the left-hand side of equation (\ref{eq:018}); the associated residue gives the weight of the bare quasiparticle in the dressed quasiparticle, $ Z_\kk = | \langle \gamma: \kk | \, || \gamma: \kk \rangle |^2 $, and must be $> 0 $.}. Likewise, the incident $ | \phi: \qq \rangle $ or emerging $ | \phi: \qq '\rangle $ phonon never ceases to interact with the phononic field; it can disintegrate virtually into two phonons, which can continue to disintegrate into pairs of phonons or, on the contrary, recombine to restore the initial phonon, etc. These processes are non-resonant, and the phonons created are virtual if the acoustic branch is initially concave [$ \gamma_\phi <0 $ in equation (\ref{eq:001})]; we thus construct, as in footnote \ref{note:defgamhab}, the true stationary states $ || \phi: \qq \rangle $ and $ || \phi: \qq '\rangle $, of eigenenergies $ \hbar \omega_\qq $ and $ \hbar \omega_{\qq '} $, dressed by captive phonons and to be used as correct asymptotic states. The convex case is of another nature, since $ | \phi: \qq \rangle $ and $ | \phi: \qq '\rangle $ are unstable and can really disintegrate into phonons going to infinity; as it is shown in \ref{app:rigor}, we are saved by the slowness (rate $ \approx q^5 $) of this decrease, which formally allows to ignore it in the calculation of the $\phi-\gamma$ scattering amplitude to leading order. 

We must therefore resume the construction of the $S$ matrix using as initial and final states the exact asymptotic states\footnote{Unlike the case of resonant scattering of a photon by a two-level atom initially in the ground state, whose perturbative series has a zero energy denominator each time the atom goes into the excited state in the presence of the radiation vacuum \cite{CCTbordeaux}, we cannot cure the naive $ S $ matrix (\ref{eq:036}) by renormalizing for example the propagator of the incident $\gamma$ quasiparticle by resummation of diagrams with $ n $ disjoint loops (bubbles) (figure \ref{fig:diag}c1 represents the case $ n = 1 $, and the loop in question is that of figure \ref{fig:diag}a). The expansion in bubbles with complex energy $ z $ leads to a geometric series of reason $ \Delta \epsilon_\kk (z-\hbar \omega_\qq^{(0)})/(z-E_{\rm i}^{(0)}) $ where $ \Delta \epsilon_\kk (z) $ is the function giving the eigenenergy shift of $ \gamma $ to order two in $ \hat{V}_{\phi \gamma} $ (for $ L =+\infty $, it is the integral in the denominator of (\ref{eq:018})). The resummation of the bubbles in the initial state $ | \gamma: \kk \rangle $ therefore transforms the amplitude of the diagram of figure \ref{fig:diag}b1 into $ \frac{z-E_{\rm i}^{(0)}}{z-E_{\rm i}^{(0)}-\Delta \epsilon_\kk (z-\hbar \omega_\qq^{(0)})} L^{-3} \langle \phi: \qq ', \gamma: \kk' | \mathcal{H}_4^{\phi \gamma} | \phi: \qq, \gamma: \kk \rangle $, which gives zero if we make $ z $ tend to $ E_{\rm i}^{(0)} $ as prescribed by equation (\ref{eq:036}), or infinity if we heuristically make $ z $ tend to $ E_{\rm i} $ written to order two in $ \hat{V}_{\phi \gamma} $. A resummation of the bubbles on the energy shell both in the final state and in the final state of $ \gamma $ leads to the same conclusion.} 
\be
\label{eq:040}
||\mathrm{i}\rangle= \hat{B}^\dagger_\qq ||\gamma:\kk\rangle\quad\mbox{and}\quad ||\mathrm{f}\rangle= \hat{B}^\dagger_{\qq'} ||\gamma:\kk'\rangle
\ee
where $ \hat{B}_\qq^\dagger $ is the creation operator of a dressed phonon $ || \phi: \qq \rangle $ of wave vector $ \qq $ \footnote{\label{note:empty} To simplify, we limit here the interaction Hamiltonian between phonons $ \hat{V}_{\phi \phi} $ to the Beliaev $ \hat{b}^\dagger \hat{b}^\dagger \hat{b} $ and Landau $ \hat{b}^\dagger \hat{b} \hat{b} $ terms, as in equation (\ref{eq:801a}), so that the phonon vacuum is stationary. This would not be the case if we kept the non-resonant terms $ \hat{b}^\dagger \hat{b}^\dagger \hat{b}^\dagger $ and $ \hat{b} \hat{b} \hat{b} $; it would then be necessary to construct a dressed vacuum $ || \mathrm{vacuum} \rangle $ on which $ \hat{B}_\qq^\dagger $ and $ \hat{B}_{\qq '}^\dagger $ would act.}. In the limit of an infinite evolution time, the transition occurs on the exact (rather than bare) energy shell 
\be
\label{eq:039}
E_{\rm i} \equiv \epsilon_\kk + \hbar \omega_\qq = E_{\rm f} \equiv \epsilon_{\kk'} + \hbar \omega_{\qq'}
\ee
with the probability amplitude 
\be
\label{eq:041}
\boxed{
A_{\rm fi}= \langle \gamma:\kk'|| \hat{B}_{\qq'} \left( [\hat{V}_{\phi\gamma},\hat{B}_\qq^\dagger]+\hat{\Delta}^\dagger_\qq\right) ||\gamma:\kk\rangle
+\langle\gamma:\kk'|| \left([\hat{B}_{\qq'},\hat{V}_{\phi\gamma}]+\hat{\Delta}_{\qq'}\right) \frac{1}{E_{\rm i}+\ii\eta-\hat{H}} 
\left([\hat{V}_{\phi\gamma},\hat{B}_\qq^\dagger]+\hat{\Delta}_\qq^\dagger\right) || \gamma:\kk\rangle
}
\ee
We had to introduce the operator 
\be
\label{eq:042}
\hat{\Delta}_\QQ^\dagger \equiv [\hat{H}_{\phi\phi},\hat{B}_\QQ^\dagger] - \hbar\omega_\QQ \hat{B}_\QQ^\dagger\, \quad\mbox{with}\quad
\hat{H}_{\phi\phi}=\sum_{\qq\neq\zero}^{q<\Lambda} \hbar \omega_\qq^{(0)} \hat{b}_\qq^\dagger \hat{b}_\qq + \hat{V}_{\phi\phi}
\ee
Here $ \hat{H}_{\phi \phi} $ is the purely phononic part of the complete Hamiltonian $ \hat{H} $ and the factor $ \omega_\QQ $ is the exact angular eigenfrequency, not the bare angular frequency. Contrary to appearances, $ \hat{\Delta}_\QQ^\dagger $ is not zero, even if its action on the phonon vacuum gives zero, $ \hat{\Delta}_\QQ^\dagger | \mbox{vacuum} \rangle = 0 $ (see however footnote \ref{note:empty}). The calculation leading to expression (\ref{eq:041}), detailed in \ref{app:rigor}, takes up that of photon-atom scattering in sections B${}_{\rm III}$.2 and B${}_{\rm III}$.3 of reference \cite{CCTbordeaux}, and generalizes it by dressing the asymptotic state of the photon (in other words, of the phonon) and not only that of the atom (in other words, of the $\gamma$ quasiparticle). It is easy to be convinced that the perturbative series of expression (\ref{eq:041}) no longer has infinite diagrams. As the dressed parts of asymptotic states dilute the bare parts in a continuum, they cannot lead to a risk of zero denominator (except on a zero-measure set) and we can, to understand what is happening, focus on the contribution to $ A_{\rm fi} $ of the bare parts $ | \phi: \qq \rangle $ and $ | \phi: \qq '\rangle $, $ | \gamma: \kk \rangle $ and $ | \gamma: \kk '\rangle $: 
\be
\label{eq:742}
A_{\rm fi}|^{\rm bare}_{\rm parts} =(Z_{\kk}Z_{\kk'}\mathcal{Z}_{\qq}\mathcal{Z}_{\qq'})^{1/2} \left[\langle \gamma:\kk'|\hat{b}_{\qq'} [\hat{V}_{\phi\gamma},\hat{b}_\qq^\dagger] |\gamma:\kk\rangle + \langle \gamma:\kk'| [\hat{b}_{\qq'},\hat{V}_{\phi\gamma}] \frac{1}{E_i+\ii\eta-\hat{H}}[\hat{V}_{\phi\gamma},\hat{b}^\dagger_{\qq}] |\gamma:\kk\rangle\right]
\ee
where the quasiparticle residues $ Z_\kk $, $ Z_{\kk'} $, $ \mathcal{Z}_\qq $ and $ \mathcal{Z}_{\qq '} $ are the weights of the bare states in the dressed states, so that $ \hat{B}_\qq^\dagger |^{\rm bare}_{\rm part} = \mathcal{Z}_\qq^{1/2} \hat{b}_\qq^\dagger $ for example. The presence of commutators in equation (\ref{eq:742}) forces the absorption of the incident phonon $ \qq $ as the first event and the emission of the phonon $ \qq '$ as the last event. Indeed, only the terms of $ \hat{V}_{\phi \gamma} $ containing at least one factor $ \hat{b}_\qq $ ($ \hat{b}_{\qq '}^\dagger $) can give a nonzero contribution to the commutator with $ \hat{b}_\qq^\dagger $ ($ \hat{b}_{\qq '} $). But, the absorption of the phonon $ \qq $ erases the memory of the initial state, and before emission of the phonon $ \qq '$, we have no knowledge of the final state, therefore (i) if the we replace $ E_{\rm i} $ by $ E_{\rm i}^{(0)} $ in (\ref{eq:742}), we cannot have zero energy denominators in the perturbative series (except inside an integral on an internal phonon wave vector), and (ii) if we keep the value of $ E_{\rm i} $ in (\ref{eq:742}), we cannot have small denominators (of the order of the difference $ O (\hat{V}^2) $ between the bare and exact eigenenergies) which would invalidate our perturbative expansion. 

Let us now analyze result (\ref{eq:041}) at leading order. The general analysis of \ref{app:rigor}, which leads us to limit (\ref{eq:036}) to order two in $ \hat{V} $, can be applied to it. As each contribution to $ A_{\rm fi} $ is at least of order one in $ \hat{V} $, it suffices to determine the dressing of the $\gamma$ quasiparticle and of the incident or emerging phonon to first order of perturbation theory: 
\bea
\label{eq:743a}
||\gamma:\kk\rangle &=&  |\gamma:\kk\rangle + \frac{1}{L^{3/2}} \sum_{\QQ\neq\zero} \frac{\langle\phi:\QQ,\gamma:\kk-\QQ|\mathcal{H}_3^{\phi\gamma}|\gamma:\kk\rangle}{\epsilon^{(0)}_\kk+\ii\eta-(\epsilon^{(0)}_{\kk-\QQ}+\hbar\omega^{(0)}_\QQ)} |\phi:\QQ,\gamma:\kk-\QQ\rangle+\ldots \\
\label{eq:743b}
\hat{B}_\QQ^\dagger &=& \hat{b}_\QQ^\dagger + \frac{1}{2L^{3/2}} \sum_{\QQ'\neq\zero,\neq\QQ} \frac{\langle \phi:\QQ',\phi:\QQ-\QQ'|\mathcal{V}_{\phi\phi}|\phi:\QQ\rangle}{\hbar\omega^{(0)}_\QQ+\ii\eta-(\hbar\omega^{(0)}_{\QQ'}+\hbar\omega^{(0)}_{\QQ-\QQ'})} \hat{b}_{\QQ'}^\dagger \hat{b}_{\QQ-\QQ'}^\dagger + \ldots
\eea
In writing (\ref{eq:743a}), we immediately neglected the correction due to $ \hat{H}_4^{\phi \gamma} $, that is to say to the double emission of phonons by $ \gamma $ (its matrix elements of a higher order in temperature and the occurrence of a double sum on the wave vectors of the phonons make it negligible in front of the single emission). From expression (\ref{eq:743b}), we get that the operator $ \hat{\Delta}_\QQ^\dagger $ is of order one in $ \hat{V} $ (just like the commutator of $ \hat{B}_\QQ $ or $ \hat{B}^\dagger_\QQ $ with $ \hat{V}_{\phi \gamma} $), and that its action on exact asymptotic state $ || \gamma: \kk \rangle $ is of order two, 
\be
\label{eq:043}
\hat{\Delta}_\QQ^\dagger ||\gamma:\kk\rangle = O(\hat{V}^2)
\ee
Indeed, the action on the bare part of the $\gamma$ quasiparticle gives zero, $ \hat{\Delta}_\QQ^\dagger | \gamma: \kk \rangle = 0 $, since $ \hat{\Delta}_\QQ^\dagger $ is purely phononic and gives zero on the phonon vacuum. In the second contribution to (\ref{eq:041}), the one containing the resolvent of $ \hat{H} $, we can then directly neglect any dressing of the phonons and of the $\gamma$ quasiparticle, that is neglect $ \hat{\Delta}_\qq^\dagger $ and $ \hat{\Delta}_{\qq '} $, replace the dressed operators $ \hat{B}_\qq^\dagger $ and $ \hat{B}_{\qq '} $ by the bare operators $ \hat{b}_\qq^\dagger $ and $ \hat{b}_{\qq'} $, the dressed states $ || \gamma: \kk \rangle $ and $ \langle \gamma: \kk '|| $ by bare states $ | \gamma: \kk \rangle $ and $ \langle \gamma: \kk' | $, and finally keep only $ \hat{H}_3^{\phi \gamma} $ in the interaction $ \hat{V}_{\phi \gamma} $ and approximate $ E_{\rm i}-\hat{H} $ by $ E_{\rm i}^{(0)}-\hat{H}^{(0)} $ in the denominator; we thus find exactly the contribution $ \mathcal{T}_2 $ of the naive theory (\ref{eq:038}), that is to say the non-crossed absorption-emission diagram of figure \ref{fig:diag}b2. In the first contribution to (\ref{eq:041}), we carry out a second order expansion to obtain directly interpretable expressions: 
\begin{multline}
\label{eq:044}
\mbox{first contribution to (\ref{eq:041})}=
\langle\gamma:\kk'| \hat{b}_{\qq'}[\hat{V}_{\phi\gamma},\hat{b}_\qq^\dagger]|\gamma:\kk\rangle+
\langle\gamma:\kk'|\hat{b}_{\qq'}[\hat{H}_3^{\phi\gamma},\hat{b}_\qq^\dagger]||\gamma:\kk\rangle^{(1)}+
{}^{(1)}\langle\gamma:\kk'||\hat{b}_{\qq'}[\hat{H}_3^{\phi\gamma},\hat{b}_\qq^\dagger]|\gamma:\kk\rangle + \\
\langle\gamma:\kk'|\hat{B}_{\qq'}^{(1)}[\hat{H}_3^{\phi\gamma},\hat{b}_\qq^\dagger]|\gamma:\kk\rangle +  
\langle\gamma:\kk'|\hat{b}_{\qq'}[\hat{H}_3^{\phi\gamma},\hat{B}_\qq^{(1)\dagger}]|\gamma:\kk\rangle 
+\langle \gamma:\kk'|\hat{b}_{\qq'}\hat{\Delta}_\qq^\dagger ||\gamma:\kk\rangle^{(1)} +\ldots
\end{multline}
where $ \hat{B}_\QQ^{(1) \dagger} $ is the first deviation, or first order deviation, between the creation operators of a dressed phonon $ \hat{B}_\QQ^\dagger $ and of a bare phonon $ \hat{b}_\QQ^\dagger $, and $ || \gamma: \kk \rangle^{(1)} $ is the first deviation, or deviation of order one, between the dressed state $ || \gamma: \kk \rangle $ and the bare state $ | \gamma: \kk \rangle $ of the $\gamma$ quasiparticle. The first term of (\ref{eq:044}) contains no dressing, and gives exactly the direct scattering diagram $ \mathcal{T}_1 $ of the naive theory (\ref{eq:038}), shown in figure \ref{fig:diag}b1, since the component $ \hat{H}_3^{\phi \gamma} $ of $ \hat{V}_{\phi \gamma} $ gives zero contrarily to $ \hat{H}_4^{\phi \gamma} $. The second term of (\ref{eq:044}) takes into account the dressing of the $\gamma$ quasiparticle in the initial state and gives the contribution $ \mathcal{T}_3 $ of the naive theory, that is to say the crossed diagram of figure \ref{fig:diag}b3; we interpret it physically by saying that the emitted phonon $ \qq '$ was a virtual phonon of the $\gamma$ quasiparticle that the arrival of the phonon $ \qq $ made real, allowing it to escape to infinity without violating conservation of energy-momentum. The third term of (\ref{eq:044}) is zero since the involved transition is scalar vis-\`a-vis the phononic variables, which allows $ \hat{b}_{\qq '} $ to act right on the phonon vacuum and give zero. For the same reason, up to the replacement of $ \hat{b}_{\qq '} $ with $ \hat{B}_{\qq'}^{(1)} $, the fourth term of (\ref{eq:044}) is zero. The fifth term of (\ref{eq:044}) gives $ \mathcal{T}_4 $, represented by the diagram of figure \ref{fig:diag}b4: it corresponds to the absorption by the $\gamma$ quasiparticle of a phonon belonging to a pair of virtual phonons dressing the incident phonon, the other phonon of the pair then being deconfined and leaving at infinity. Finally, the sixth term of (\ref{eq:044}) gives again $ \mathcal{T}_5 $, represented by the diagram of figure \ref{fig:diag}b5: it results from the scattering of the incident phonon on a halo phonon dressing the $\gamma$ quasiparticle. \footnote{To complete these calculations, we only need, with regard to the dressing of phonons, the properties (\ref{eq:043}) and $ \langle \phi: \qq '| \hat{B}_{\qq}^{(1) \dagger} | \phi: \qq '-\qq \rangle = 0 $ and the expression of $ \hat{B}_\qq^{(1) \dagger} | \mbox{vacuum} \rangle $. We can then pretend that $ \hat{B}_{\qq}^{(1)^\dagger} = \hat{Q} \, G_0^\phi (\hbar \omega_\qq^{(0)}+\ii \eta) \hat{V}_{\phi \phi} \hat{b}_\qq^\dagger $, by restriction to the subspace of $ n \leq 1 $ phonon. Here $ \hat{Q} $ is the orthogonal projector on the single phonon space and $ G_0^\phi (z) = (z-\hat{H}_0^{\phi \phi})^{-1} $ is the resolvent of the non interacting part of the purely phononic Hamiltonian. We also give the identity $ \hat{\Delta}^{(1) \dagger} = L^{-3/2} \sum_{\qq '} \langle \phi: \qq, \phi: \qq'-\qq | \mathcal{V}_{\phi \phi} | \phi: \qq '\rangle \hat{b}_{\qq'}^\dagger \hat{b}_{\qq '-\qq} $.} 

To leading order at low temperature, $ q, q '= O (T), T \to 0 $, there is therefore perfect agreement between the phonon-$\gamma$  scattering amplitude (\ref{eq:036}) predicted by ordinary quantum mechanics theory and that (\ref{eq:041}) of field theory with exact asymptotic states, both leading to expression (\ref{eq:038}), which reproduces that of reference \cite{PRLerr} after replacement (legitimate at this order, as we have seen) of the bare eigenenergies by the effective eigenenergies. Consequently, our disagreement with the scattering amplitude of reference \cite{Penco} for a $\gamma$ quasiparticle of zero group velocity $ v_k $, raised by reference \cite{PRLerr} and in contradiction with the hasty note 8 of the same reference \cite{Penco}, persists and remains unexplained. 

\paragraph{Final result at order $ q $} Our calculation of the $\phi-\gamma$ scattering amplitude is based on the \og hydrodynamic\fg\, low energy effective Hamiltonian (\ref{eq:033}) and is only valid to leading order in $ q $. It is therefore useless to keep a dependence on all orders in $ q $ in the energy denominators and in the matrix elements in the numerator of the different terms of expression (\ref{eq:038}). We then pass to the limit $ q \to $ with fixed wave vector $ \kk $ of the $\gamma$ quasiparticle, without assuming (as did references \cite{Penco, PRLphigam}) that its speed $ v_k $ tends to zero; a calculation a little long but without particular difficulty thus gives on the energy shell the (to our knowledge) original result\footnote{\label{note:lourdeur} The heaviness of the calculation comes from the fact that the terms $ \mathcal{T}_2 $ and $ \mathcal{T}_3 $ are of order $ q^0 $ and compensate in the sum (\ref{eq:038}) to leading order, which obliges to expand their numerator and their denominator up to the sub-leading order, that is to say with a relative error $ O (q^2) $ and absolute error $ O (q^3) $ \cite{PRLphigam}. We arrive at a notable simplification for a more symmetrical configuration of the incoming and outgoing wave vectors, by expanding the amplitude $ \mathcal{A} (\gamma: \kk-\frac{\qq-\qq '}{2}, \phi: \qq \to \gamma: \kk+\frac{\qq-\qq '}{2}, \phi: \qq') $ at $ \kk $ fixed, which remains equivalent to the original amplitude to order $ q $. We can thus reduce ourselves to expressions to be expanded that are even or odd in $ \qq $ or $ \qq '$, for example in the energy denominator $ \hbar \omega_\qq+\epsilon_{\kk+\frac{\qq'-\qq}{2}}-\epsilon_{\kk+\frac{\qq '+\qq}{2}} $ of $ \mathcal{T}_2 $, which automatically cancels the odd or even order. In particular, the expression of $ q '$ as a function of $ q $ on the energy shell given in equation (\ref{eq:045}) now presents, as it is, the required sub-sub-leading error $ O (q^3) $. We must also think about replacing $ \epsilon_{\kk_{\rm i}}-\epsilon_{\kk_{\rm f}} $ with $ \hbar \omega_{\qq '}-\hbar \omega_{\qq} $ in the denominator of $ \mathcal{T}_5 $. We then notice that to leading order, we go from $ \mathcal{T}_4 $ to $ \mathcal{T}_5 $ by changing $ | \qq-\qq '| $ to $-| \qq-\qq '| $; to this order, $ \mathcal{T}_4+\mathcal{T}_5 $ is in fact a rational function of $ | \qq-\qq '|^2 $, which makes it possible to eliminate the square root contained in $ | \qq-\qq '| $.}: 
\be
\label{eq:045}
\boxed{\mathcal{A}_{\rm fi}=\mathcal{A}(\gamma:\kk,\phi:\qq\to \gamma:\kk',\phi:\qq')\stackrel{\kk\,\mbox{\scriptsize fixed}}{\underset{q\to 0}{\sim}} 
\frac{\hbar cq}{\rho} \left(\frac{1-u e_k}{1-u' e_k}\right)^{1/2} R_k(u,u',w) \quad\mbox{with}\quad q'\stackrel{\kk\,\mbox{\scriptsize fixed}}{\underset{q\to 0}{=}}\frac{1-u e_k}{1-u' e_k}q+O(q^2)}
\ee
We had to introduce some notations: the volume transition amplitude $ \mathcal{A}_{\rm fi} = L^3 A_{\rm fi} $ (i.e.\ for a unit quantization volume), the cosines of the angles between the three vectors $ \qq $, $ \qq '$ and $ \kk $ (with directions $ \hat{\qq} = \qq/q $, etc), 
\be
\label{eq:046}
u=\hat{\qq}\cdot\hat{\kk}, u'=\hat{\qq}'\cdot\hat{\kk}, w=\hat{\qq}\cdot\hat{\qq}'
\ee
the following dimensionless expression (a symmetric function of $ u $ and $ u '$), called reduced scattering amplitude, 
\begin{multline}
\label{eq:052}
R_k(u,u',w)=\frac{\hbar k}{2 m c} \Bigg\{e_x-w e_\rho -e_k (1+e_k e_\rho) (u-u')^2 \frac{w+\frac{1+\lambda}{2}}{(1-u e_k) (1-u' e_k) (1-w)}\\
+\frac{(u+e_\rho)(u'+e_\rho)[e_k(w-u u')+u u' e_{kk}]+(u+e_\rho) (1-u' e_k)(w+u e_{\rho k}) + (u'+e_\rho) (1-u e_k)(w+u' e_{\rho k})}{(1-u e_k) (1-u' e_k)}
\Bigg\}
\end{multline}
and the dimensionless physical parameters constructed from the chemical potential $ \mu (\rho) $ of the superfluid at zero temperature, from the dispersion relation $ \epsilon_k (\rho) $ of the $\gamma$ quasiparticle and their derivatives with respect to the wave number $ k $ of the quasiparticle or to the density $ \rho $ of the superfluid (at fixed interaction potential between the atoms of mass $ m $ of the superfluid): 
\be
\label{eq:047}
\begin{array}{lllllll}
\displaystyle e_k= \frac{v_k}{c}=\frac{1}{\hbar c}\frac{\partial\epsilon_k}{\partial k} &\hspace{5mm}  & \displaystyle e_\rho= \frac{\rho}{\hbar kc} \frac{\partial\epsilon_k}{\partial\rho} &\hspace{5mm} & \displaystyle e_{\rho\rho}=\frac{\rho^2}{\hbar k c} \frac{\partial^2\epsilon_k}{\partial\rho^2} & \hspace{5mm} & 
\displaystyle \lambda=\frac{\rho\, \dd^2\mu/\dd\rho^2}{\dd\mu/\dd\rho}=2\frac{\dd\ln c}{\dd\ln \rho}-1 \\
\displaystyle e_{kk}=\frac{k}{c}\frac{\partial v_k}{\partial k}=\frac{k}{\hbar c} \frac{\partial^2 \epsilon_k}{\partial k^2} & \hspace{5mm} & \displaystyle e_{\rho k} = \frac{\rho}{c}\frac{\partial v_k}{\partial\rho}=\frac{\rho}{\hbar c}\frac{\partial^2\epsilon_k}{\partial\rho\partial k} & \hspace{5mm}  & e_x=e_{\rho\rho}-\lambda e_\rho &\hspace{5mm} &
\end{array}
\ee
We have related the quantity $ \lambda $, a bit apart because the only one not to depend on the wave number $ k $, to the Gr\"uneisen parameter $ \dd \ln c/\dd \ln \rho $ by means of the exact hydrodynamic relation on the speed of sound, $ mc^2 = \rho \, \dd \mu/\dd \rho $. The quantity $ e_{\rho \rho} $ is only an intermediate of calculation and contributes to the scattering amplitude only through $ e_x $; it will hardly appear in the following. Contrary to appearances, the function $ R_k $ therefore the scattering amplitude have no divergence in $ w = 1 $, taking into account the inequality $ (u-u ')^2 = [\hat{\kk} \cdot (\hat{\qq}-\hat{\qq} ')]^2 \leq (\hat{\qq}-\hat{\qq}')^2 = 2 (1-w) $. From expression (\ref{eq:045}), it is easy to verify that our scattering amplitude obeys the principle of microreversibility to leading order in $ q $ on the energy shell:
\be
\label{eq:053}
\mathcal{A}(\gamma:\kk,\phi:\qq\to \gamma:\kk',\phi:\qq')=
\mathcal{A}(\gamma:\kk',\phi:\qq'\to \gamma:\kk,\phi:\qq)
\ee
taking into account the link between $ q $ and $ q '$ which exists there and the invariance of the function $ R_k $ under the exchange of $ u $ and $ u' $. 

\paragraph{At zero speed} The function $ R_k $ is much simplified in the limit of a zero group velocity for the $\gamma$ quasiparticle. The calculation is simple: we use the expansion (\ref{eq:002}) of the dispersion relation of $ \gamma $ in the vicinity of an extremum to second order to calculate the behavior of quantities (\ref{eq:047}) in this limit. We must take into account the fact that the coefficients $ \Delta_* $, $ k_0 $ and $ m_* $ in equation (\ref{eq:002}) depend on the density $ \rho $, and separate the cases $ k_0 (\rho)> 0 $ and $ k_0 (\rho) \equiv 0 $. (i) If $ k_0 (\rho)> 0 $, only coefficient $ e_k $ {\sl a priori} tends to zero when $ k $ tends to $ k_0 $; the limits of the other coefficients are identified by a tilde: 
\be
\label{eq:048}
\begin{array}{lllll}
\displaystyle e_k \stackrel{k_0>0}{\underset{k\to k_0}{\sim}} \frac{\hbar(k-k_0)}{m_*c}  &\hspace{5mm} & \displaystyle  e_\rho \stackrel{k_0>0}{\underset{k\to k_0}{\to}} \tilde{e}_\rho=\frac{\rho}{\hbar c k_0}  \frac{\dd\Delta_*}{\dd\rho} & \hspace{5mm} & \displaystyle e_x \stackrel{k_0>0}{\underset{k\to k_0}{\to}} \tilde{e}_x=  \frac{\rho^2}{\hbar c k_0} \left[\frac{\dd^2\Delta_*}{\dd\rho^2}-\frac{\dd^2\mu/\dd\rho^2}{\dd\mu/\dd\rho} \frac{\dd\Delta_*}{\dd\rho}+\frac{\hbar^2}{m_*}\left(\frac{\dd k_0}{\dd\rho}\right)^{2}\right] \\
\displaystyle e_{kk} \stackrel{k_0>0}{\underset{k\to k_0}{\to}} \tilde{e}_{kk}=\frac{\hbar k_0}{m_* c} &\hspace{5mm} & \displaystyle e_{\rho k} \stackrel{k_0>0}{\underset{k\to k_0}{\to}}\tilde{e}_{\rho k}=-\frac{\rho\hbar}{m_*c}\frac{\dd k_0}{\dd\rho} &\hspace{5mm} & 
\end{array}
\ee
and we obtain the reduced scattering amplitude in $ k = k_0 $ in agreement with reference \cite{PRLerr} but not with \cite{Penco}: 
\be
\label{eq:049}
R_{k}(u,u',w) \stackrel{k_0>0}{\underset{k\to k_0}{\to}}\frac{\hbar k_0}{2mc} \left[(u+u'+\tilde{e}_\rho)\, w + \tilde{e}_x + \tilde{e}_{\rho k}\, u\, (u+\tilde{e}_\rho) + \tilde{e}_{\rho k}\, u'\, (u'+\tilde{e}_\rho)+ \tilde{e}_{kk}\, u u'\, (u+\tilde{e}_\rho) (u'+\tilde{e}_\rho) \right]
\ee
In the problem of fermion gas which interests us, this corresponds to a positive chemical potential $ \mu $ in the BCS dispersion relation (\ref{eq:011}), considered in the vicinity of its minimum. (ii) If $ k_0 (\rho) \equiv 0 $, coefficients $ e_k $, $ e_{kk} $ and $ e_{\rho k} $ tend linearly to zero when $ k \to k_0 $, and the others diverge as $ 1/k $, with coefficients (marked by a Czech accent) which must be kept track of: 
\be
\label{eq:050}
\begin{array}{lllll}
\displaystyle e_k \stackrel{k_0\equiv 0}{\underset{k\to 0}{\sim}} \frac{\hbar k}{m_* c} &\hspace{5mm} &\displaystyle  \frac{\hbar k}{m c} e_{\rho} \stackrel{k_0\equiv 0}{\underset{k\to 0}{\to}} \check{e}_\rho = \frac{\rho}{m c^2} \frac{\dd\Delta_*}{\dd\rho} & \hspace{5mm} & \displaystyle \frac{\hbar k}{m c} e_{x} \stackrel{k_0\equiv 0}{\underset{k\to 0}{\to}} \check{e}_x = \frac{\rho^2}{mc^2}\left(\frac{\dd^2\Delta_*}{\dd\rho^2}-\frac{\dd^2\mu/\dd\rho^2}{\dd\mu/\dd\rho} \frac{\dd\Delta_*}{\dd\rho}\right) \\
\displaystyle e_{kk}\stackrel{k_0\equiv 0}{\underset{k\to 0}{\sim}}  \frac{\hbar k}{m_* c} & \hspace{5mm} &\displaystyle e_{\rho k}\stackrel{k_0\equiv 0}{\underset{k\to 0}{\sim}}  -\frac{\hbar k\rho}{m_*^2 c}\frac{\dd m_*}{\dd\rho} & \hspace{5mm} &
\end{array}
\ee
to write the final result, which corrects a hasty affirmation of reference \cite{PRLphigam} \footnote{Reference \cite{PRLphigam} thought wrongly to deduce the case $ k \to k_0 \equiv 0 $ from the case $ k \to k_0> 0 $ by making $ k_0 $ tend to zero in its equation (10), therefore in our equation (\ref{eq:049}), see the sentence after this equation (10). In reality, to obtain the equivalent (\ref{eq:049}), we made $ e_k $ tend to zero at fixed $ e_{kk} $, while $ e_{kk} \sim e_k \to 0 $ in the limit $ k \to k_0 \equiv 0 $. As a result, the cancellation of the two terms $ u u '$ in the subexpression $ e_k (wu u')+uu 'e_{kk} $ in square brackets in (\ref{eq:052}) cannot be taken into account (the subexpression is $ \sim e_k w $ and not $ \sim uu 'e_{kk} $ if $ k \to k_0 \equiv 0 $); however, this sub-expression is assigned a divergent weight and gives a nonzero contribution to the final result (\ref{eq:051}). Let us note that footnote \ref{note:plus} extends (\ref{eq:051}) to order one in $ k $.}: 
\be
\label{eq:051}
R_k(u,u',w) \stackrel{k_0\equiv 0}{\underset{k\to k_0}{\to}}  \frac{1}{2} \left[\check{e}_x + w\, \check{e}_\rho \left(1+\frac{m}{m_*}\check{e}_\rho\right)\right]
\ee
In our fermionic problem, this case corresponds to a negative chemical potential $ \mu $ in the BCS dispersion relation (\ref{eq:011}) considered in the vicinity of its minimum, or again to a positive chemical potential $ \mu $ in this same relation, considered this time in the vicinity of its relative maximum in $ k = 0 $; this second situation, of effective mass $ m_* <0 $, is that of the {\it maxon}, seen in figure \ref{fig:map}a to be stable according to BCS theory in a fairly strong interaction regime. 

\paragraph{4-phonon scattering} To end this section, let us emphasize that the central result (\ref{eq:045}) is very general. It applies whatever the quantum statistics of the $\gamma$ quasiparticle or its isotropic dispersion relation $ \epsilon_\kk $. In particular, as reference \cite{Penco} emphasizes, it must describe the scattering of a soft phonon (of infinitesimal wave vector $ \qq $) on a hard phonon (of fixed wave vector $ \kk $, in particular $ k \gg q $), with the production of a soft phonon $ \qq '$ and a hard phonon $ \kk' $. This four-phonon problem has been studied in the context of superfluid helium 4, and the corresponding scattering amplitude can be calculated exactly with quantum hydrodynamics in the limit $ \hbar k \ll mc $ where $ \epsilon_k \sim \hbar ck $ \cite{LandauKhalatnikov}. In this case, only the quantity $ e_{kk} $ tends to zero; $ e_k $ tends to one (sonic limit), $ e_\rho $ and $ e_{\rho k} $ tend to $ (1+\lambda)/2 $, $ e_{\rho \rho} $ tends to $ (\rho^2/c) \, \dd^2c/\dd \rho^2 $, and our scattering amplitude (\ref{eq:045}) effectively reproduces that of equation (111) of reference \cite{Annalen} applied to the linear dispersion relation $ \omega_q = cq $, which constitutes a good test. The four-phonon problem can also be studied in a weakly interacting gas of bosons with the Bogoliubov method, which makes it possible to get rid of the constraint $ \hbar k \ll mc $. At the location of a rotonic minimum ($ k \to k_0> 0 $), reference \cite{halphigamv3} was thus able to successfully test the result of reference \cite{PRLerr}, that is to say {\sl in fine} (\ref{eq:049}), in a situation where it differs from that of reference \cite{Penco}. We have extended this verification to any wave number $ k $ and find a reduced scattering amplitude in perfect agreement with our general prediction (\ref{eq:052}). \footnote{In the notations of reference \cite{Annalen}, the volume transition amplitude $\mathcal{A}_{\rm fi} $ is written $ (4mc^2/ \rho) \mathcal{A}^{2 \leftrightarrow 2, {\rm eff}} (\qq_1, \qq_2, \qq_3, \qq_4) $ where the mass of a boson $ m $ is also denoted $ m_{\rm B} $ and the $ (\qq_j)_{1\leq j\leq 4} $ are the wave vectors of the incident and emerging bosons. Instructed by footnote \ref{note:lourdeur}, we set $\qq_1 = \qq $, $\qq_2 =\kk + (\qq'-\qq) / 2 $, $ \qq_3 = \qq' $, $ \qq_4 = \kk + (\qq-\qq')/2 $. Equation (105) of \cite {Annalen} gives the effective amplitude $ \mathcal{A}^{2\leftrightarrow 2, {\rm eff}} $ on the energy shell ($\epsilon_{q_2}-\epsilon_{q_4} $ has been replaced by $ \epsilon_{q_3} - \epsilon_{q_1} $ in the energy denominators) in terms of the amplitudes of the elementary processes (equations (E18-E20) of \cite{Annalen} in Bogoliubov theory); it remains to expand it to order one in $ q \to 0 $ with wave vector $ \kk $ and directions $ \hat {\qq} $ and $ \hat {\qq}'$ fixed, using computer algebra software.  Let us indicate some simplifying tips: (i) we place ourselves in a system of units such that $ \hbar = m = c = 1 $, (ii) the Bogoliubov spectrum $ \epsilon_Q $ naturally shows the Fourier transform $ \tilde{V}_Q $ of the interaction potential between bosons ($ V(\rr)$ is arbitrary, isotropic, short range), $ \epsilon_Q = [E_Q (E_Q + 2 \rho \tilde{V}_Q)]^{1/2} $ with $ E_Q = \hbar^2 Q^2 / 2m $, but it is better to eliminate $ \tilde{V}_Q $ in terms of $ \epsilon_Q $ as follows, $ \tilde{V}_Q = (\epsilon_Q^2-E_Q^2)/2 \rho E_Q $, (iii) for soft phonons, we can replace $ \epsilon_Q $ by its Taylor expansion (\ref{eq:001}) of order 3 in $ Q = 0 $, and for hard phonons, we can replace $ \epsilon_Q $ by its Taylor expansion of order 3 in $ Q = k $, $ \epsilon_Q \simeq \sum_{n=0}^{3} e_n (k) (Q-k)^n / n! $ (indeed, the numerators and denominators of $ \mathcal{A}^{2 \leftrightarrow 2, {\rm eff}}$ must be expanded to the sub-sub-leading order, since the most divergent diagrams are of order $ q^{-1} $), (iv) if we introduce the internal wave vectors $ \qq_0 = \kk + (\qq + \qq') / 2 $, $ \qq_5 = \qq- \qq'$, $\qq_6 = \kk - (\qq + \qq') / 2 $, $ \mathcal{A}^{2\leftrightarrow 2, {\rm eff}} $ only depends on the moduli of $ (\qq_j)_{0 \leq j \leq 6} $, (v) we provisionally set $ q'= \eta q $ and we expand the $ q_j $ of hard phonons (indices $ j $ even) at fixed $\eta > 0 $ to order $ q^3 $ (without using the value of $\eta $ imposed by energy conservation, which should in principle be determined at the order $q^2 $). 
Remarkably, we then find that the terms of $\mathcal{A}^{2\leftrightarrow 2,{\rm eff}}$ of order $q^{-1}$ and $q^0$ are zero for all $\eta$; in the term of order $ q $, we can finally replace $ \eta $ by its value at order 0, $\eta^{(0)}=(1-u e_1)/(1-u' e_1)$. 
To compare the result to (\ref{eq:052}), it remains to calculate the coefficients (\ref{eq:047}) for the mean field equation of state $ \mu = \rho\tilde{V}_0 = mc^2$ and the Bogoliubov dispersion relation: $e_k=e_1$, $e_\rho=(e_0^2-E_k^2)/2 k e_0$, $e_{\rho\rho}=-(e_0^2-E_k^2)^2/4k e_0^3$, $\lambda=0$, $e_{kk}=k e_2$, $e_{\rho k}=[e_1+(E_k/e_0)^2(e_1-4 e_0/k)]/2$ and $e_x=e_{\rho\rho}$.}

\section{Characterizing the random walk: mean force, momentum diffusion, spatial diffusion} 
\label{sec:carac} 

Our $\gamma$ quasiparticle is now immersed in the phonon thermal gas of the superfluid of arbitrarily low temperature $ T $; by incessant interaction with phonons, it undergoes a stochastic dynamic in the momentum and position space, which we will now describe. 

\paragraph{Master Equation} 
The $\gamma$ quasiparticle is initially prepared in a stable wave vector $ \kk $ in the meaning of section \ref{sec:stab}, in particular with a subsonic group velocity $ v_k $. It cannot therefore emit phonons, it would be an endo-energetic process. For the same reason, it cannot absorb phonons, it would be an exo-energetic process, see equation (\ref{eq:031}). There remains therefore, to leading order in temperature, the process of scattering of a phonon of wave vector $ \qq $ studied in section \ref{sec:ampdiff}. Let us write a master equation on the wave vector distribution $ \Pi (\kk, t) $ of the $\gamma$ quasiparticle, negatively counting the departure processes $ (\kk)+(\qq) \to (\kk ')+(\qq') $ from the mode $ | \gamma: \kk \rangle $ 
and positively the feeding processes $ (\kk ')+(\qq') \to (\kk)+(\qq) $ of this mode, sum being taken at fixed $ \kk $ on all the distinct wave vectors $ \qq $ and $ \qq '$ of the phonons, the wave vector $ \kk' $ being deduced by momentum conservation: 
\bea
\nonumber
\frac{\partial}{\partial t}\Pi(\kk,t) &=& -\frac{1}{L^6} \sum_{\qq,\qq'}
\frac{2\pi}{\hbar} \left|\mathcal{A}(\gamma:\kk,\phi:\qq\to
\gamma:\kk',\phi:\qq')\right|^2
\delta(\epsilon_\kk+\hbar\omega_\qq-\epsilon_{\kk'}-\hbar\omega_{\qq'})\,
\bar{n}_\qq (1+\bar{n}_{\qq'}) \Pi(\kk,t) \\
\label{eq:100}
&& + \frac{1}{L^6} \sum_{\qq,\qq'} \frac{2\pi}{\hbar} \left|\mathcal{A}(\gamma:\kk',\phi:\qq'\to \gamma:\kk,\phi:\qq)\right|^2
\delta(\epsilon_{\kk'}+\hbar\omega_{\qq'}-\epsilon_\kk-\hbar\omega_\qq)\,
\bar{n}_{\qq'} (1+\bar{n}_\qq) \Pi(\kk',t)
\eea
To calculate the rate of the processes, we used Fermi's golden rule, which involves the factor $ 2 \pi/\hbar $ and the Dirac distribution reflecting energy conservation (\ref{eq:039}), including the bosonic amplification factors accompanying absorption (factor $ \bar{n} $) or emission (factor $ 1+\bar{n} $) of a phonon in a mode with thermal mean occupation number $ \bar{n}_\qq = [\exp (\beta \hbar \omega_\qq)-1]^{-1} $, where $ \beta = 1/k_B T $. \footnote{We have taken here a Bose law of zero chemical potential because the total number of phonons $ N_\phi $ is not a constant of motion of the purely phononic dynamics. When the acoustic branch is initially convex, the dominant collisional processes conserving the energy-momentum are the three-phonon ones of Beliaev-Landau with a rate of order $T^5$; they obviously do not conserve $ N_\phi $. In the concave case, the dominant collisional processes are the four-phonon ones $ \phi \phi \to \phi \phi $ of Landau-Khalatnikov, with a rate of order $T^7$, and we must invoke sub-leading processes (the five-phonon ones of \cite{Khalat5,Khalat5bis} with a rate of order $T^{11}$ or the three-phonon ones allowed by non-conservation of energy between unstable states \cite{Castincohfer} with a rate of order $T^9$) to change $ N_\phi $; the phonon gas could then present a thermal pseudo-equilibrium state of chemical potential $ \mu_\phi <0 $ for a duration of order between $T^{-7}$ and $T^{-9}$ or $T^{-11} $.} We recall that the golden rule implicitly deals with the scattering process in the Born approximation. We cannot therefore apply it directly to the interaction Hamiltonian $ \hat{V} $, in this case to its component $ \hat{H}_{4}^{\phi \gamma} $ in equation (\ref{eq:033}), which would amount to taking into account only the direct scattering diagram of figure \ref{fig:diag}b1. Rather, it is applied to an effective interaction Hamiltonian $ \hat{W} $ of the same form as the part $ \hat{b}_{\qq '}^\dagger \hat{b}_\qq $ of $ \hat{H}_{4}^{\phi \gamma} $ in equation (\ref{eq:801c}) but with volumic matrix elements given directly for $ \qq \neq \qq ' $ by the true volume transition amplitude $\mathcal{A} (\gamma:\kk,\phi:\qq\to\gamma:\kk',\phi:\qq')$. \footnote{The more powerful approach of the quantum master equation in the Born-Markov approximation \cite{CCTbordeaux} would also predict a reactive effect of the phonon reservoir on the $\gamma$ quasiparticle, namely a change of its temperature-dependent dispersion relation, in particular under the effect of the diagonal terms $ \qq = \qq '$ and the off-energy-shell terms of $ \hat{W} $, not given here. We can assume in the following that $ \veps_\kk $ is this modified dispersion relation, without this affecting the reduced scattering amplitude $ R_k (u, u ', w) $ to leading order in temperature.} $ \hat{W} $ is a Hermitian operator, due to the microreversibility property (\ref{eq:053}); by construction, it leads in Born's approximation to the desired $\phi-\gamma$ scattering amplitude. The same microreversibility and the relation $ 1+\bar{n} = \exp (\beta \hbar \omega) \, \bar{n} $ ensure that $ \Pi_{\rm st} (\kk) \propto \exp (-\beta \epsilon_\kk) $ is an exact stationary solution of the master equation (\ref{eq:100}): the $\gamma$ quasiparticle is well thermalized by the phonon gas phonons at long times.

\paragraph{Fokker-Planck equation} 
In this work, we always assume that the wave number distribution of the $\gamma$ quasiparticle has a width $ \Delta k $ much larger than that of the thermal phonons, like in equation (\ref{eq:003}). This condition is moreover automatically satisfied when $ \gamma $ is at thermal equilibrium (at sufficiently low temperature), the parabolic dispersion relation (\ref{eq:002}) of $ \gamma $ imposing on $ \Pi_{\rm st} (\kk) $ a width $ \propto T^{1/2} $ much larger than that $ \propto T $ typical of a linear dispersion relation (\ref{eq:001}). In the regime (\ref{eq:003}), the $\gamma$ quasiparticle undergoes with each scattering a change of wave vector $ \delta \kk = \pm (\qq-\qq ') $ very weak compared to $ \Delta k $ and the finite difference equation (\ref{eq:100}) can be simplified into a partial differential equation by second order expansion in $ \delta \kk/\Delta k \approx T^{1/2} $. To do this, we introduce the probability flux leaving $ \kk $ by $\qq\to\qq'$ scattering (a functional of $ \Pi $):
\be
\label{eq:101}
\Phi_t(\kk|\qq,\qq')\equiv \frac{2\pi}{\hbar} 
\left|\mathcal{A}(\gamma:\kk,\phi:\qq\to\gamma:\kk',\phi:\qq')\right|^2
\delta(\epsilon_\kk+\hbar\omega_\qq-\epsilon_{\kk'}-\hbar\omega_{\qq'})
\bar{n}_\qq (1+\bar{n}_{\qq'}) \Pi(\kk,t)
\ee
and, by simple exchange of the dummy variables $ \qq $ and $ \qq '$ in the feeding term then passage to the thermodynamic limit, we rewrite master equation (\ref{eq:100}) as follows: 
\be
\label{eq:102}
\frac{\partial}{\partial t}\Pi(\kk,t) =
\int_{\mathbb{R}^3}\frac{\dd^3q}{(2\pi)^3}\int_{\mathbb{R}^3}\frac{\dd^3q'}{(2\pi)^3} 
\left[\Phi_t(\kk+\qq'-\qq|\qq,\qq')-\Phi_t(\kk|\qq,\qq')\right]
\ee
This form makes obvious conservation of the total probability $ \int \dd^3k \, \Pi (\kk, t)/(2 \pi)^3 $, equal to one, and leads to order two in $ \qq-\qq '$, to the three-dimensional Fokker-Planck equation 
\be
\label{eq:103}
\frac{\partial}{\partial t} \Pi(\kk,t) = -\frac{1}{\hbar}\sum_{i=x,y,z}
\frac{\partial}{\partial k_i}[F_i(\kk) \Pi(\kk,t)] + 
\frac{1}{\hbar^2} \sum_{i=x,y,z} \sum_{j=x,y,z}
\frac{\partial^2}{\partial k_i \partial k_j} [D_{ij}(\kk) \Pi(\kk,t)]
\ee
with the coefficients of the mean force $ \mathbf{F} (\kk) $ and of the momentum diffusion matrix $ \underline{\underline{D}} (\kk) $ in the Cartesian basis, 
\be
\label{eq:104}
F_i(\kk)=\int\frac{\dd^3q\,\dd^3q'}{(2\pi)^6} \hbar(q_i-q_i') 
\frac{\Phi_t(\kk|\qq,\qq')}{\Pi(\kk,t)} \quad\mbox{and}\quad
D_{ij}(\kk)= \int\frac{\dd^3q\,\dd^3q'}{(2\pi)^6} \frac{1}{2}\hbar^2(q_i-q'_i)(q_j-q'_j)
\frac{\Phi_t(\kk|\qq,\qq')}{\Pi(\kk,t)}
\ee
independent of $ \Pi (\kk, t) $ and time. In this form, the matrix $ \underline{\underline{D}} (\kk) $ is clearly symmetric real positive. 

\begin{figure} [t] 
\centerline{\includegraphics [width = 0.23 \textwidth, clip =]{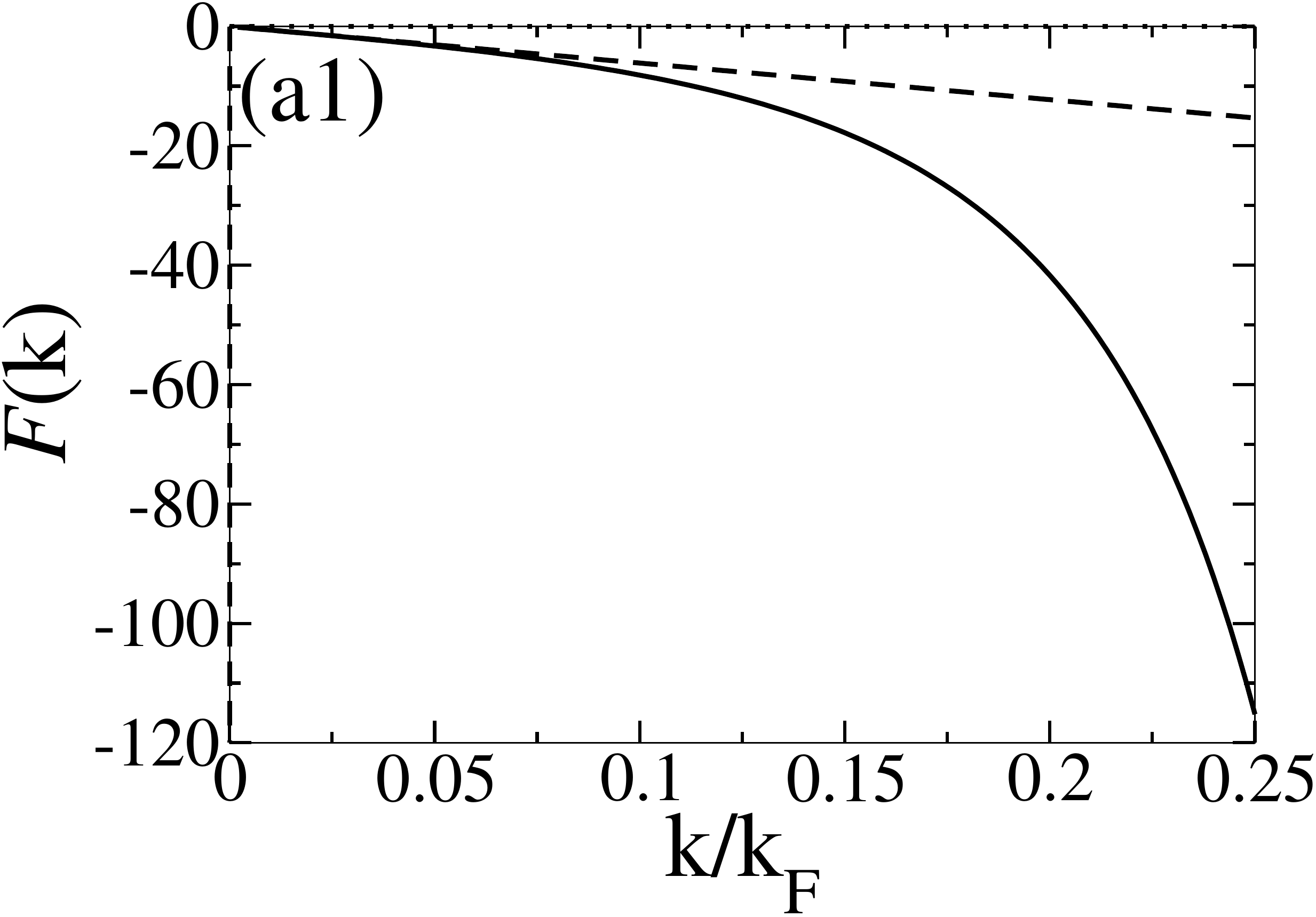} \hspace{3mm} \includegraphics [width = 0.23 \textwidth, clip =]{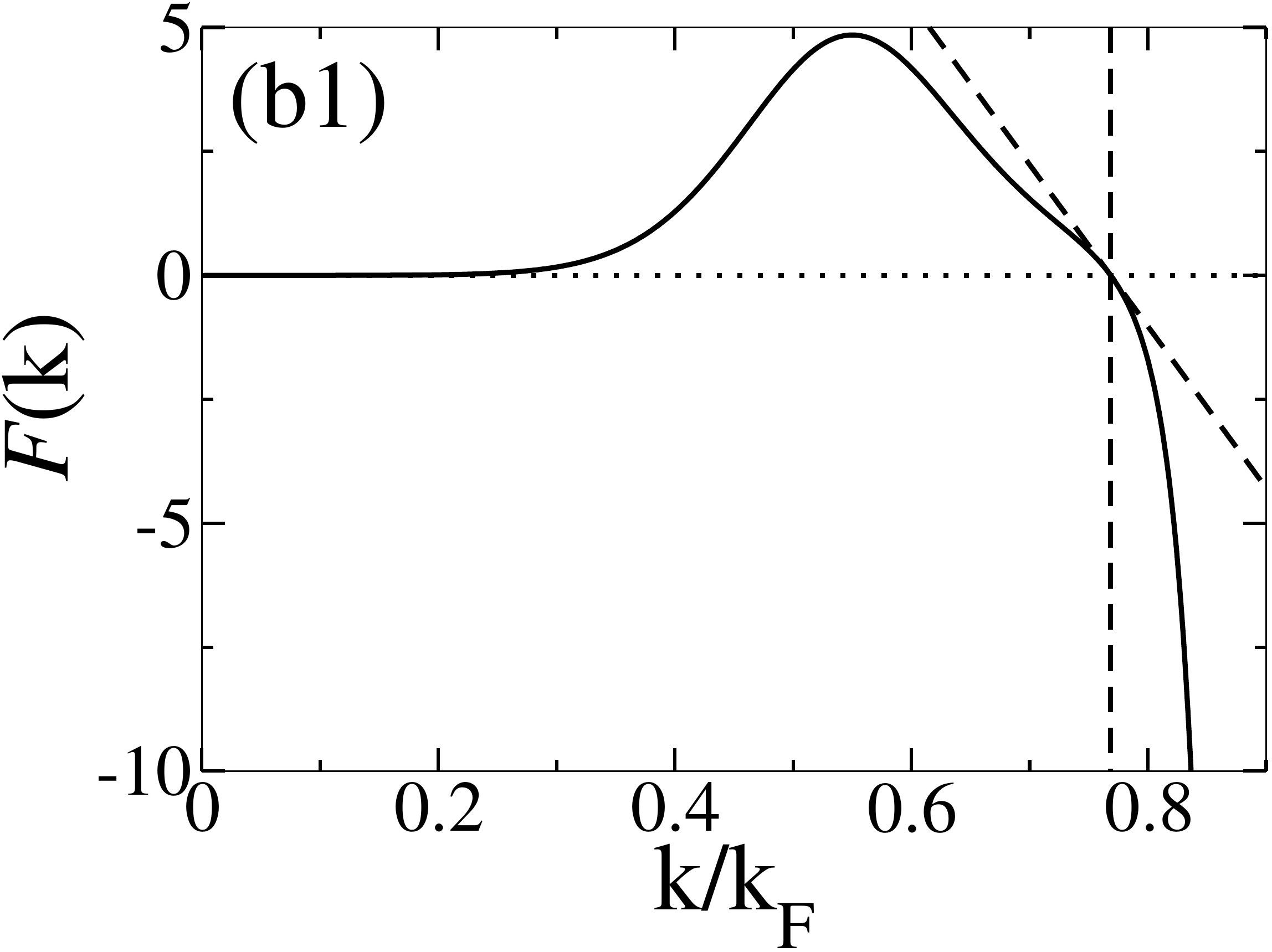} \hspace{3mm} \includegraphics [width = 0.23 \textwidth, clip =]{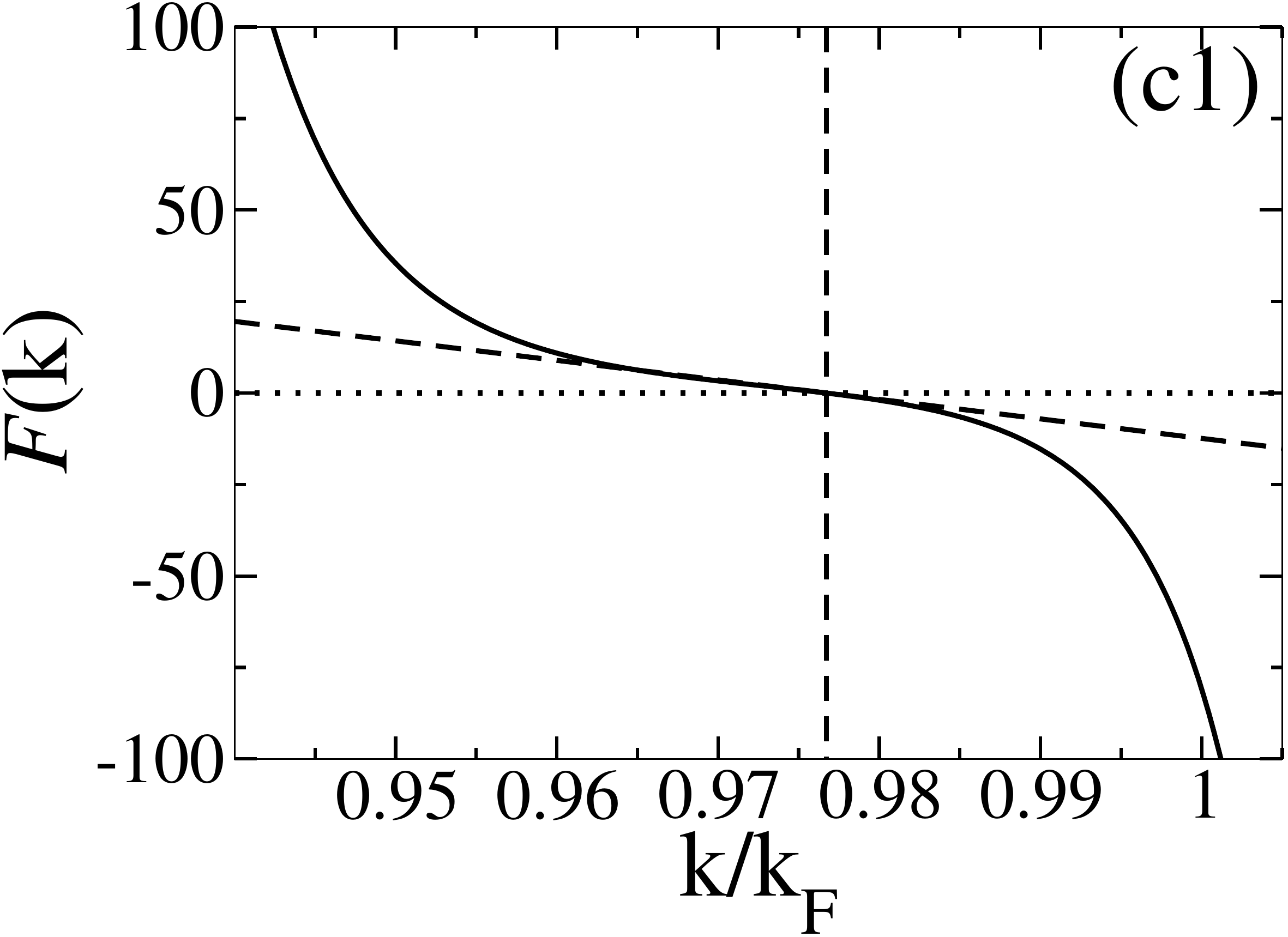} \hspace{3mm} \includegraphics [width = 0.23 \textwidth, clip =]{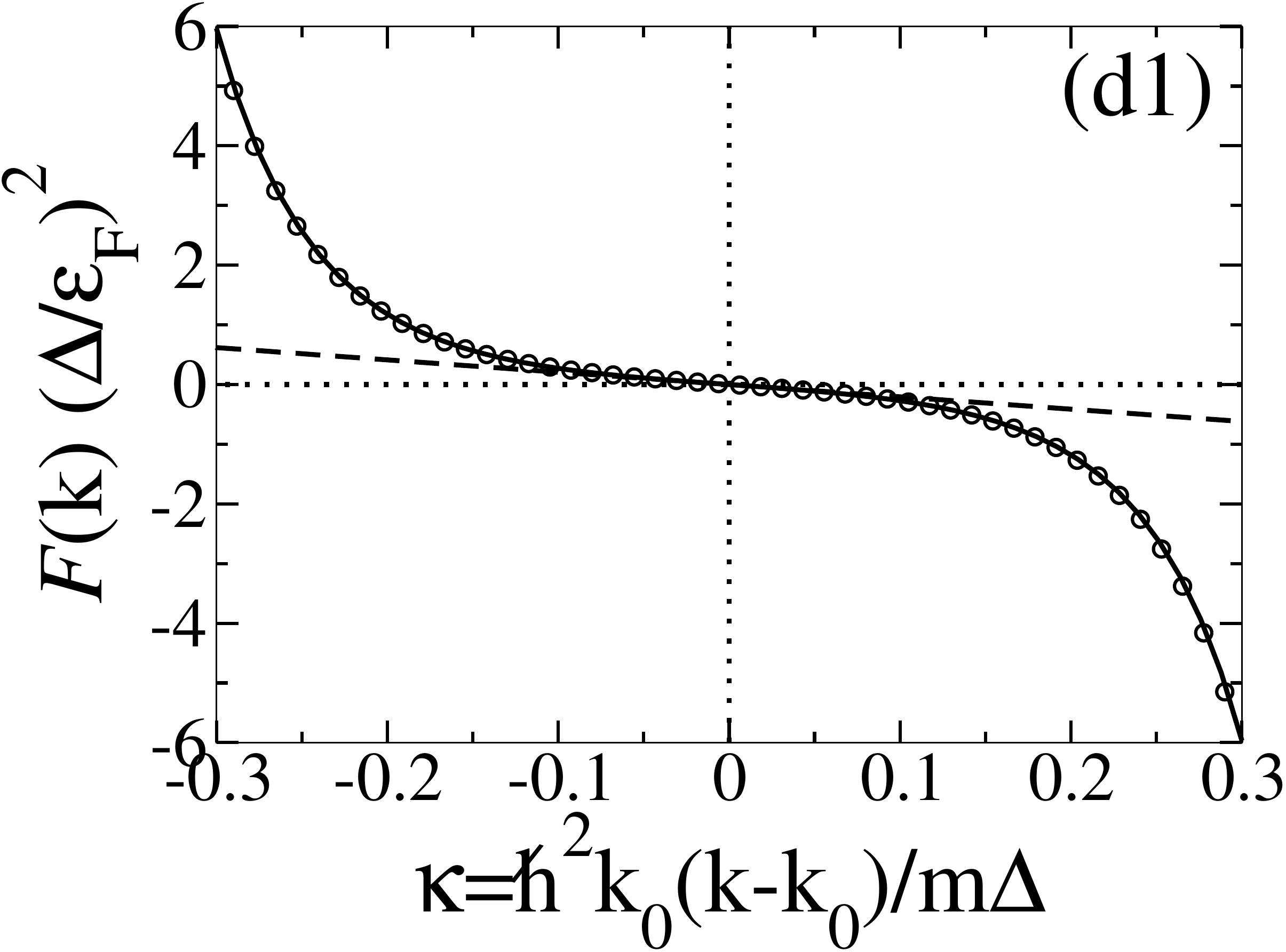}} 
\centerline{\includegraphics [width = 0.23 \textwidth, clip =]{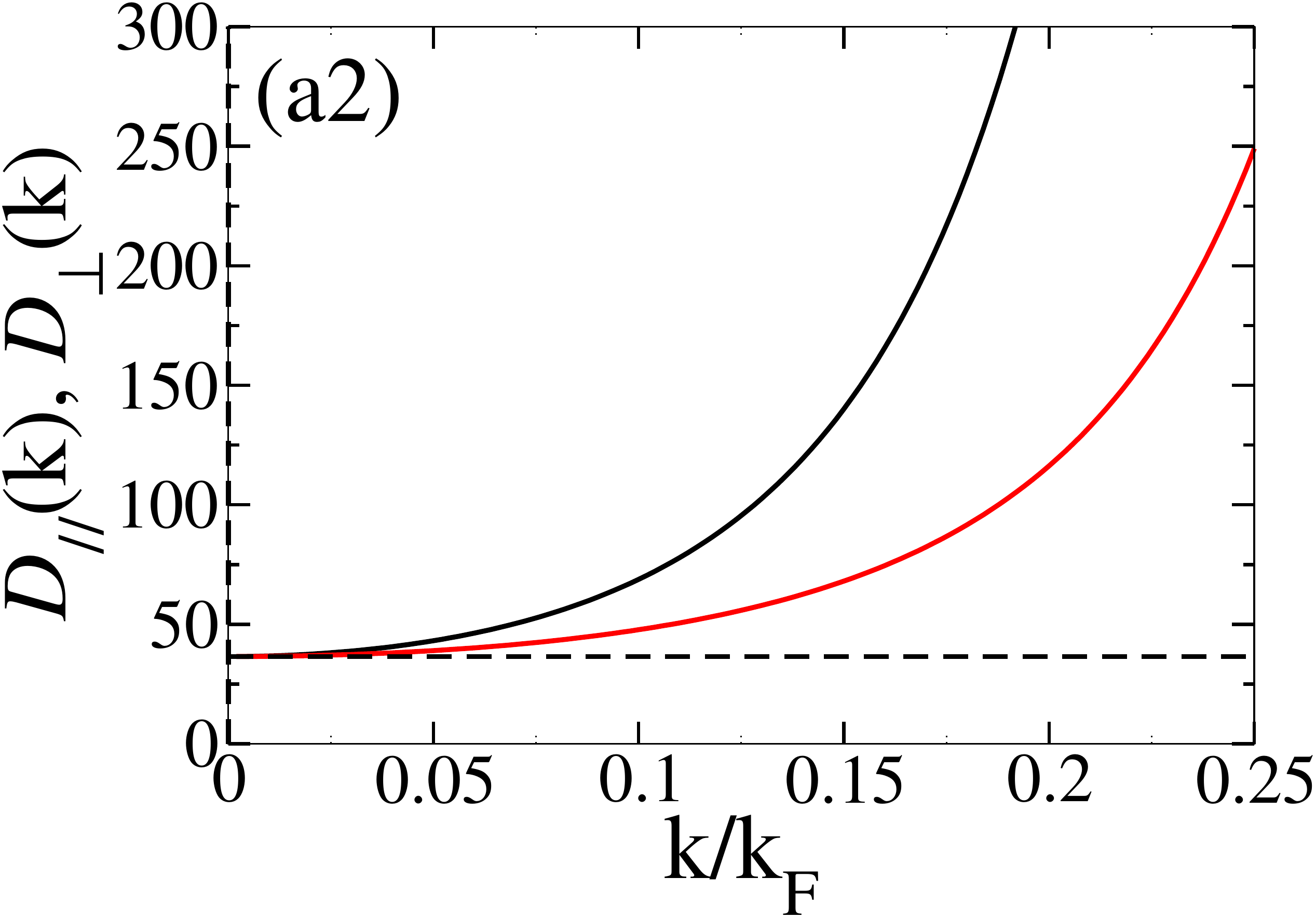} \hspace{3mm} \includegraphics [width = 0.23 \textwidth, clip =]{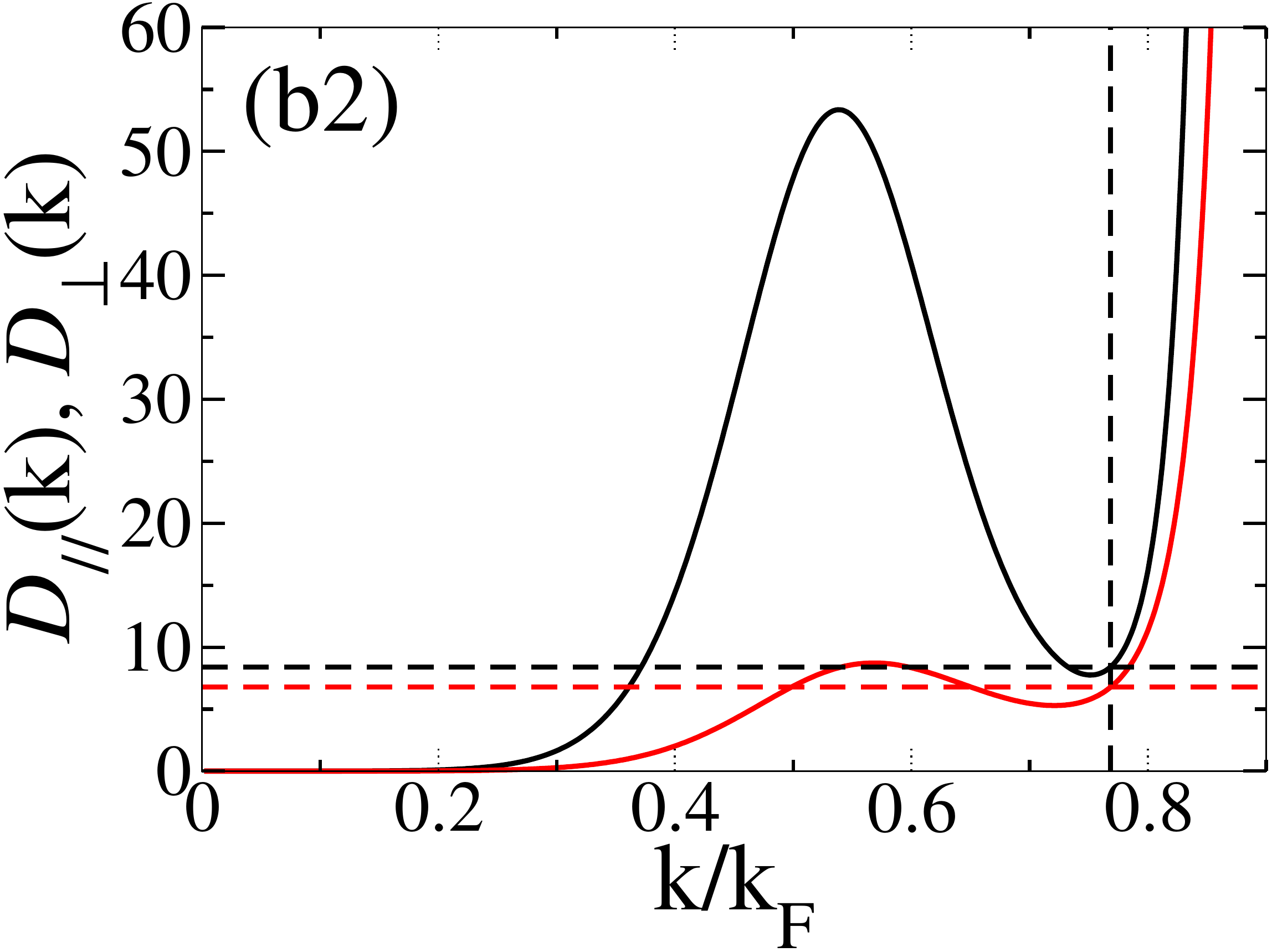} \hspace{3mm} \includegraphics [width = 0.23 \textwidth, clip =]{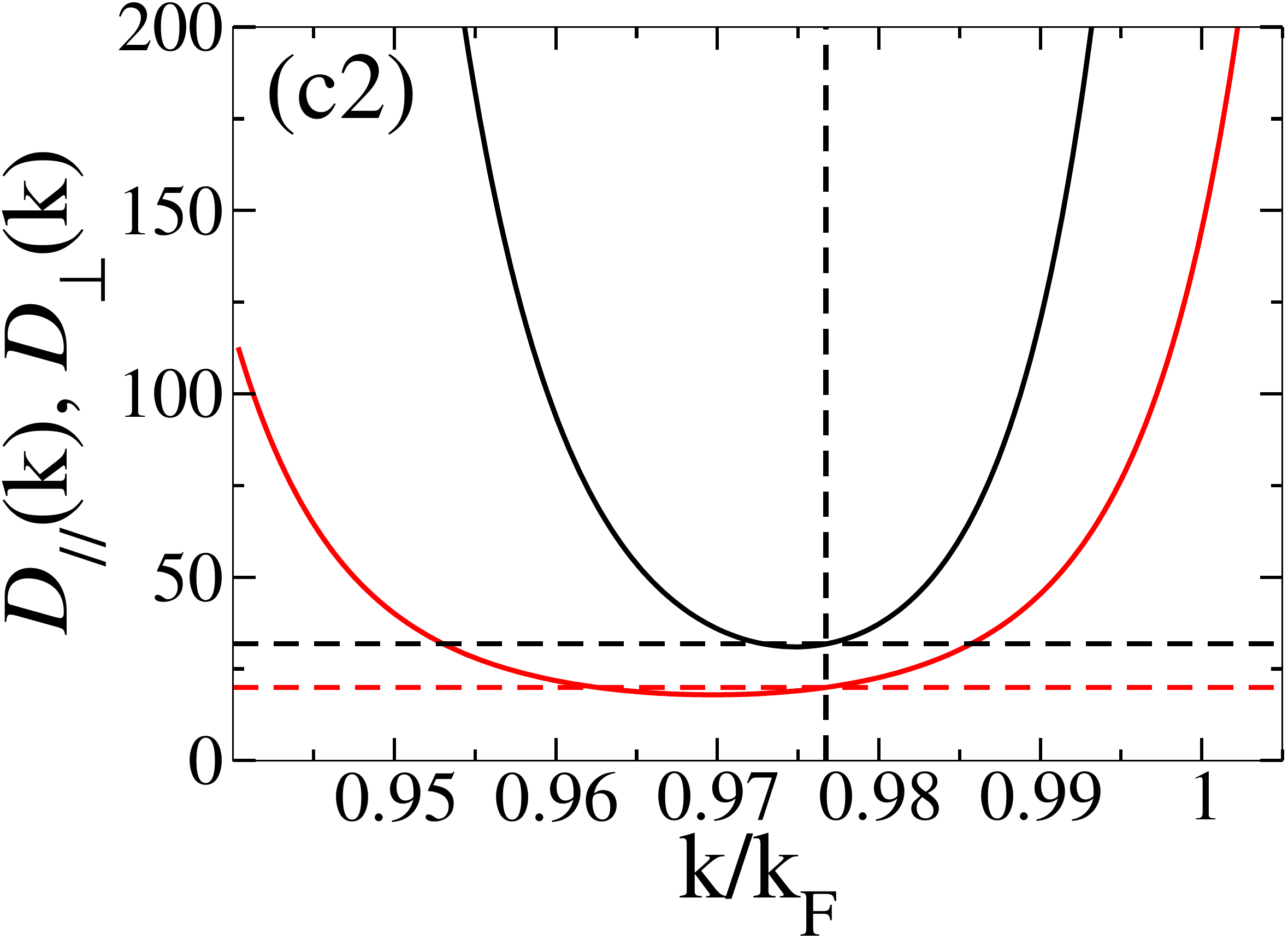} \hspace{3mm} \includegraphics [width = 0.23 \textwidth, clip =]{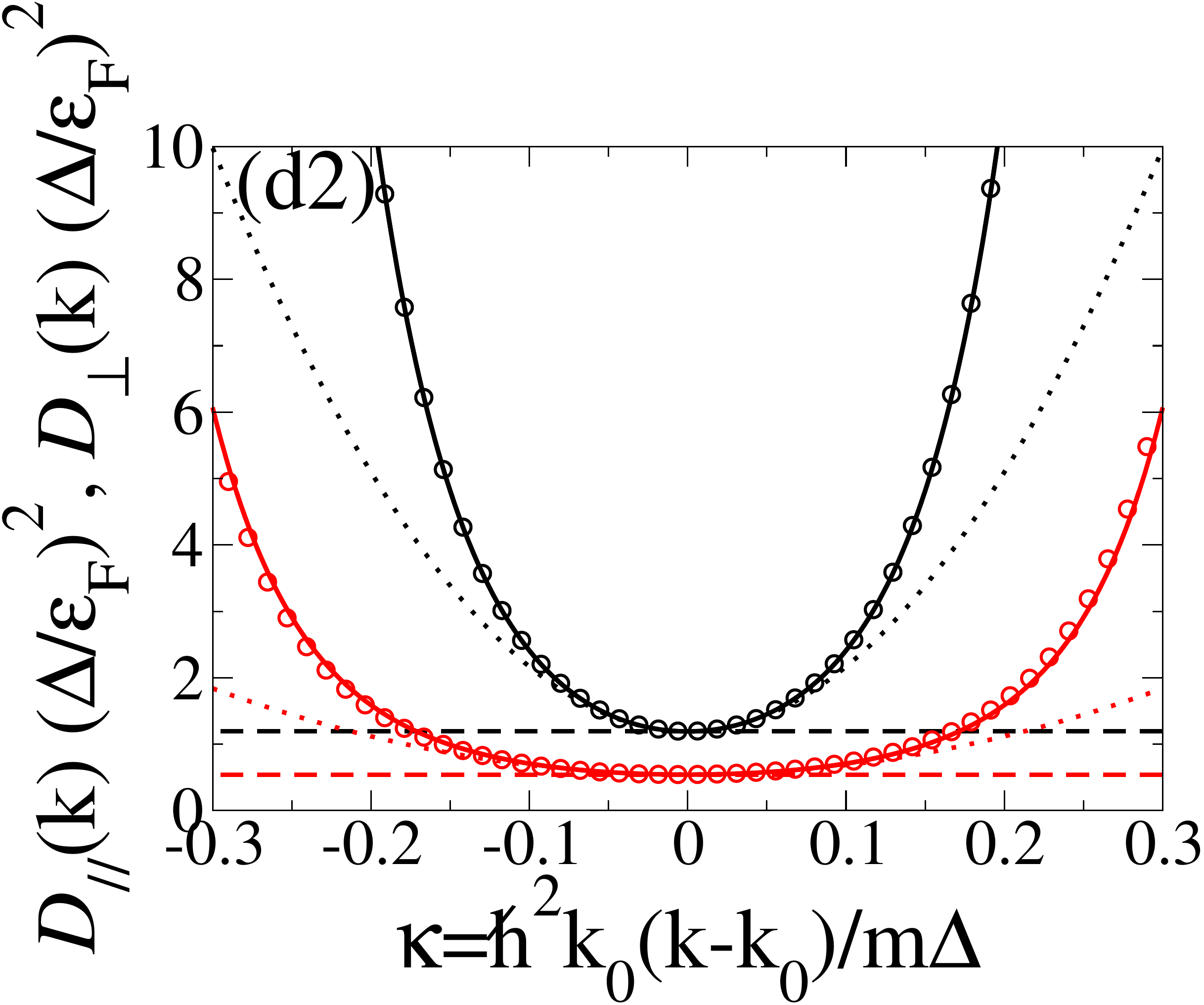}} 
\caption{Mean force $ \mathcal{F} $ (black) (upper row), longitudinal $ \mathcal{D}_{\sslash} $ (black) and transverse $ \mathcal{D}_{\perp} $ (red) momentum diffusion coefficients (lower row), after extraction of the leading order in temperature and rescaling as in equation (\ref{eq:106}), for a fermionic $\gamma$ quasiparticle  in an unpolarized pair-condensed gas of fermions at very low temperature, as functions of the quasiparticle wave number $ k $. We have deduced the coefficients (\ref{eq:047}) from the gas equation of state (see footnote \ref{note:equet}) and from the dispersion relation (\ref{eq:011}) of $ \gamma $ in the BCS approximation, then we have numerically integrated equations (\ref{eq:108a}, \ref{eq:108b}). In the BEC-BCS crossover: (a) on the BEC side $ 1/k_{\rm F} a = 1 $ ($ \mu/\epsilon_{\rm F} \simeq-0.801 $, $ \Delta/\epsilon_{\rm F} \simeq $ 1.332), (b) in the unitary limit $ 1/k_{\rm F} a = 0 $ ($ \mu/\epsilon_{\rm F} \simeq $ 0.591, $ \Delta/\epsilon_{\rm F} \simeq $ 0.686), (c) on the BCS side $ 1/k_{\rm F} a =-1 $ ($ \mu/\epsilon_{\rm F} \simeq 0.954 $, $ \Delta/\epsilon_{\rm F} \simeq $ 0.208). Vertical dashed line: position $ k_0 $ of the minimum of the dispersion relation. Oblique or horizontal dashed line: analytical predictions (\ref{eq:115}, \ref{eq:116}, \ref{eq:117a}, \ref{eq:117b}, \ref{eq:119}) and linear approximation $ e_k \simeq \hbar (k-k_0)/m_* $ in equation (\ref{eq:115}); in (b1), the nodal point of $ \mathcal{F} $ is not an inflection point. $ k_{\rm F} = (3 \pi^2 \rho)^{1/3} $ is the Fermi wave number of the gas, $ \epsilon_{\rm F} = \hbar^2 k_{\rm F}^2/2m $ its Fermi energy, $ \rho $ its uniform density, $ \mu $ its chemical potential and $ \Delta> 0 $ its order parameter at zero temperature, $ a $ the $s$-wave scattering length between opposite spin fermions of mass $ m $. In the BCS limit (d) $ k_{\rm F} a \to 0-$, universal limiting behaviors deduced from equation (\ref{eq:109}) after multiplication of the force, diffusion and deviation of $ k $ from $ k_0 $ by well-chosen powers of $ \Delta $, in solid line [the limiting law for $ \mathcal{F} $ is explicit, see equation (\ref{eq:199}), the one for $ \mathcal{D}_{\sslash} $ or $ \mathcal{D}_\perp $ requires numerical integration]; oblique or horizontal dashed lines: as in (a), (b) and (c) but for $ \Delta/\epsilon_{\rm F} \to 0 $; dashed curves for diffusion: quadratic approximations (\ref{eq:199}); circles: $ 1/k_{\rm F} a =-3 $ ($ \Delta/\epsilon_{\rm F} \simeq 9.72 \times 10^{-3} $) taken numerically from equations (\ref{eq:108a}, \ref{eq:108b}).} 
\label{fig:fdk} 
\end{figure}

\paragraph{Force and momentum diffusion to leading order in $ T $} 
As the $\gamma$ quasiparticle of wave vector $ \kk $ evolves in a homogeneous and isotropic medium, the mean force and the momentum diffusion which it undergoes must be rotationally invariant around the direction $ \hat{\kk} $ of the wave vector. They are therefore characterized by functions of the wave number $ k $ only, namely the longitudinal component $ F (k) $ of the force, the longitudinal $ D_{\sslash} (k) $ and transverse $ D_\perp (k) $ momentum diffusion coefficients: 
\be
\label{eq:105}
F_i(\kk) = \frac{k_i}{k} F(k) \quad\mbox{and}\quad D_{ij}(\kk)=\frac{k_i k_j}{k^2}
D_{\sslash}(k) + \left(\delta_{ij}-\frac{k_i k_j}{k^2}\right) D_\perp(k)
\ee
To leading order, we can easily get the temperature dependence of these coefficients by expressing the phonon wave vectors in units of $ k_B T/\hbar c $ in the six-fold integrals of equation (\ref{eq:104}) and using the equivalent (\ref{eq:045}) of $ \mathcal{A} $ to obtain the power laws (the numerical factor $ \pi^5/15 $ is pure convenience) 
\be
\label{eq:106}
F(k) \underset{k_B T/mc^2\to 0}{\sim} 
\frac{\pi^5}{15} \frac{\hbar c}{\rho^2} 
\left(\frac{k_B T}{\hbar c}\right)^8\mathcal{F}(k)
\quad\mbox{and}\quad
D_{\perp,\sslash}(k) \underset{k_B T/mc^2\to 0}{\sim}
\frac{\pi^5}{15} \frac{\hbar^2c}{\rho^2}\left(\frac{k_B T}{\hbar c}\right)^9
\mathcal{D}_{\perp,\sslash}(k)
\ee
The remaining functions $ \mathcal{F} (k) $ and $ \mathcal{D}_{\perp, \sslash} (k) $ are dimensionless. To calculate them, we move to spherical coordinates of polar axis $ \hat{\kk} $ on the integration variables $ \qq $ and $ \qq '$. Energy-conserving Dirac distributions allow easy integration on the modulus $ q '$, {\sl in fine} linked to $ q $ by (\ref{eq:045}). The integral on the modulus $ q $ of the contributions that are linear in the occupation numbers pulls out the Riemann $\zeta$ function; that of quadratic contributions brings out less studied functions ($ s $ is an integer $> 2 $, $ x $ a real number $> 0 $): 
\be
\label{eq:107}
\Phi_s(x) = \int_0^{+\infty} \frac{\dd Q \, Q^s}
{\left(\eee^{Q\sqrt{x}}-1\right)\left(\eee^{Q/\sqrt{x}}-1\right)}=
\sum_{m=1}^{\infty}\sum_{n=1}^{+\infty} 
\frac{s!}{(m\sqrt{x}+n/\sqrt{x})^{s+1}}= \Phi_s(1/x)
\ee
Due to invariance of the integrand by joint rotation of $ \qq $ and $ \qq '$ around $ \kk $, one can reduce to an integral on a single azimuthal angle, the relative angle $ \varphi = \phi-\phi' $. In the force, a difficult contribution involving the function $ \Phi_7 $ turns out to be an antisymmetric function of the cosines $ u $ and $ u '$ of the polar angles, of zero integral \footnote{Up to a factor, it is $ (u-u') \, \Phi_7 (\frac{1-u e_k}{1-u 'e_k})/[(1-u e_k) (1-u' e_k)] $.}. Such simplification does not occur in momentum diffusion. It remains with the notations (\ref{eq:046}, \ref{eq:052}, \ref{eq:047}) \footnote{In particular, $ w = u \, u '+[(1-u^2) (1-u '^ 2)]^{1/2} \cos \varphi $. \label{note:w}}: 
\bea
\label{eq:108a}
\mathcal{F}(k) &=& \int_{-1}^1 \dd u \int_{-1}^1 \dd u'  (u-u') 
\frac{(1-u e_k)^3}{(1-u' e_k)^5} 
\int_0^{2\pi} \frac{\dd\varphi}{2\pi} [R_k(u,u',w)]^2 \\
\binom{\mathcal{D}_\sslash(k)}{\mathcal{D}_\perp(k)} 
&=& \frac{15}{16\pi^8} \int_{-1}^1 \dd u \int_{-1}^1 \dd u' 
\left[\frac{(1-u e_k)^3}{(1-u'e_k)^6} 8!\, \zeta(9) 
+\frac{\Phi_8(\frac{1-u e_k}{1-u'e_k})}{[(1-u e_k)(1-u' e_k)]^{3/2}} \right]  
\nonumber\\
\label{eq:108b}
&\times& \int_0^{2\pi} \frac{\dd\varphi}{2\pi} [R_k(u,u',w)]^2
\binom{(u'-u)^2}{(u'-u)^2 \frac{e_k^2-1}{2}+(1-w) (1-u e_k)(1-u' e_k)}
\eea
Triple integration in (\ref{eq:108a}) can be done analytically\footnote{\label{note:diffi} The only slightly difficult integrals are $ \int_0^{2 \pi} \dd \varphi/(1-w) = 2 \pi/| u-u '| $ and $ \int_0^{2 \pi} \dd \varphi/(1-w)^2 = 2 \pi (1-u u ')/| u-u' |^3 $ where $ w $ is written as in footnote \ref{note:w}.}; the mean force thus has an explicit expression in the parameters (\ref{eq:047}), combination of rational functions and a logarithm, unfortunately too long to be written here. The momentum diffusion coefficients must be evaluated numerically. In the BCS approximation, where the dispersion relation $ \epsilon_k $ and the equation of state at zero temperature have an explicit analytical form (see our equation (\ref{eq:011}) and reference \cite{Strinati}) \footnote{\label{note:equet} We have made the expressions of reference \cite{Strinati} more compact using the properties of the elliptical integrals $ E (\ii x) = | 1+\ii x | E (x/| 1+\ii x |) $ and $ K (\ii x) = K (x/| 1+\ii x |)/| 1+\ii x | $ valid for all $ x \in \mathbb{R} $. By setting $ \mu/\Delta = \sh \tau $, it comes then $-\pi/2a = (2m \Delta/\hbar^2)^{1/2} I_1 $ and $ \rho = (2m \Delta/\hbar^2)^{3/2} I_2/(2 \pi^2) $ with $ I_1 = (2 \eee^{-\tau})^{1/2} [\eee^\tau \ch \tau \, K (\ii \eee^\tau)-E (\ii \eee^\tau)] $ and $ I_2 = (2 \eee^{-\tau}/9)^{1/2} [\sh \tau \, E (\ii \eee^\tau)+\ch \tau \, K (\ii \eee^\tau)] $. In addition, $ e_k = (\hbar k/mc) (\xi_k/\epsilon_k) $, $ e_{kk} = (\hbar k/mc) [\xi_k^3+\Delta^2 (3 \xi_k+2 \mu)]/\epsilon_k^3 $, $ e_{\rho k} =-(\hbar k/mc) \Delta (\xi_k \rho \Delta '+\Delta \rho \mu')/\epsilon_k^3 $, $ e_\rho = (\Delta \rho \Delta '-\xi_k \rho \mu')/(\hbar ck \epsilon_k) $, $ e_{\rho \rho} = [(\Delta \rho \mu '+\xi_k \rho \Delta')^2+\epsilon_k^2 (\Delta \rho^2 \Delta ''-\xi_k \rho^2 \mu '')]/(\hbar ck \epsilon_k^3) $. We noted with a prime the derivate with respect to the density $ \rho $ at fixed  scattering length $ a $ and we setted $ \xi_k = \hbar^2 k^2/(2m)-\mu $.}, we show in figure \ref{fig:fdk} the dependence in wave number of the force and the diffusion coefficients for various interaction regimes. 

In the so-called BCS limit $ k_{\rm F} a \to 0^-$, where the BCS approximation is the most quantitative, simple results can be obtained by making $ \Delta/\epsilon_{\rm F} $ tend to zero at fixed $ \kappa = \hbar^2 k_0 (k-k_0)/m \Delta $. 
The speed of sound becomes proportional to the Fermi velocity, $ c \sim \hbar k_{\rm F}/m \sqrt{3} $, the wave number of the minimum of the dispersion relation merges with the Fermi wave number $ k_0 \sim k_{\rm F} $, the $\gamma$ quasiparticle is subsonic and stable as long as $ | \kappa | <1/\sqrt{2} $ and the reduced $\phi-\gamma$ scattering amplitude, dominated by $ \mathcal{T}_1+\mathcal{T}_2+\mathcal{T}_3 $ in expression (\ref{eq:038}) (the diagrams omitted in reference \cite{PRLphigam} become negligible), takes the very simple $w$-independent form: \footnote{To be complete, we give $ e_\rho \sim-e_k/3 $, $ \lambda \to-1/3 $, $ e_{kk} \sim (2 \sqrt{3} \epsilon_{\rm F}/\Delta) (1+\kappa^2)^{-3/2} $, $ e_{\rho k} \sim-e_{kk}/3 $, $ e_{\rho \rho} \sim e_{kk}/9 $.} 
\be
\label{eq:109}
R_k(u,u',w) \stackrel{\kappa\ \mbox{\scriptsize fixed}}
{\underset{k_{\rm F} a\to 0^-}{\sim}} 
\frac{3\epsilon_{\rm F}/\Delta}{(1+\kappa^2)^{3/2}}
\frac{(u^2-1/3)(u'^2-1/3)}{(1-u e_k)(1-u' e_k)}
\quad\mbox{with}\quad e_k \sim \frac{\sqrt{3}\kappa}{(1+\kappa^2)^{1/2}}
\ee
It then becomes reasonable to give the analytical expression of the force and an expansion of the momentum diffusion coefficients at low speed: \footnote{In the limit (\ref{eq:109}), one can also reduce the contribution of $ \Phi_8 $ in (\ref{eq:108b}) to a single integral at the cost of introducing Bose functions or polylogarithms $ g_s(z) = \sum_{n \geq 1} z^n/n^s$ (the remainder, analytically calculable, is a rational function of $ e_k $): by setting $ f_s (u) \equiv g_s (\exp (-Q (1-e_k u)))/(Q e_k)^s $, it comes $ \int_{-1}^{1} \dd u \, \dd u '(u^2-1/3)^2 (u'^2-1/3)^2 (u'-u)^2 \Phi_8 (\frac{1-u e_k}{1-u 'e_k})/[(1-u e_k) (1-u' e_k)]^{7/2} = 2 \int_0^{+\infty} \dd Q \, Q^8 [A_0 (Q) A_2 (Q)-A_1 (Q)^2] $ with $ A_0 (Q) = [120 e_k f_6 (u)+24 (1-5 e_k u) f_5 (u)+8 (7 e_k-3u) f_4 (u)+\frac{16}{3} (2-3 e_k u) f_3 (u)+\frac{4}{9} (7 e_k-6 u) f_2 (u)+\frac{4}{9} (1-e_k u) f_1 (u)]_{-1}^{1} $, $ A_1 (Q) = [-720 e_k f_7 (u)-120 (1-6 e_k u) f_6 (u)+8 (15 u-43 e_k) f_5 (u)+8 (13 e_k u-7) f_4 (u)+\frac{8}{9} (18 u-25 e_k) f_3 (u)+\frac{4}{9} (8 e_k u-7) f_2 (u)+\frac{4}{9} (u-e_k) f_1 (u)]]_{-1}^{1} $ and $ A_2 (Q) = [5040 e_k f_8 (u)+720 (1-7 e_k u) f_7 (u)+40 (61 e_k-18 u) f_6 (u)+8 (43-95 u e_k) f_5 (u)+\frac{8}{3} (64 e_k-39 u) f_4 (u)+\frac{8}{9} (25-33 e_k u) f_3 (u)+\frac{4}{9} (9 e_k-8u) f_2 (u)+\frac{4}{9} (1-e_k u) f_1 (u)]_{-1}^{1} $. Similarly, in $ \mbox{Tr} \, \underline{\underline{D}} $, $ \int_{-1}^{1} \dd u \, \dd u '(u^2-1/3)^2 (u '^ 2-1/3)^2 [(1-u e_k)^2+(1-u' e_k)^2-2 uu '(1-u e_k) (1-u' e_k)] \Phi_8 (\frac{1-u e_k}{1-u 'e_k})/[(1-u e_k) (1-u' e_k)]^{7/2} = 2 \int_0^{+\infty} \dd Q \, Q^8 [B_0 (Q) B_2 (Q)-B_1 (Q)^2] $ with $ B_0 (Q) = A_0 (Q) $, $ B_1 (Q) = A_1 (Q)-e_k A_2 (Q) $ and $ B_2 (Q) = A_0 (Q)-2 e_k A_1 (Q)+e_k^2 A_2 (Q) $. These expressions are numerically ill-posed for $ Q $ or $ | e_k | $ too small, the terms $ f_s (u) $ becoming separately very large.} 
\be
\label{eq:199}
\mathcal{F}(k)\sim-\frac{512}{14175} 
\frac{(\epsilon_{\rm F}/\Delta)^2}{(1+\kappa^2)^3}
\frac{e_k(1+3e_k^2)(33+47 e_k^2)}{(1-e_k^2)^6}
\quad\mbox{and}\quad
\binom{\mathcal{D}_{\sslash}(k)}{\mathcal{D}_{\perp}(k)}
\sim \left(\frac{\epsilon_{\rm F}}{\Delta}\right)^2 
\binom{\frac{5632}{4725}+(\frac{159232}{4725}+\frac{94208\pi^2}{14553})\,\kappa^2
+O(\kappa^4)}{\frac{512}{945}+(\frac{61952}{11025}+\frac{38912\pi^2}{43659})\,\kappa^2
+O(\kappa^4)}
\ee

\paragraph{Langevin forces} 
The Fokker-Planck equation (\ref{eq:103}) is a deterministic equation on the probability distribution of the $\gamma$-quasiparticle wave vector $ \kk $. It is however more intuitive, in particular when the notion of temporal correlation comes into play, to use the stochastic reformulation given by Langevin \cite{Langevin}, in terms of a random walk of wave vector $ \kk $ in Fourier space: 
\be
\label{eq:110}
\hbar\, \dd\kk = \mathbf{F}(\kk)\, \dd t + 
[2\dd t\, \underline{\underline{D}}(\kk)]^{1/2}\mbox{\etab}
\ee
Here we use Ito calculus: between the times $ t $ and $ t+\dd t $ we randomly select a real Gaussian vector \etab, with zero mean, identity covariance matrix, $ \langle \eta_i \eta_j \rangle = \delta_{ij} $, statistically independent of the vectors $ \mbox{\etab} $ selected at other times. This shows in particular that $ \mathbf{F} (\kk) $ is indeed the mean force, since $ \dd \langle \hbar \kk \rangle/\dd t = \langle \mathbf{F} (\kk) \rangle $. The usual method of the test function \footnote{We express in two different ways the variation of order $ \dd t $ of the expectation $ \langle f (\kk (t)) \rangle $ where $ f $ is an arbitrary rapidly decreasing smooth function from $\mathbb{R}^3$ to $\mathbb{R}$.} allows to immediately get equation (\ref{eq:103}) from (\ref{eq:110}). The rotational invariance having led to the forms (\ref{eq:105}) suggests to transform (\ref{eq:110}) into stochastic equations on the modulus $ k $ and the direction $ \hat{\kk} $ of the wave vector: 
\bea
\label{eq:111a}
\hbar\,\dd k &=& F(k)\,\dd t + \frac{2 D_\perp(k)}{\hbar k} \dd t +
[2\dd t\, D_{\sslash}(k)]^{1/2} \eta_\sslash \\
\label{eq:111b}
\dd\hat{\kk} &=& - \frac{2 D_{\perp}(k)}{\hbar^2 k^2}  \dd t\, \hat{\kk}
+\frac{[2\dd t\, D_\perp(k)]^{1/2}}{\hbar k} \mbox{\etab}_\perp
\eea
where $ \eta_\sslash $ and \etab${}_\perp$ are the components of the vector \etab \, parallel and orthogonal to $ \kk $. This gives a physical meaning to $ D_\perp (k)/(\hbar k)^2 $, that of a diffusion coefficient of the direction of $ \kk $ on the unit sphere. 

\paragraph{Near zero speed} 
Let $ k_0 $ be a nodal point of the $\gamma$-quasiparticle group velocity $ v_k $, corresponding to a minimum or to a maximum of the dispersion relation [effective mass $ m_*> 0 $ or $ <0 $ in expansion (\ref{eq:002})]. In this point, at the order $ T^8 $ in temperature, the mean force tends to zero linearly with $ v_k $, which gives for the reduced form (\ref{eq:106}): 
\be
\label{eq:115}
\mathcal{F}(k)\underset{k\to k_0}{\sim} -\alpha e_k \quad \mbox{with} \quad
\alpha=\int_{-1}^1 \dd u \int_{-1}^1 \dd u' \, 4(u-u')^2 
\int_0^{2\pi} \frac{\dd\varphi}{2\pi} \left[\lim_{k\to k_0} R_{k}(u,u',w)\right]^2 \geq 0 
\ee
The quasiparticle then undergoes a viscous friction force, of reduced friction coefficient $ \alpha $. The longitudinal momentum diffusion coefficient at order $ T^9 $, in the reduced form (\ref{eq:106}), simply tends to $ \alpha $: 
\be
\label{eq:116}
\mathcal{D}_{\sslash}(k) \underset{k\to k_0}{\to} \alpha
\ee
This emerges from equation (\ref{eq:108b}), with the value $ \Phi_8 (1) = 8! [\zeta (8)-\zeta (9)] $ and the expression of $ \alpha $ in (\ref{eq:115}). The physical explanation is moreover very simple in the case of $ m_*> 0 $: it is, in dimensionless form, the Einstein relation linking equilibrium temperature, momentum diffusion coefficient and friction coefficient. We write it later in dimensioned form, see equation (\ref{eq:149}). It is well known when $ k_0 = 0 $, but it therefore also holds in the more unusual case $ k_0> 0 $ of an anisotropic momentum diffusion, provided that it is applied to $ \mathcal{D}_{\sslash} (k_0) $. Indeed, the reduced transverse diffusion coefficient has a nonzero limit in $k=k_0$ {\sl different} from $ \alpha $ if $ k_0> 0 $, but {\sl equal} to $ \alpha $ if $ k_0 = 0 $ (matrix $ \underline{\underline{D}} $ then becomes scalar). To write explicit expressions, we must distinguish these two cases. If $ k_0> 0 $, we use the limiting expression (\ref{eq:049}) of the reduced $\phi-\gamma$ scattering amplitude; after triple angular integration, it comes 
\begin{multline}
\label{eq:117a}
\alpha \stackrel{k_0>0}{=}\left(\frac{\hbar k_0}{2 m c}\right)^2 \Bigg[\frac{128}{225} +\frac{32}{35} \tilde{e}_{kk}^2 +\frac{2944}{525} \tilde{e}_{kk} \tilde{e}_{\rho k} +\frac{384}{35} \tilde{e}_{\rho k}^2 +\left(\frac{64}{15} \tilde{e}_{kk} +\frac{896}{45} \tilde{e}_{\rho k}\right) \tilde{e}_x +\frac{32}{3} \tilde{e}_x^2 \\
-\left(\frac{1088}{525} \tilde{e}_{kk} +\frac{128}{15}  \tilde{e}_{\rho k} +\frac{64}{9} \tilde{e}_x\right) \tilde{e}_\rho +\left(\frac{32}{9}-\frac{1216}{525} \tilde{e}_{kk}^2 -\frac{128}{15} \tilde{e}_{kk} \tilde{e}_{\rho k} +\frac{128}{45} \tilde{e}_{\rho k}^2 -\frac{64}{9} \tilde{e}_{kk} \tilde{e}_x\right) \tilde{e}_{\rho}^2 +\frac{64}{15} \tilde{e}_{kk} \tilde{e}_{\rho}^3 +\frac{32}{15} \tilde{e}_{kk}^2 \tilde{e}_{\rho}^4\Bigg] 
\end{multline}
\vspace{-8mm}
\begin{multline}
\label{eq:117b}
\mathcal{D}_{\perp}(k) \stackrel{k_0>0}{\underset{k\to k_0}{\to}}
\left(\frac{\hbar k_0}{2 m c}\right)^2 \Bigg[ \frac{256}{225} +\frac{32}{175} \tilde{e}_{kk}^2 +\frac{256}{175} \tilde{e}_{kk} \tilde{e}_{\rho k} +\frac{1408}{315} \tilde{e}_{\rho k}^2 +\left(\frac{64}{45} \tilde{e}_{kk} +\frac{512}{45} \tilde{e}_{\rho k}\right) \tilde{e}_x\\
 +\frac{32}{3} \tilde{e}_x^2 -\left(-\frac{64}{105} \tilde{e}_{kk} +\frac{128}{45}  \tilde{e}_{\rho k} +\frac{64}{9} \tilde{e}_x\right) \tilde{e}_\rho+\left(\frac{32}{9}+\frac{128}{175} \tilde{e}_{kk}^2 +\frac{128}{45} \tilde{e}_{kk} \tilde{e}_{\rho k} +\frac{256}{45} \tilde{e}_{\rho k}^2 \right) \tilde{e}_{\rho}^2 +\frac{64}{45} \tilde{e}_{kk} \tilde{e}_{\rho}^3 +\frac{32}{45} \tilde{e}_{kk}^2 \tilde{e}_{\rho}^4\Bigg]
\end{multline}
As we see on the modulus-direction Langevin form (\ref{eq:111a}, \ref{eq:111b}), near the extremum ($ e_k \simeq \hbar (k-k_0)/m_* c $), the average wave number $ \langle k (t) \rangle $ exponentially tends towards $ k_0 $ if $ m_*> 0 $ \footnote{\label{notegauch} The stationary value of $ \langle k \rangle $ generally differs slightly from $ k_0> 0 $. On the one hand, $ \epsilon_k $ contains a small cubic term $ \propto (k-k_0)^3 $ which distorts $ \Pi_{\rm st} (k) $ and leads to $ \langle k-k_0 \rangle \approx T $; on the other hand, even in the absence of this cubic term, the three-dimensional Jacobian $ k^2 $ distorts the wave number probability distribution $ k^2 \Pi_{\rm st} (k) $ and leads to a difference of the same order. In total, $\langle k-k_0\rangle\underset{T\to 0}{\sim} (m_*k_B T/\hbar^2 k_0)(2-k_0 b)$.} or deviates from it exponentially if $ m_* <0 $ ($\alpha$ is always $\geq 0$), and the mean wave vector direction tends exponentially to zero, with rates 
\be
\label{eq:118}
\Gamma_k = \frac{\pi^5}{15}\alpha \frac{\hbar}{m_*\rho^2}
\left(\frac{k_B T}{\hbar c}\right)^8 \quad \mbox{and}\quad
\Gamma_{\hat{\kk}} = \frac{2\pi^5}{15} \mathcal{D}_\perp(k_0)
\frac{c}{\rho^2 k_0^2} \left(\frac{k_B T}{\hbar c}\right)^9
\ee
If $ k_0 \equiv 0 $ you should rather use the limiting expression (\ref{eq:051}) to get \footnote{To show that $ \mathcal{D}_\perp (k)-\mathcal{D}_{\sslash} (k) \to 0 $ when $ k \to k_0 \equiv 0 $, we apply the remark of footnote \ref{note:diff} to the function $ f (u, u ', w) = g (w) [2 (1-w)-3 (u-u')^2] $ where $ g (w) $ is arbitrary. We then have $ \int_{-1}^{1} \dd u \int_{-1}^{1} \dd u '\int_0^{2 \pi} \dd \varphi f (u, u', w) = 0 $ because $ 2 (1-w)-3 (u-u ')^2 = 2 (1-\hat{\qq} \cdot \hat{\qq}')-3 [\hat{\kk} \cdot (\hat{\qq}-\hat{\qq} ')]^2 $ has zero mean on $\hat{\kk}$ at fixed $\hat{\qq}$ and $\hat{\qq}'$ therefore at fixed $w$.} 
\be
\label{eq:119}
\mathcal{D}_{\sslash}(k) \stackrel{k_0\equiv 0}{\underset{k\to k_0}{\sim}} 
\mathcal{D}_\perp(k) \stackrel{k_0\equiv 0}{\underset{k\to k_0}{\to}} 
\alpha\stackrel{k_0\equiv 0}{=}\frac{8}{9}\left\{\left[\check{e}_\rho 
\left(1+\frac{m}{m_*}\check{e}_\rho\right)-\check{e}_x\right]^2
+2\check{e}_x^2\right\}
\ee
As the Cartesian form (\ref{eq:110}) of the Langevin equation shows, the mean wave vector $ \langle \kk (t) \rangle $ then relaxes towards zero with a rate $ \Gamma_\kk $ of the same formal expression as $ \Gamma_k $ in the vicinity of a minimum of the dispersion relation, or on the contrary deviates from it exponentially with this same rate near a maximum. To be complete, let us note that, compared to the case $ k_0> 0 $, we gain one order in precision in the expansions (\ref{eq:115}, \ref{eq:116}, \ref{eq:119}), the relative deviation from the leading term now being $ \approx e_k^2 $. \footnote{\label{note:plus} We show this by pushing expansion (\ref{eq:051}) one step further, $ R_k (u, u ', w) \stackrel{k_0 \equiv 0}{\underset{k \to 0}{=}} \frac{1}{2} (\check{e}_x+f \check{e}_\rho w)+\frac{\hbar k}{2mc} (u+u ') (wf^2+\check{e}_\rho \check{e}_{\rho k})+O (k^2) $ with $ f = 1+(m/m_*) \, \check{e}_\rho $ and $ \check{e}_{\rho k} = \lim_{k \to 0} (mc/\hbar k) e_{\rho k} = \rho \dd (m/m_*)/\dd \rho $, and using in the angular integrals on $ (u, u ', \varphi) $ the antisymmetry of the integrand under the transformation $ (u, u ', \varphi) \to (-u,-u', \varphi) $, which preserves $ w $ and $ (u-u ')^2 $.} 

\begin{figure} [t] 
\centerline{\includegraphics [width = 0.45 \textwidth, clip =]{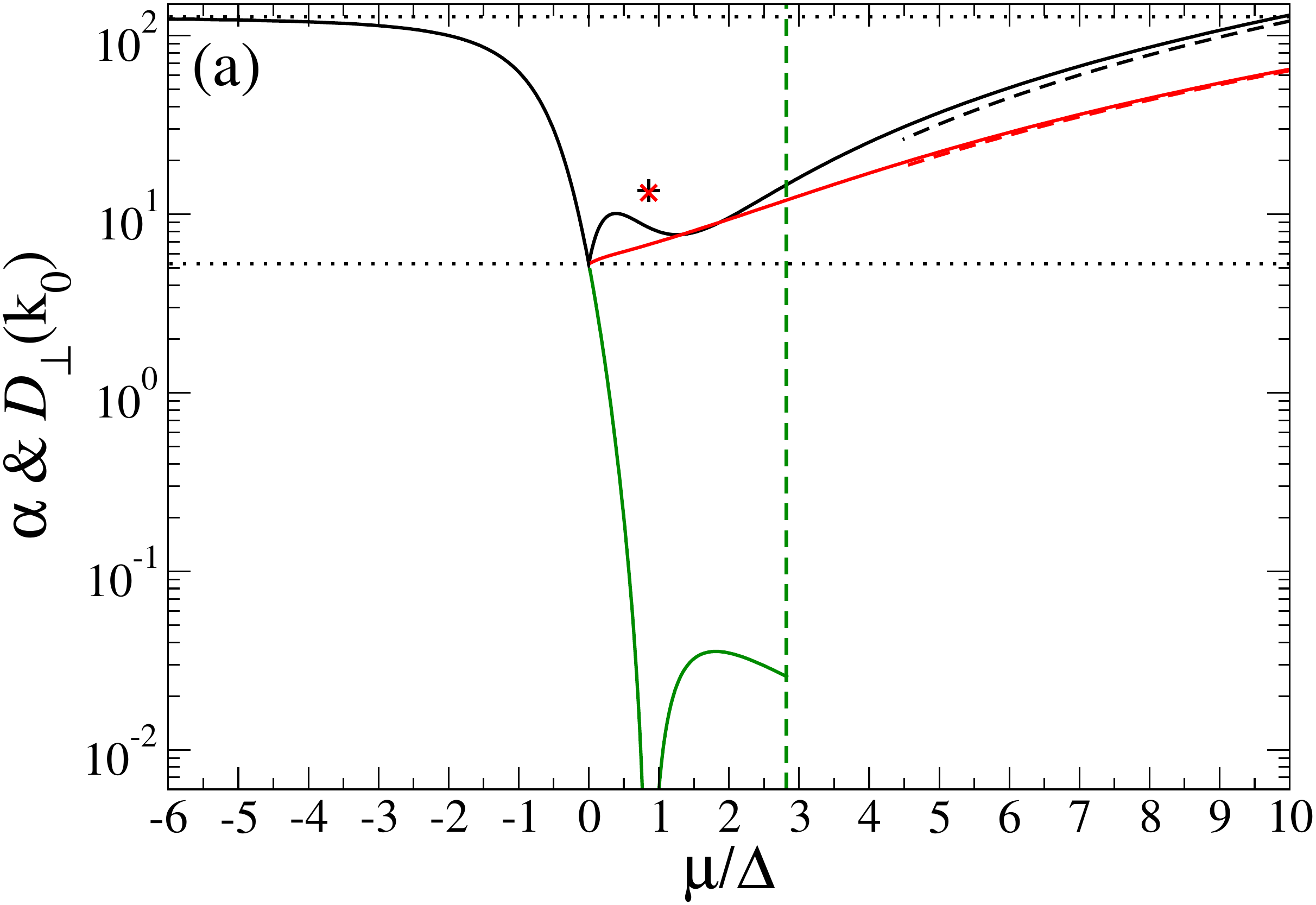} \hspace{3mm} \includegraphics [width = 0.45 \textwidth, clip =]{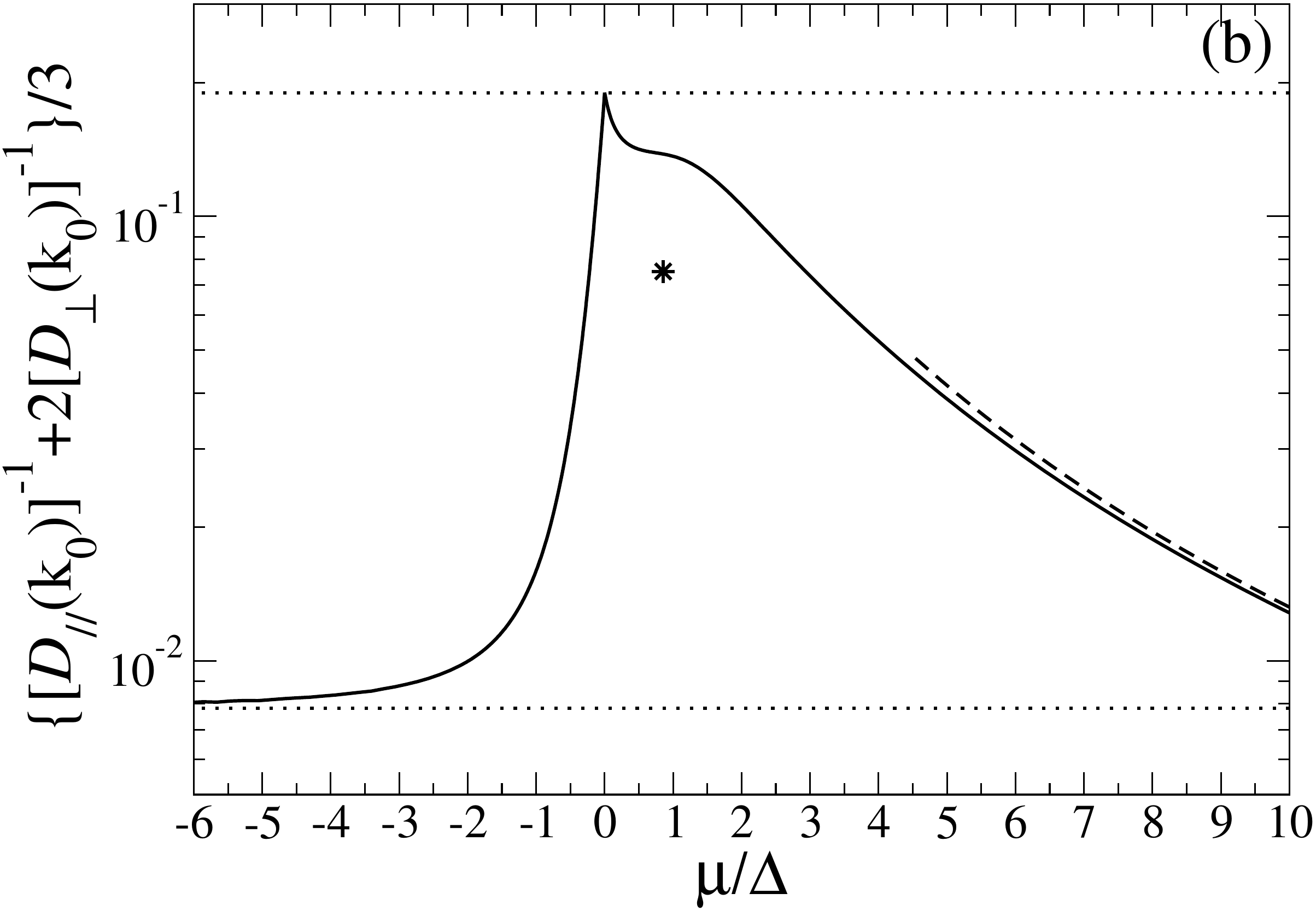}} 
\caption{
(a) Reduced friction coefficient $ \alpha $ and reduced transverse momentum diffusion coefficient $ \mathcal{D}_\perp (k_0) $ at the place $ k_0 $ of an extremum of the dispersion relation $ \epsilon_k $, for a fermionic $\gamma$ quasiparticle in an unpolarized pair-condensed gas of fermions of mass $ m $ at very low temperature, as functions of the interaction strength in the gas identified by the ratio between its chemical potential $ \mu $ and its order parameter $ \Delta $. The reduced longitudinal momentum diffusion coefficient $ \mathcal{D}_\sslash (k_0) $ is identically equal to $ \alpha $, see equation (\ref{eq:116}). Solid black line: $ \alpha $ at the minimum and in the BCS approximation ($ k_0 = 0 $ if $ \mu <0 $, $ k_0 = (2m \mu)^{1/2}/\hbar $ if $ \mu> 0 $). Solid green line: $ \alpha $ at the location of the relative maximum (maxon) and in the BCS approximation ($ k_0 = 0 $, $ \mu> 0 $). Solid red line: $ \mathcal{D}_\perp (k_0) $ in the BCS approximation where $ k_0> 0 $ (elsewhere, it coincides with $ \alpha $); it crosses the black line so it overcomes $ \mathcal{D}_\sslash (k_0) $ around $ \mu/\Delta = 1.6 $. We used equations (\ref{eq:117a}), (\ref{eq:117b}) or (\ref{eq:119}) depending on the case. Green vertical dashed line: maxon stability limit as in figure \ref{fig:map}a. Symbols: more precise coordinates (beyond the BCS approximation) of the coefficients $ \alpha $ (black sign) and $ \mathcal{D}_\perp (k_0) $ (red cross) at the location of the minimum in the unitary limit ($ \mu/\Delta \simeq $ 0.85, $ \alpha \simeq $ 13.6, $ \mathcal{D}_\perp (k_0) \simeq $ 13.2). Upper horizontal dotted line: value of $ \alpha $ in the BEC limit according to BCS theory [$ \simeq 5.8 $ times lower than the exact value (\ref{eq:120}) not shown here]. Lower horizontal dotted line: limit (\ref{eq:121}) of the different branches of $ \alpha $ and $ \mathcal{D}_\perp (k_0) $ in $ \mu = 0 $ according to BCS theory. The dashed curves in the weak coupling regime $ \mu/\Delta \gg 1 $ come from the dispersion relation (\ref{eq:011}) and from BCS limiting forms $ \Delta \simeq 8 \eee^{-2} \epsilon_{\rm F} \exp [-\pi/(2 k_{\rm F} | a |)] $, $ \mu \simeq \epsilon_{\rm F} $. (b) Reduced spatial diffusion coefficient of the $\gamma$ quasiparticle at equilibrium in the phonon gas, that is to say numerator of the last side of expression (\ref{eq:146}). Solid black line: BCS approximation. Horizontal dotted lines: its limit in $ \mu/\Delta = 0 $ or $ \mu/\Delta=-\infty $. Dashed line: its limiting form in weak coupling [dispersion relation (\ref{eq:011}) and limiting BCS forms $ \Delta \simeq 8 \eee^{-2} \epsilon_{\rm F} \exp [-\pi/(2 k_{\rm F} | a |)] $, $ \mu \simeq \epsilon_{\rm F} $]. Star: more precise estimate than BCS in the unitary limit.} 
\label{fig:fricdiff} 
\end{figure}

In the BCS approximation, $ \alpha $ and $ \mathcal{D}_\perp (k_0) $ are shown as functions of the interaction strength in figure \ref{fig:fricdiff}a. The case $ k_0> 0 $ corresponds to the minimum of the BCS dispersion relation (\ref{eq:011}) for a chemical potential $ \mu> 0 $; the case $ k_0 \equiv 0 $ corresponds either to the minimum of the BCS dispersion relation for $ \mu <0 $, or to the maxon, that is to say to the relative maximum of the dispersion relation, for $ \mu> $ 0. It is noted that the maxon reduced friction coefficient $\alpha_{\rm maxon}$ is connected to the branch $ \mu <0 $ of the reduced friction coefficient at the minimum $ \alpha_{\rm minon} $ in a differentiable way in $ \mu = 0 $, while $ \alpha_{\rm minon} $ exhibits a kink. When $ \mu = 0 $, the quasiparticle dispersion relation varies quartically at the location $ k = 0 $ of its minimum; relation (\ref{eq:115}) applies, with a simple analytical expression of the friction coefficient deduced from the BCS equation of state given in footnote \ref{note:equet}, \footnote{At fixed scattering length $ a> 0 $, the limit $ \mu \to 0^+$ in BCS theory corresponds to $ \rho \to \rho_0^+$, with $ \rho_0 a^3 = [\Gamma (1/4)]^8/(1536 \pi^4) $; here we only give the power laws $ k_0 \approx (\rho-\rho_0)^{1/2} $ and $ 1/m_* \approx \rho-\rho_0 $, to show that $ (\hbar \rho \dd k_0/\dd \rho)^2/m_* $ has a finite and nonzero limit.} 
\be
\label{eq:121}
\lim_{\mu\to 0} \alpha  \stackrel{\rm BCS\, approx.}{=} 
\left\{36\pi^4 \left[
\frac{1}{\Gamma(1/4)^{8}}+\frac{1}{24\pi^4-\Gamma(1/4)^8/4+3\Gamma(1/4)^{16}/(512\pi^4)}
\right]\right\}^{-1}=5. 269\, 833 \ldots
\ee
but the reduced group velocity $ e_k $ and therefore the mean force tend cubically to zero with $ k $; as for momentum diffusion, it is isotropic in $ k = 0 $, as equation (\ref{eq:119}) says. In the unitary limit, as we see in figure \ref{fig:fricdiff}a, $ \alpha_{\rm maxon} $ vanishes, which seems to us to be an artifact of the BCS approximation \footnote{According to BCS theory, there would be cancellation in the unitary limit in $ R_k (u, u', w) $ up to order $ k $ when $ k \to 0 $, $ R_k \sim 2 \delta^2 k^2 (w+2 u u ')/(9 | 3+2 \ii \delta |) $ where $ \delta = \Delta/mc^2 = \mbox{const} $ by scale invariance, so that $ \mathcal{F} (k) \approx (\hbar k/mc)^5 $ and $ \mathcal{D}_{\sslash, \perp} (k) \approx (\hbar k/mc)^4 $, the momentum diffusion matrix remaining non-scalar at this order. In an exact theory, we would have $ \epsilon_k = mc^2 G ((\hbar k/mc)^2) $, where the function $ G $ is smooth but unknown, assuming that the maxon exists and is stable. Then $ R_k (u, u ', w) \stackrel{1/a = 0}{\underset{k \to 0}{=}} \frac{w}{9} G (0) [3+4 G (0) G '(0)] \{1+\frac{\hbar k}{2 mc G (0)} (u+u') [3+4G (0) G '(0)] \}+O (k^2) $. In the usual notations, $ \check{e}_x = 0 $, $ G (0) = 3 \check{e}_\rho/2 = \Delta_*/mc^2 $ and $ G '(0) = m/2 m_* $ so that $ G (0) G '(0) = \Delta_*/2 m_* c^2 $, equal to $-3/4 $ in BCS theory}, and a more precise value of $ \alpha_{\rm minon} $, going beyond BCS, was added from the measured values $ k_0 \simeq 0.92 k_{\rm F} $ and $ \Delta/\epsilon_{\rm F} \simeq $ 0.44 \cite{KetterleGap}, $ \mu \simeq 0.376 \epsilon_{\rm F} $ \cite{Zwierlein2012} and the theoretical value resulting from dimensional expansion in $ \epsilon = 4-d $, $ m_*/m = 0.56 $ \cite{Nishida}. In the BEC limit $ k_{\rm F} a \to 0^+$, the fermion gas is reduced to a condensate of dimers of mass $ 2m $ and of scattering length $ a_{\rm dd} \simeq 0.60 a $ \cite{Petrov, MKagan}, and the $\gamma$ quasiparticle to an unpaired supernumerary fermion of mass $ m_* = m $, interacting with dimers with a scattering length $ a_{\rm ad} \simeq 1.18 a $ \cite{Skorniakov, Levinsen,StringariRMP,Leyronas}, hence the exact limit (not shown in figure \ref{fig:fricdiff}a): \footnote{Indeed, in the BEC limit, the ground state of the gas has an energy $ E_0 \simeq (N/2) (-\hbar^2/ma^2)+(N/2) (N/2-1) g_{\rm dd}/(2 L^3) $ and the excited state obtained by breaking a dimer into two free atoms of wave vectors $ \pm \kk $ has an energy $ E_1 \simeq (N/2-1) (-\hbar^2/ma^2)+(N/2-1) (N/2-2) g_{\rm dd}/(2 L^3)+2 \times \hbar^2 k^2/2m+2 \times g_{\rm ad} (N/2-1)/L^3 $, where $ g_{\rm dd} = 4 \pi \hbar^2 a_{\rm dd}/m_{\rm d} $ and $ g_{\rm ad} = 2 \pi \hbar^2 a_{\rm ad}/m_{\rm r} $ are the dimer-dimer and atom-dimer coupling constants, $ m_{\rm r} = m m_{\rm d}/(m+m_{\rm d}) = 2m/3 $ is the reduced mass of an atom and a dimer, $ N $ is the total number of fermions. Needless to say, the fermionic atoms and the bosonic dimers being distinguishable particles, there is no Hanbury-Brown and Twiss bunching effect in their mean field coupling energy. In the thermodynamic limit, $ \dd E_0/\dd N \to \mu $ and $ (E_1-E_0)/2 \to \epsilon_\kk \simeq \Delta_*+\hbar^2 k^2/2m_* $. Note that $ \check{e}_x = O (\rho^{1/2}). $} 
\be
\label{eq:120}
\left\{
\begin{array}{l}
\displaystyle
\mu\underset{k_{\rm F}a\to 0^+}{=} \frac{\hbar^2}{2 m a^2} 
\left[-1+\pi\rho a^2 a_{\rm dd} + O(\rho a^3)^{3/2}\right] \\
\displaystyle
\Delta_* \underset{k_{\rm F}a\to 0^+}{=} \frac{\hbar^2}{2 m a^2} 
\left[1 + \pi\rho a^2 (3 a_{\rm ad}-a_{\rm dd}) + O(\rho a^3)^{3/2}\right]
\end{array}
\right.
\Longrightarrow \alpha \underset{k_{\rm F}a\to 0^+}{\to} 
8 \left(\frac{a_{\rm ad}}{a_{\rm dd}}\right)^2 
\left(1-\frac{3 a_{\rm ad}}{a_{\rm dd}}\right)^2 \simeq 750
\ee
BCS theory is far from it (it underestimates the limit of $ \alpha $ by a factor $ \simeq 6 $) because it estimates the dimer-dimer and atom-dimer scattering lengths very badly, $ a_{\rm dd}^{\rm BCS} = 2a $ and $ a_{\rm ad}^{\rm BCS} = 8 a/3 $. 

\paragraph{Mean force at zero speed} 
At the location $ k_0 $ of the minimum of the dispersion relation $ \epsilon_k $, the $\gamma$-quasiparticle group velocity is zero. If $ k_0 = 0 $, the mean force undergone by $ \gamma $ is then strictly zero, due to parity invariance. If $ k_0> 0 $, on the other hand, the mean force undergone has no reason to be zero, $ F (k_0) \neq 0 $. More precisely, at low temperature, it vanishes at the order $ T^8 $, as we saw in the previous paragraph, but not at the order $ T^9 $, and we can obtain its exact expression at this order without needing to know the phonon scattering amplitude beyond its leading order $ \approx T $, 
\be
\label{eq:299}
F(k_0) \stackrel{k_0>0}{\underset{T\to 0}{\sim}} 
\frac{1}{\hbar} \frac{\dd}{\dd k} D_\sslash(k_0) 
+ \frac{2}{\hbar k_0} \left[D_\sslash(k_0)
-D_\perp(k_0)\right] \stackrel{k_0>0}{\underset{T\to 0}{\sim}}
\frac{\pi^5}{15} \left(\frac{k_B T}{\hbar c}\right)^9 \hbar c \rho^{-2}
\left[\frac{\dd}{\dd k} \mathcal{D}_\sslash(k_0) + \frac{2}{k_0}
\left(\mathcal{D}_\sslash(k_0)-\mathcal{D}_\perp(k_0)\right)\right]
\ee
The order $ T^9 $ of the momentum diffusion coefficients is indeed already known. It leads to a third nonzero side of the equation deducible from equations (\ref{eq:116}, \ref{eq:117a}, \ref{eq:117b}) and from the following expression, 
\begin{multline}
\label{eq:298}
k_0 \frac{\dd}{\dd k} \mathcal{D}_{\sslash}(k_0)= \left(\frac{\hbar k_0}{2 mc}\right)^2 \frac{64}{4725}\{630c_1\tilde{e}_{kk}^2\tilde{e}_{\rho}^4\!-\!3(228c_1\!-\!210c_2\!-\!175c_3\!-\!81)\tilde{e}_{kk}^2\tilde{e}_{\rho}^2\!+\!420(\tilde{e}_{\rho}\!+\!7c_1\tilde{e}_{kk}^{-1})\tilde{e}_{\rho k}^3\\ \!+\!42[3\tilde{e}_{kk}^3 \!+\!5(3c_1\!+\!1)\tilde{e}_{kk}]\tilde{e}_{\rho}^3 \!+\!3(90c_1\!-\!138c_2\!-\!105c_3\!+\!19)\tilde{e}_{kk}^2\!-\!3(102c_1\!-\!210c_2\!-\!175c_3\!-\!47)\tilde{e}_{kk}\tilde{e}_{\rho} \\ \!-\!30[7c_1\tilde{e}_{\rho}^2\!-\!129c_1\!+\!98c_2\!-\!33 \!-\!7(6\tilde{e}_{kk}\!-\!5c_1\tilde{e}_{kk}^{-1})\tilde{e}_{\rho}]\tilde{e}_{\rho k}^2\!-\!3[105(8c_1\!-\!2c_2\!+\!1)\tilde{e}_{kk}\tilde{e}_{\rho}^2\!-\!(552c_1\!-\!750c_2\!-\!490c_3\!+\!45)\tilde{e}_{kk} \\ \!-\!7(6\tilde{e}_{kk}^2\!-\!60c_1\!+\!50c_2\!-\!5)\tilde{e}_{\rho}]\tilde{e}_{\rho k}\!-\!105[10c_1\tilde{e}_{kk}\tilde{e}_{\rho}^2\!-\!(6c_1\!-\!14c_2\!-\!15c_3\!-\!3)\tilde{e}_{kk}\!-\!30c_1\tilde{e}_{kk}^{-1}\tilde{e}_{\rho k}^2\!-\!(10\tilde{e}_{kk}\tilde{e}_{\rho}\!+\!28c_1\!-\!30c_2\!+\!9)\tilde{e}_{\rho k}]\tilde{e}_{x}\\ \!+\!5\lambda[81\tilde{e}_{kk}^2\tilde{e}_{\rho}^2\!-\!49\tilde{e}_{kk}^2\!+\!63\tilde{e}_{kk}\tilde{e}_{\rho}\!-\!294\tilde{e}_{\rho k}^2\!-\!63(3\tilde{e}_{kk}\!+\!5\tilde{e}_{\rho k})\tilde{e}_{x}\!+\!3(35\tilde{e}_{kk}\tilde{e}_{\rho}^2\!-\!87\tilde{e}_{kk}\!+\!35\tilde{e}_{\rho})\tilde{e}_{\rho k}]\!+\!84\}
\end{multline}
which has been obtained by taking the derivative with respect to $ k $ of the first component of the identity (\ref{eq:108b}), and involves the additional dimensionless coefficients $ c_1 = k_0 b $, $ c_2 = \rho m_*^{-1} \dd m_*/\dd \rho $ and $ c_3 = \rho^2 k_0^{-1} \dd^2 k_0/\dd \rho^2 $, where the length $ b $ appears at order $ (k-k_0)^3 $ in the expansion (\ref{eq:002}) of $ \epsilon_k $. \footnote{We first calculate the derivative of the functions $ e_\rho $, etc, in $ k = k_0 $: by noting $ \tilde{\tilde{e}}_\rho = k_0 \dd e_\rho (k_0)/\dd k $, etc, we find $ \tilde{\tilde{e}}_\rho =-\tilde{e}_\rho+\tilde{e}_{\rho k} $, $ \tilde{\tilde{e}}_{k} = \tilde{e}_{kk} $, $ \tilde{\tilde{e}}_{kk} = (1+2 c_1) \tilde{e}_{kk} $, $ \tilde{\tilde{e}}_{\rho k} = 2 c_1 \tilde{e}_{\rho k}-c_2 \tilde{e}_{kk} $, $ \tilde{\tilde{e}}_x =-\tilde{e}_x-(\lambda+2c_2) \tilde{e}_{\rho k}-c_3 \tilde{e}_{kk}+2 c_1 \tilde{e}_{\rho k}^2/\tilde{e}_{kk} $; the values $ \tilde{e}_\rho $, etc, in $ k = k_0 $ are given in equation (\ref{eq:048}). Then we expand the reduced amplitude (\ref{eq:052}) around $ k_0 $ to first order in $ k-k_0 $. Finally, we insert the expansion obtained in the first component of equation (\ref{eq:108b}), which we expand in the same way, and we integrate on $ \varphi $, $ u $ and $ u '$, using the first integral of footnote \ref{note:diffi}.} In the BCS approximation, we find that the expression in square brackets in the third side of equation (\ref{eq:299}) is positive and tends to zero ($+\infty $) in units of $ 1/k_F $ when $ \mu/\Delta \to 0 \, (+\infty) $; we did not find it useful to plot it here. An interesting consequence of the results (\ref{eq:115}, \ref{eq:299}) is that, in the case of $ k_0> 0 $, the mean force undergone by the $\gamma$ quasiparticle at low temperature vanishes a little away from $ k_0 $, at a wave number deviation $ k-k_0 $ or at a speed $ v_k $ proportional to temperature; need it be specified, this deviation has no simple relation with the stationary mean $ \langle k-k_0 \rangle $ (see notes \ref{notegauch} and \ref{noteelab}). 

The simplest way to establish equation (\ref{eq:299}) is to go through the stationary solution $ \Pi_0 $ of the Fokker-Planck equation (\ref{eq:103}) seen as an equation of continuity in Fourier space. This rotationally invariant distribution $ \Pi_0 (k) $ is easily obtained by expressing that the current of probability $ \JJ (\kk) $ which it carries is zero. From the general expression of the component along $ Oi $, $ J_i (\kk) = F_i (\kk) \Pi (\kk)/\hbar-\sum_j \frac{\partial}{\partial k_j} [D_{ij} (\kk) \Pi (\kk)]/\hbar^2 $, we get after replacement of $ \Pi(\kk) $ by $ \Pi_0(k)$, division by $ \Pi_0 (k)/\hbar $ and use of spherical forms (\ref{eq:105}): 
\be
\label{eq:300}
F(k) = \frac{2}{\hbar k} [D_\sslash(k)-D_\perp(k)]+\frac{1}{\hbar} 
\frac{\dd}{\dd k}D_\sslash(k) + 
D_\sslash(k) \frac{1}{\hbar}\frac{\dd}{\dd k}\ln \Pi_0(k)
\ee
As we said, $ \Pi_{\rm st} (\kk) \propto \exp (-\beta \epsilon_\kk) $ is an exact stationary solution of the master equation. However, this is not an exact stationary solution of the Fokker-Planck equation, since the latter results from a truncated expansion of the master equation at low temperature. By estimating the error made, knowing that $ \Pi_0 (k) $ has a width of order $ T^{1/2} $ around $ k_0 $, \footnote{Note that the first terms neglected in the expression of the current along $ Oi $ are of the form $ \frac{\partial^2}{\partial k_j \partial k_k} [E_{ijk} (\kk) \Pi_0 (k)]/\hbar^3 $ and $ \frac{\partial^3}{\partial k_j \partial k_k \partial k_l} [G_{ijkl} (\kk) \Pi_0 (k)]/\hbar^4 $, where the tensors $ E_{ijk} $ and $ G_{ijkl} $ are moments of order 3 and 4 of the momentum change $ \hbar(\qq-\qq')$ accompanying phonon scattering, while $ F $ and $ D $ are the moments of order 1 and 2, see equation (\ref{eq:104}). By proceeding as for force and diffusion, we find that to leading order in $ T $, $ E_{ijk} \propto T^{10} \mathcal{E}_{ijk} $ and $ G_{ijkl} \propto T^{11} \mathcal{G}_{ijkl} $, where the reduced functions $ \mathcal{E} $ and $ \mathcal{G} $, independent of temperature, have a width in $ k $ space of order $ T^0 $, $ \mathcal{E} $ vanishing linearly in $ k = k_0 $ ($ \mathcal{E} \approx e_k $) and $ \mathcal{G} $ having a finite limit. Then each differentiation of $ \Pi_0 (k) $ with respect to $ \kk $ leaves a factor $ \approx T^{-1/2} $, and the factor $ e_k $ is of the order of $ T^{1/2} $ over the width of $ \Pi_0 (k) $, so that $ \frac{\partial^2}{\partial k_j \partial k_k} [E_{ijk} (\kk) \Pi_0 (k)] \approx T^{-1+10+1/2} \Pi_0 $ and $ \frac{\partial^3}{\partial k_j \partial k_k \partial k_l} [G_{ijkl} (\kk) \Pi_0 (k)] \approx T^{-3/2+11} \Pi_0 $ and we get equation (\ref{eq:301}).} we arrive at the exact relation within the width of the distribution: 
\be
\label{eq:301}
F(k)\stackrel{k-k_0=O(T^{1/2})}{\underset{T\to 0}{=}}
\frac{2}{\hbar k} [D_\sslash(k)-D_\perp(k)]+\frac{1}{\hbar}
\frac{\dd}{\dd k}D_\sslash(k) 
-\beta D_\sslash(k) \frac{1}{\hbar}\frac{\dd}{\dd k}\epsilon_k+O(T^{9+1/2})
\ee
It suffices to apply it in $ k = k_0 $, where the derivative of $ \epsilon_k $ vanishes, to find result (\ref{eq:299}) \footnote{\label{noteiden} By identifying the terms of order $ k-k_0 $ then of order $ (k-k_0)^2 $ in relation (\ref{eq:301}), we show that there is a coefficient $ \alpha $ already introduced and a new coefficient $ \zeta $ such as $ \mathcal{F} (k) =-(\alpha e_k+\zeta e_k^2+O (e_k^3)) $ and $ \mathcal{D}_\sslash (k) = \alpha+\zeta e_k+O (e_k^2) $, so that $ \mathcal{F} (k)/\mathcal{D}_\sslash (k) =-e_k+O (e_k^3) $ when $ k \to k_0 $. We arrive at the same results by Taylor expansion of expressions (\ref{eq:108a}, \ref{eq:108b}) to sub-leading order in $ e_k $, considering $ R_k $ as a function of $ e_k $ rather than $ k $. The value of $ \zeta $ is easily deduced from equation (\ref{eq:298}).}. 

We can also establish (\ref{eq:299}) by a fairly long calculation, starting from the general expression (\ref{eq:104}) of the mean force and expanding it to order $ T^9$ in the particular case $ k = k_0 $. We must express $ q '$ as a function of $ q $ using energy conservation one step further than in the calculation of $ \mathcal{F} (k) $, $ q' = q-\hbar q^2 (u'-u)^2/(2m_* c)+O (q^3) $, and above all use an exactly microreversible expression of the $\phi-\gamma$ scattering amplitude on the energy shell: 
\be
\label{eq:303}
\mathcal{A}(\gamma:\kk,\phi:\qq\to \gamma:\kk',\phi:\qq') = \frac{\hbar c}{\rho}(q q')^{1/2} R_{||\frac{\kk+\kk'}{2}||} \left(\hat{\qq}\cdot \frac{\kk+\kk'}{||\kk+\kk'||}, \hat{\qq}'\cdot \frac{\kk+\kk'}{||\kk+\kk'||}, \hat{\qq}\cdot\hat{\qq}'\right)
\ee
where $ \kk+\qq = \kk '+\qq' $ and the function $ R_k (u, u ', w) $ is that of equation (\ref{eq:045}). By noting $ R_{k_0}^2 $ and $ \frac{\dd}{\dd k} R_{k_0}^2 $ the values of the function $ R_k^2 $ and its derivative with respect to $ k $ in $ k = k_0 $, we obtain for the left-hand side of equation (\ref{eq:299}): 
\begin{multline}
\label{eq:305}
F(k_0) \stackrel{k_0>0}{\underset{T\to 0}{\sim}} \frac{4\pi^5}{15} \frac{\hbar c}{\rho^2 k_0} \left(\frac{k_B T}{\hbar c} \right)^9 \int_{-1}^{1} \dd u \int_{-1}^{1} \dd u' \int_0^{2\pi} \frac{\dd\varphi}{2\pi} \Bigg[\frac{3\hbar k_0}{m_*c} u' (u'-u)^2 R_{k_0}^2+(u'-u)^2 k_0 \frac{\dd}{\dd k} R_{k_0}^2 + \\
(1-w) (u-u')\left(\frac{\partial R_{k_0}^2}{\partial u}- \frac{\partial R_{k_0}^2}{\partial u'}\right)-(u-u')^2 \left(u\frac{\partial R_{k_0}^2}{\partial u}+u' \frac{\partial R_{k_0}^2}{\partial u'}\right)\Bigg]
\end{multline}
The different terms in the right-hand side of (\ref{eq:299}) are deduced from the order $ T^9 $ of expression (\ref{eq:108b}) and the derivative with respect to $ k $ of its first component, the whole taken in $ k = k_0 $ ($ e_k = 0 $). For form (\ref{eq:052}) of the function $ R_k $, there is indeed agreement at order $ T^9 $ between the first two sides of equation (\ref{eq:299}). \footnote{\label{note:diff} This conclusion is not specific to form (\ref{eq:052}) but is ultimately based on the identity $ \int_{-1}^{1} \dd u \int_{-1}^{1} \dd u '\int_0^{2 \pi} \frac{\dd \varphi}{2 \pi} \{(u-u') [(1-w) (\partial_u-\partial_{u '})-(u-u') (u \partial_u+u '\partial_{u'})] R_{k_0}^2-[3 (u-u ')^2-2 (1-w)] R_{k_0}^2 \} = $ 0. To establish it, we first notice that $ \frac{1}{4} \int_{-1}^{1} \dd u \int_{-1}^{1} \dd u '\int_0^{2 \pi} \frac{\dd \varphi}{2 \pi} f (u, u ', w) = \langle f (\hat{\qq}, \hat{\qq}', \hat{\kk}) \rangle $ where $ \langle \ldots \rangle $ is the average taken uniformly on the unit sphere for the three directions $ \hat{\qq}, \hat{\qq} '$ and $ \hat{\kk} $, $ f $ being a rotationally invariant function or, which amounts to the same thing, a function of the three scalar products $ u = \hat{\kk} \cdot \hat{\qq} $, $ u '= \hat{\kk} \cdot \hat{\qq}' $ and $ w = \hat{\qq} \cdot \hat{\qq} '$. So if $ \hat{\kk} \mapsto \mathcal{T}_{\hat{\qq}, \hat{\qq} '} (\hat{\kk}) $ is a diffeomorphism from the unit sphere to the unit sphere parametrically depending on $ (\hat{\qq}, \hat{\qq} ') $ and of Jacobian $ J_{\hat{\qq}, \hat{\qq}'} (\hat{\kk}) $, we have $ \langle f (\hat{\qq}, \hat{\qq} ', \hat{\kk}) \rangle = \langle J_{\hat{\qq}, \hat{\qq} '} (\hat{\kk}) f (\hat{\qq}, \hat{\qq}', \mathcal{T}_{\hat{\qq}, \hat{\qq} '} (\hat{\kk})) \rangle $. It remains to apply this identity to the function $ f (\hat{\qq}, \hat{\qq} ', \hat{\kk}) = \hat{\kk} \cdot (\hat{\qq}-\hat{\qq} ') \, R_{k_0}^2 (\hat{\kk} \cdot \hat{\qq}, \hat{\kk} \cdot \hat{\qq}', \hat{\qq} \cdot \hat{\qq} ') $ and diffeomorphism $ \mathcal{T}_{\hat{\qq}, \hat{\qq}'} (\hat{\kk}) = (\hat{\kk}+\eta \frac{\hat{\qq}-\hat{\qq} '}{2})/|| \hat{\kk}+\eta \frac{\hat{\qq}-\hat{\qq} '}{2} || = \hat{\kk} [1-\eta (u-u')/2]+\eta (\hat{\qq}-\hat{\qq} ')/2+O (\eta^2) $ of Jacobian $ J_{\hat{\qq}, \hat{\qq}'} (\hat{\kk}) = 1-\eta (u-u ')+O (\eta^2) $ where $\eta$ is infinitesimal.} This explicit calculation allows us to understand how we can obtain exactly the mean force to sub-leading order $ T^9 $ without knowing the scattering amplitude $ \mathcal{A} $ to sub-leading order $ T^2 $. This is because we restricted to the case $ k = k_0 $: at this point, the contribution to $ F (k) $ of the first correction to $ \mathcal{A} $ vanishes due to antisymmetry of the integrand under the exchange of $ u $ and $ u '$. 

\paragraph{Near the speed of sound} 
At fixed atomic interaction strength in the superfluid, one of the edges of the $\gamma$-quasiparticle stability domain in the wave vector space (red line in figure \ref{fig:map}) corresponds to the sonic regime of a group velocity equal in absolute value to the speed of sound. However, our expressions (\ref{eq:108a}, \ref{eq:108b}) of the reduced mean force and momentum diffusion diverge in the sonic limit $ e_k \to \pm 1 $, as expected from the denominators $ 1-u e_k $ and $ 1-u 'e_k $.

To see it on the mean force, we carry out in the integrals the changes of variables $ u = 1-x $ and $ u '= 1-(1-e_k) x' $ if $ e_k \to 1^-$, $ u = x-1 $ and $ u '=-1+(1+e_k) x' $ if $ e_k \to-1^+$, then we take the limit in the integral at fixed $ x, x '$, which eliminates the dependence on the azimuthal angle $ \varphi $, to obtain after integration on $ x '\in \mathbb{R}^+$ (performing integration on $ x $ is elementary but does not lead to a more compact expression): \footnote{\label{note:limson} Suppose that $ e_k \to 1^-$ to fix ideas and split the integration domain $ [-1, \, 1]^2$ on the cosines $ (u, u ') $ of the polar angles in a large square $ [-1, \, 1-\epsilon] \times [-1, \, 1-\epsilon] $, a small square $ [1-\epsilon, 1] \times [1-\epsilon, 1] $ and two very elongated rectangles, one lying $ [-1,1-\epsilon] \times [1-\epsilon, 1] $ and the other standing $ [1-\epsilon, 1] \times [-1,1-\epsilon] $ where $ 0 <\epsilon \ll 1 $ is fixed when $ e_k \to 1^-$, then tends to zero at end of calculations. The contribution of the large square has a finite limit when $ e_k \to 1^-$ therefore is always negligible. For the force, the contribution of the small square is small $ \approx \epsilon^3/(1-e_k)^6 $, that of the lying rectangle is dominant $ \approx \epsilon^0/(1-e_k)^6 $, that of the standing rectangle is negligible $ \approx (1-e_k)^0/\epsilon^3 $. This justifies the procedure used in the text. For the diffusion, there are two types of contributions, that of the same type as in $ \mathcal{F} (k) $ and that involving the function $ \Phi_8 $. The first one is treated as before. In the second one, we find that the dominant contribution comes from the small square and that it diverges in $ \mathcal{D}_\sslash $ as $ (1-e_k)^{-3} $ with a prefactor containing the integral $ \int_0^{+\infty} \! \! \dd x \, \int_0^{+\infty} \! \! \dd x '\, (x-x')^2 \Phi_8 (\frac{1+x}{1+x '})/[(1+x) (1+x')]^{7/2} $, and in $ \mathcal{D}_\perp $ as $ (1-e_k)^{-2} $ with a prefactor containing the integral $ \int_0^{+\infty} \! \! \dd x \, \int_0^{+\infty} \! \! \dd x '\, [(1+x) (1+x') (x+x ')-(x-x')^2] \Phi_8 (\frac{1+x}{1+x '})/[(1+x) (1+x ')]^{7/2} $ (this results from changes of variables $ u = 1-x (1-e_k) $ and $ u' = 1-x ' (1-e_k) $ and passage to the limit $ e_k \to 1^-$ at fixed $ x $ and $ x '$ in the integral); the contributions of the rectangles are negligible by virtue of the upper bound $ \Phi_8 (x) \leq (8 \pi^8/15) x^{-7/2} $ on $ \mathbb{R}^{+*} $ and the equivalent $ \Phi_8 (x) \sim (8 \pi^8/15) x^{7/2} $ when $ x \to 0^+$, which also show that $ \Phi_8 (x) $ is bounded and the prefactor integrals are finite.} 
\bea
\label{eq:130a}
\mathcal{F}(k) &\underset{e_k\to 1^-}{\sim} &
-\frac{1}{24} \left(\frac{\hbar k}{mc}\right)^2 \frac{(1+e_\rho)^2}
{(1-e_k)^6} \int_0^2 \dd x\, x^2 \left\{e_{kk}(1+e_\rho) 
+\left[e_{\rho k}-\frac{1+\lambda}{2}-(2+e_\rho) e_{kk}\right]x
 +e_{kk} x^2\right\}^2 \\
\label{eq:130b}
\mathcal{F}(k) &\underset{e_k\to -1^+}{\sim} &
\phantom{+}\frac{1}{24} \left(\frac{\hbar k}{mc}\right)^2 \frac{(1-e_\rho)^2}
{(1+e_k)^6} \int_0^2 \dd x\, x^2 \left\{e_{kk}(1-e_\rho) 
+\left[e_{\rho k}+\frac{1+\lambda}{2}-(2-e_\rho) e_{kk}\right]x
 +e_{kk} x^2\right\}^2
\eea
The mean force is negative in the limit $ e_k \to 1-$. However, this limit always corresponds to the upper edge $ k_{\rm sup} $ of the wave number stability domain of the $\gamma$ quasiparticle, as we can see in figure \ref{fig:map}. The force undergone by the $\gamma$ quasiparticle near the edge therefore tends systematically to move its wave vector $ \kk $ away from the supersonic zone \footnote{We arrive at the same conclusion by reasoning on the mean wave number: according to equation (\ref{eq:111a}), $ \dd \langle \hbar k \rangle/\dd t = \langle F (k)+2 D_\perp (k)/\hbar k \rangle $, but the transverse diffusion term, of the same sonic-divergence exponent $-6 $ as the force term, is negligible because it is sub-leading in temperature.}. Conversely, when the stability domain of the $\gamma$ quasiparticle has an edge $ e_k =-1 $, it is always a lower edge $ k_{\rm inf} $; the mean force is positive in the limit $ e_k \to-1^+$ and there also pushes the quasiparticle away from the supersonic zone. We could however have come to the opposite (and physically incorrect) conclusion if we had ignored the subsonic destabilization line SC (green line in figure \ref{fig:map}c): for $ \Delta/\mu <(\Delta/\mu)_{\rm S} $, that is to say below the summit point S in the figure, we would then have $ e_k =-1 $ at an {\it upper} edge  of the stability domain, and the mean force near the edge would bring $\kk$ closer to the sonic boundary. 

In the case of momentum diffusion, we first show that the contribution of the function $ \Phi_8 $ in expression (\ref{eq:108b}), although divergent, is a negligible fraction in the sonic limit (see footnote \ref{note:limson}), then we apply to the rest the same procedure as for the force, which gives 
\bea
\label{eq:131a}
\binom{\mathcal{D}_{\sslash}(k)}{\mathcal{D}_\perp(k)}
\underset{e_k\to 1^-}{\sim} 
\frac{225\,\zeta(9)}{\pi^8} \left(\frac{\hbar k}{mc}\right)^2 (1+e_\rho)^2
\binom{6(1-e_k)^{-7}}{(1-e_k)^{-6}}
\int_0^2 \dd x\, x^3 \left\{e_{kk}(1\!+\!e_\rho)
\!+\!\left[e_{\rho k}\!-\!\frac{1\!+\!\lambda}{2}\!-\!(2\!+\!e_\rho) e_{kk}\right]x
 \!+\!e_{kk} x^2\right\}^2  \\
\label{eq:131b}
\binom{\mathcal{D}_{\sslash}(k)}{\mathcal{D}_\perp(k)}
\underset{e_k\to -1^+}{\sim} 
\frac{225\,\zeta(9)}{\pi^8} \left(\frac{\hbar k}{mc}\right)^2 (1-e_\rho)^2 
\binom{6(1+e_k)^{-7}}{(1+e_k)^{-6}}
\int_0^2 \dd x\, x^3 \left\{e_{kk}(1\!-\!e_\rho)
\!+\!\left[e_{\rho k}\!+\!\frac{1\!+\!\lambda}{2}\!-\!(2\!-\!e_\rho) e_{kk}\right]x
 \!+\!e_{kk} x^2\right\}^2 
\eea
The longitudinal momentum diffusion coefficient therefore diverges more quickly than the mean force. Should it be specified, the results (\ref{eq:130a}, \ref{eq:130b}, \ref{eq:131a}, \ref{eq:131b}) are only valid if we make first the gas temperature tend to zero, as in equation (\ref{eq:106}), then the speed of the $\gamma$ quasiparticle tend to $ \pm c $. There is nothing to indicate that the divergence remains in an exchange of the limits. 

\paragraph{Spatial diffusion} 
We now consider times long enough for the $\gamma$-quasiparticle wave vector distribution to have reached its thermal equilibrium. The direction $ \hat{\kk} $ of the wave vector is uniformly distributed on the unit sphere. The quasiparticle wave number $ k $ deviates weakly from the minimum position $ k_0 $ of the dispersion relation, $ k-k_0 \approx T^{1/2} $, so we can replace the mean force $ F (k) $ to order $ T^8 $ by its linear approximation (\ref{eq:115}) and the momentum diffusion coefficients by their value in $ k_0 $. There is, however, an unsteady degree of freedom which we have not discussed so far, and whose probability distribution spreads indefinitely in the thermodynamic limit. It is the $\gamma$-quasiparticle position $ \rr (t) $, whose time derivative $ \vv (t) $ is simply the $\gamma$-quasiparticle group velocity:
\be
\label{eq:140}
\frac{\dd}{\dd t} \rr(t) \equiv \vv(t) = \frac{1}{\hbar} \frac{\dd \epsilon_k}
{\dd k} \hat{\kk}
\ee
The change of position between $ 0 $ and $ t $ is given by the integral of the velocity between these two times; its variance is therefore given by a double time integral of the two-time velocity correlation function. At equilibrium, this matrix correlation function $C_{ij}$ only depends on the time difference. We thus set in full generality (without taking into account the fact that the average velocities $ \langle v_i \rangle $ and $ \langle v_j \rangle $ are obviously zero due to the equilibrium isotropy): 
\be
\label{eq:141}
C_{ij}(\tau)= \langle v_i(\tau) v_j(0)\rangle 
-\langle v_i\rangle \langle v_j\rangle
\ee
The covariance of the $\gamma$-quasiparticle displacements during $ t $ along the directions $ i $ and $ j $ is then \footnote{We are facing $ I_{ij} (t) = \int_0^t \dd \tau \int_0^t \dd \tau'\, f_{ij} (| \tau-\tau' |) $ with $ f_{ij} (| \tau-\tau'|) = [C_{ij} (\tau-\tau')+C_{ji} (\tau-\tau')]/2 $; indeed, $C_{ji}(\tau-\tau')=C_{ij}(\tau'-\tau)$ even in a quantum treatment of motion, because $ v_i $ and $ v_j $ are Hermitian operators (hence $ C_{ij}^*(\tau) = C_{ji} (-\tau) $) commuting with the stationary density operator of $ \gamma$ (hence $C_{ij}(\tau) \in \mathbb{R}$). We then successively have $ I_{ij} (t) = 2 \int_0^t \dd \tau \int_0^\tau \dd \tau' f_{ij} (\tau-\tau') = 2 \int_0^t \dd \tau \int_0^{\tau} \dd \tau'' f_{ij} (\tau'') = 2 \int_0^t \dd \tau (t-\tau) f_{ij} (\tau) $. We put $ \tau'= \tau-\tau''$ in the internal integral, then we integrated by parts on $ \tau $.} 
\be
\label{eq:142}
\mbox{Cov}\,(r_i(t)-r_i(0),r_j(t)-r_j(0))
= \int_0^{t} \dd\tau\, (t-\tau)\, [C_{ij}(\tau)+C_{ji}(\tau)]
\ee
As we will see by distinguishing the cases $ k_0 \equiv 0 $ and $ k_0> 0 $, we find that the velocity correlation matrix, scalar because of the rotational invariance, decreases rapidly (exponentially) in time. Also the $\gamma$-quasiparticle position always undergoes asymptotically in time an isotropic diffusive spreading, the variance of displacement during $ t $ having a linear divergence independent of the direction of unitary vector $ \nn $ considered: 
\be
\label{eq:143}
\mbox{Var}\, ([\rr(t)-\rr(0)]\cdot \nn) 
\underset{t\to +\infty}{\sim} 2 \mathcal{D^{\rm spa}} t \quad
\mbox{with}\quad \mathcal{D}^{\rm spa} = \int_0^{+\infty} \dd\tau\,
C_{zz}(\tau)
\ee
In other words, at times much longer than the velocity correlation time, the $\gamma$ quasiparticle performs in position space a Brownian motion with a spatial diffusion coefficient $ \mathcal{D}^{\rm spa} $, which we have to calculate to leading order in temperature. For this, as we will see, we can limit ourselves in $ C_{zz} (\tau) $, if $ k_0 \equiv 0 $, to a single exponential contribution, of amplitude $ \approx T $ and decay rate $ \Gamma_k $ given by equation (\ref{eq:118}): 
\be
\label{eq:144}
C_{ij}(\tau) \stackrel{k_0\equiv 0}{\simeq}\delta_{ij} 
\frac{k_B T}{m_*} \eee^{-\Gamma_k |\tau|}
\ee
The asymptotic law (\ref{eq:143}) then applies for $ \Gamma_k t \gg 1 $. If $ k_0> 0 $, on the other hand, you have to keep {\it two} contributions in $ C_{zz} (\tau) $, one of amplitude $ \approx T $ and rate $ \Gamma_k $, which comes from the faster decorrelation of the wave number $ k $, and the other of amplitude $ \approx T^2 $ and rate $ \Gamma_{\hat{\kk}} $ given by equation (\ref{eq:118}), which comes from the slower decorrelation of the angular variables $ \hat{\kk} $: 
\be
\label{eq:145}
C_{ij}(\tau) \stackrel{k_0>0}{\simeq} \delta_{ij} \frac{k_B T}{3 m_*}
\left[\eee^{-\Gamma_k |\tau|} + \frac{4 m_* k_B T}{\hbar^2 k_0^2} 
\eee^{-\Gamma_{\hat{\kk}}|\tau|} \right]
\ee
This second contribution is initially weaker than the first by a factor $ \approx T $ but it decreases more slowly by a factor $ \approx T $: it therefore ultimately contributes to the spatial diffusion coefficient to the same order in $ T $ as the first contribution; the asymptotic law (\ref{eq:143}) then only applies for $ \Gamma_{\hat{\kk}} t \gg 1 $. Let us collect the two cases in a synthetic expression of the spatial diffusion coefficient: 
\be
\label{eq:146}
\boxed{
\mathcal{D}^{\rm spa} \underset{T\to 0}{\sim} \frac{k_B T}{3 m_*\Gamma_k}
\left[1+\frac{2D_\sslash(k_0)}{D_\perp(k_0)}\right]
\underset{T\to 0}{\sim} \frac{1}{3} (k_B T)^2 \left[\frac{1}{D_\sslash(k_0)}
+\frac{2}{D_\perp(k_0)}\right]
\underset{T\to 0}{\sim} \frac{\hbar}{m}
\frac{\frac{1}{3}\left[\frac{1}{\mathcal{D}_\sslash(k_0)}+
\frac{2}{\mathcal{D}_\perp(k_0)}\right]}
{\frac{\pi^5}{15}\left(\frac{mc}{\hbar\rho^{1/3}}\right)^6
\left(\frac{k_B T}{mc^2}\right)^7}
}
\ee
using first Einstein's relation (\ref{eq:116}) then the equivalents (\ref{eq:106}). It is recalled that $ \mathcal{D}_\sslash (k_0) $ coincides with the reduced friction coefficient $ \alpha $, always defined by equation (\ref{eq:115}) and explicit expression (\ref{eq:119}) for $ k_0 \equiv 0 $, (\ref{eq:117a}) for $ k_0> 0 $. Similarly, $ \mathcal{D}_\perp (k_0) $, which we can always deduce from the limit of equation (\ref{eq:108b}) in $k=k_0$, is equal to $ \alpha $ if $ k_0 \equiv 0 $ and has the explicit expression (\ref{eq:117b}) otherwise. The spatial diffusion coefficient (\ref{eq:146}) is shown in figure \ref{fig:fricdiff}b in dimensionless form in the BCS approximation, as a function of the fermion gas interaction strength. Note that it is continuous at the transition between the cases $ k_0 \equiv 0 $ and $ k_0> 0 $ [$ \mu = 0 $ in the BCS dispersion relation (\ref{eq:011})] but presents a kink there. 

Let us briefly expose the calculations leading to equations (\ref{eq:144},\ref{eq:145}). Let us start with the case $ k_0 \equiv 0 $, abundantly treated in the literature. From the leading order approximations $ \vv \simeq \hbar \kk/m_* $, $ \FF (\kk) \simeq-\Gamma_k \hbar \kk $, $ \underline{\underline{D}} (\kk) \simeq \underline{\underline{D}} (\mathbf{0}) = D_\sslash (0) \mathrm{Id} $ and $ \Pi_{\rm st} (\kk) \propto \exp (-\beta \hbar^2 k^2/2 m_*) $, and from the Cartesian Langevin equation (\ref{eq:110}), we derive the initial value and the evolution equation of the velocity correlation function: 
\be 
\label{eq:148} 
C_{ij} (0) \simeq \delta_{ij} \frac{k_B T}{m_*} \quad \mbox{and} \quad 
\dd C_{ij} (\tau) \underset{\tau> 0}{=} \langle \dd v_i (\tau) v_j (0) \rangle 
\simeq-\Gamma_k \dd \tau \, C_{ij} (\tau) 
\ee 
Integration of (\ref{eq:148}) and use of the relation $ C_{ij} (-\tau) = C_{ji} (\tau) $ give expression (\ref{eq:144}) where the coefficient $ \alpha $ is that of equation (\ref{eq:119}) and, after insertion in expression (\ref{eq:143}) of $ \mathcal{D}^{\rm spa} $, lead to the announced result (\ref{eq:146}) with $ \mathcal{D}_\sslash (k_0) = \mathcal{D}_\perp (k_0) = \alpha $. 

The case $ k_0> 0 $ is more unusual. To build a minimal model, we now take $ \vv = \hbar [(k-k_0)/m_*] \hat{\kk} $, $ D_{\sslash, \perp} (k) \equiv D_{\sslash, \perp} (k_0) $ and a stationary distribution $ \Pi_0 (k) \propto \exp [-\beta \hbar^2 (k-k_0)^2/2 m_*] $ (the direction $ \hat{\kk} $ of $ \kk $ has a uniform distribution on the unit sphere) \footnote{As the direction of $ \kk $ thermalizes more slowly than the modulus by one order in temperature according to equation (\ref{eq:118}), there exists, for an initial state out of equilibrium, a transient regime in which the distribution of $ k $ has reached equilibrium, but that of $ \hat{\kk} $ not yet ($ \hat{\kk} $ has not had time to diffuse on the unity sphere, the constant term in force (\ref{eq:149}) has not yet really acted), so that $ C_{ij} (t) \simeq (k_B T/m_*) \langle \hat{k}_i \hat{k}_j \rangle_0 \exp (-\Gamma_k t) $ where the average $ \langle \ldots \rangle_0 $ is taken over the initial distribution of $ \hat{\kk} $. This single-exponential form resembles the case $ k_0 \equiv 0 $, up to a factor 3 on the trace (the effective dimension $ d_{\rm eff} $, i.e.\ the number of quadratic degrees of freedom in the approximate form of the energy $ \epsilon_k $ near $ k = k_0 $, goes from $ d_{\rm eff} = 3 $ for $ k_0 \equiv 0 $ to $ d_{\rm eff} = 1 $ for $ k_0> 0 $). At these intermediate times $ \Gamma_k^{-1} \! \ll \! t \! \ll \! \Gamma_{\hat{\kk}}^{-1} $, the $\gamma$ quasiparticle performs a transient Brownian motion of spatial diffusion matrix $ \mathcal{D}_{ij}^{\rm spa, \, trans} = (k_B T/m_* \Gamma_k) \langle \hat{k}_i \hat{k}_j \rangle_0 $, maybe anisotropic but of fixed trace.}. The relation (\ref{eq:300}) then obliges us to include in the mean force not only the expected viscous friction term, but also the sub-leading-in-temperature contribution at zero speed, which takes a simpler form here because we neglect the dependence of $ D_\sslash $ on $k$: 
\be
\label{eq:149}
F(k)=-\hbar\Gamma (k-k_0)+\frac{2}{\hbar k_0}\left[D_\sslash(k_0)-D_\perp(k_0)\right]
\quad\mbox{with}\quad 
\Gamma \equiv\frac{D_\sslash(k_0)}{m_* k_B T} \underset{T\to 0}{\sim}
\Gamma_k
\ee
By inserting these elements of the model in the stochastic equations for modulus and direction (\ref{eq:111a}, \ref{eq:111b}), and by replacing as in (\ref{eq:149}) the wave number $ k $ by $ k_0 $ in the denominators, we obtain the minimum Langevin formulation \footnote{\label{noteelab} In a more elaborate formulation taking into account the cubic distorsion of $\veps_k$ and note \ref{noteiden}, we replace equation (\ref{eq:150a}) by $\hbar\dd k=-\Gamma\dd t\, \hbar(k-\langle k\rangle)+[2\dd t D_\sslash(k_0)]^{1/2}\eta_\sslash$ where the stationary mean $\langle k\rangle$ is given by note \ref{notegauch}, assuming that $m_* v_k=\hbar(k-k_0)+\hbar(k-k_0)^2b+\ldots\simeq \hbar(k-k_0)+b m_* k_B T/\hbar\simeq \hbar(k-\langle k\rangle)+2m_*k_B T/\hbar k_0$ ($(k-k_0)^2$ was approximated by its stationary mean) ; equation (\ref{eq:150b}) remains unchanged. As in the minimum formulation, $k(t)$ and $\hat{\kk}(t)$ are independant stochastic processes, so that $C_{zz}(\tau)\simeq\langle v_k(\tau)v_k(0)\rangle \langle\hat{k}_z(\tau)\hat{k}_z(0)\rangle\simeq (\hbar/m_*)^2[\langle (k-\langle k\rangle)^2\rangle \exp(-\Gamma|\tau|)+(2 m_* k_B T/\hbar^2 k_0)^2](1/3)\exp(-\Gamma_\perp|\tau|)$. This gives again result (\ref{eq:146}). This also shows that it is the \og effective mean force on the modulus \fg \, i.e.\, the deterministic part of $\dd k$, rather than $F(k)$, which vanishes for $ k-k_0 \simeq \langle k-k_0 \rangle $.}
\bea
\label{eq:150a}
\hbar\,\dd k &=&-\Gamma\dd t\,\hbar(k-k_0)+\frac{2 D_\sslash(k_0)}{\hbar k_0}\dd t
+ \left[2\dd t D_\sslash(k_0)\right]^{1/2} \eta_\sslash \\
\label{eq:150b}
\dd\hat{\kk} &=& -\Gamma_\perp \dd t\,\hat{\kk} +(\Gamma_\perp \dd t)^{1/2}
\mbox{\etab}_\perp \quad \mbox{with}\quad \Gamma_\perp \equiv \frac{2 D_\perp(k_0)}
{\hbar^2 k_0^2} \underset{T\to 0}{\sim} \Gamma_{\hat{\kk}}
\eea
The corresponding variation of the velocity along $ Oi $, 
\be
\label{eq:151}
\dd v_i = \frac{\hbar \hat{k}_i}{m_*} \dd k + \frac{\hbar (k-k_0)}{m_*} 
\dd\hat{k}_i + \frac{\hbar}{m_*} \dd k\, \dd \hat{k}_i=
-(\Gamma+\Gamma_{\perp})\dd t \, v_i +\frac{2\Gamma\dd t\, k_B T}{\hbar k_0}
\hat{k}_i + \mbox{noise}
\ee
no longer leads as in relation (\ref{eq:148}) to a closed equation on the velocity correlation function, but couples to a mixed direction-velocity correlation function $ \tilde{C}_{ij} (\tau) = \langle \hat{k}_i (\tau) v_j (0) \rangle $, hence the differential system at $ \tau \geq 0 $: 
\be
\label{eq:152}
\frac{\dd}{\dd \tau} \binom{C_{ij}(\tau)}{\tilde{C}_{ij}(\tau)} = 
\begin{pmatrix} 
-(\Gamma+\Gamma_\perp) & \frac{2 k_B T}{\hbar k_0}\Gamma \\
0 & - \Gamma_\perp 
\end{pmatrix}
\binom{C_{ij}(\tau)}{\tilde{C}_{ij}(\tau)}
\ee
The variations $ \dd (k-k_0) $ and $ \dd [(k-k_0)^2] $ have a zero mean at equilibrium, and isotropy imposes $ \langle \hat{k}_i \hat{k}_j \rangle = \frac{1}{3} \delta_{ij} $, so we obtain the initial values in the Langevin model (\ref{eq:150a}, \ref{eq:150b}), $ C_{ij} (0) = \delta_{ij} (k_B T/3 m_*) [1+4 m_* k_B T/(\hbar k_0)^2] $ and $ \tilde{C}_{ij} (0) = \delta_{ij} (2 k_B T/3 \hbar k_0) $. Integration of the system (\ref{eq:152}), together with the relation $ C_{ij} (-\tau) = C_{ji} (\tau) $, gives the announced result (\ref{eq:145}) if we neglect in the first exponential the rate $ \Gamma_\perp $ in comparison with $ \Gamma $, which is legitimate since $ \Gamma_\perp/\Gamma = O (T) $. 

The minimal model that we have just used for $ k_0> 0 $, which neglects the wave number dependence of $ D_{\sslash, \perp} (k) $ and the corrections $ \approx (k-k_0)^3 $ to the dispersion relation, does not provide a mathematical proof. A more solid study is carried out in \ref{app:Fokkergen}, using a controlled expansion at low temperature. It leads to the same results (\ref{eq:145}, \ref{eq:146}). Finally, these calculations of spatial diffusion do not apply to the transition point between the cases $ k_0 \equiv 0 $ and $ k_0> 0 $, where the mean force is of cubic departure, $ F (k) \approx-k^3$. A separate study of the transition zone, exposed in the same \ref{app:Fokkergen}, confirms however the validity of the low temperature equivalent (\ref{eq:146}). 

\appendix 
\section{Complements on $\phi-\gamma$ scattering} 
\label{app:rigor} 

The main lines of the computation of the scattering amplitude of a phonon of wave vector $ \qq $ on a stable $ \gamma $ quasiparticle of wave vector $ \kk $ from an effective low energy Hamiltonian were presented in section \ref{sec:ampdiff}. We give here some details and additional points of rigor. 

\paragraph{The Hamiltonian} The explicit form of the different contributions to the Hamiltonian (\ref{eq:033}) is available in the literature, see for example the most recent references \cite{PRLphigam, PRLerr}. We recall it for convenience (and to correct a venial error in the last term in square brackets of equation (4) of \cite{PRLphigam}, wrongly written as a Hermitian conjugate), limiting ourselves in $ \hat{V}_{\phi \phi} $ to the three-phonon processes of Beliaev and Landau, the only ones to be useful here: 
\bea
\label{eq:801a}
\hat{V}_{\phi\phi} &=& \frac{1}{2L^{3/2}} \sum_{\qq_1,\qq_2,\qq_3}^{q_1,q_2,q_3<\Lambda} \langle\phi:\qq_1,\phi:\qq_2|\mathcal{V}_{\phi\phi}|\phi:\qq_3\rangle\, (\hat{b}_{\qq_1}^\dagger\hat{b}_{\qq_2}^\dagger \hat{b}_{\qq_3}+\hat{b}_{\qq_3}^\dagger \hat{b}_{\qq_1} \hat{b}_{\qq_2}) \\
\label{eq:801b}
\hat{H}_3^{\phi\gamma} &=& \frac{1}{L^{3/2}} \sum_{\kk,\kk',\qq}^{q<\Lambda} \langle \phi:\qq,\gamma:\kk|\mathcal{H}_3^{\phi\gamma}|\gamma:\kk'\rangle\, (\hat{\gamma}_{\kk'}^\dagger \hat{\gamma}_\kk \hat{b}_\qq + \hat{b}_\qq^\dagger \hat{\gamma}_\kk^\dagger \hat{\gamma}_{\kk'}) \\
\label{eq:801c}
\hat{H}_4^{\phi\gamma} &=& \frac{1}{L^3} \sum_{\kk,\kk',\qq,\qq'}^{q,q'<\Lambda} \langle\phi:\qq',\gamma:\kk'|\mathcal{H}_4^{\phi\gamma} |\phi:\qq,\gamma:\kk\rangle\, \hat{\gamma}_{\kk'}^\dagger \hat{\gamma}_{\kk} \left(\hat{b}_{\qq'}^\dagger\hat{b}_{\qq}+\frac{1}{2} \hat{b}_{\qq}\hat{b}_{-\qq'}+\frac{1}{2} \hat{b}_{-\qq}^\dagger\hat{b}_{\qq'}^\dagger\right)
\eea
with the matrix elements for a unit volume between states with one or two phonons, zero or one $\gamma$ quasiparticle: 
\bea
\label{eq:802a}
\langle\phi:\qq_1,\phi:\qq_2|\mathcal{V}_{\phi\phi}|\phi:\qq_3\rangle &=& \delta_{\qq_1+\qq_2,\qq_3} \frac{mc^2}{\rho^2} \rho_{q_1}\rho_{q_2}\rho_{q_3} \left(\lambda+\frac{\qq_1\cdot\qq_2}{q_1 q_2}+ \frac{\qq_1\cdot\qq_3}{q_1 q_3}+\frac{\qq_2\cdot\qq_3}{q_2 q_3}\right) \\
\label{eq:802b}
\langle\phi:\qq,\gamma:\kk|\mathcal{H}_3^{\phi\gamma}|\gamma:\kk'\rangle &=& \delta_{\qq+\kk,\kk'} \frac{1}{2}\rho_q\left[\frac{\partial\epsilon^{(0)}_\kk}{\partial\rho}+\frac{\partial\epsilon^{(0)}_{\kk'}}{\partial\rho}+ \frac{\hbar c \qq}{\rho q} \cdot (\kk+\kk')\right] \\
\label{eq:802c}
\langle\phi:\qq',\gamma:\kk'|\mathcal{H}_4^{\phi\gamma} |\phi:\qq,\gamma:\kk\rangle &=& \delta_{\qq+\kk,\qq'+\kk'} \frac{1}{2}\rho_q \rho_{q'} \left(\frac{\partial^2\epsilon^{(0)}_\kk}{\partial\rho^2}+\frac{\partial^2\epsilon^{(0)}_{\kk'}}{\partial\rho^2}\right)
\eea
To clarify the notations, we decide that the wave vectors $ \qq $ are always those of the phonons, and the wave vectors $ \kk $ always those of the $\gamma$ quasiparticle. Parameter $ \lambda $ is given by equation (\ref{eq:047}) (see however footnote \ref{note:c0}). Kronecker's deltas conserve momentum. One introduced the amplitude of the quantum fluctuations of the density of the superfluid in the phonon mode of wave vector $ \qq $, \footnote{\label{note:c0} In all rigor, it would be necessary at this stage to use in $ \rho_q $ and in $ \lambda $ the bare equation of state and the bare speed of sound, which depend on the ultraviolet cutoff $ \Lambda $ introduced in equations (\ref{eq:033}) and (\ref{eq:801a},\ref{eq:801b},\ref{eq:801c}), see appendix B of reference \cite{CRASbrou}.} 
\be
\label{eq:803}
\rho_q=\left(\frac{\hbar\rho q}{2 m c}\right)^{1/2}
\ee
to show that the density fluctuations govern both the interaction between phonons and the interaction between phonons and $\gamma$ quasiparticle. In expressions (\ref{eq:802b}) and (\ref{eq:802c}), the derivative $ \partial/\partial \rho $ of the bare dispersion relation of the $\gamma$ quasiparticle with respect to the density $ \rho $ of the superfluid is carried out at fixed wave vector $ \kk $ and interaction potential between the atoms of the superfluid.

\paragraph{Analysis of orders> 2 in $ \hat{V} $} Let us justify the fact that to leading order in temperature ($ q, q '= O (T) \to 0 $ at fixed $ \kk $), we can limit the scattering amplitude (\ref{eq:036}) to order two in $ \hat{V} $. The term of arbitrary order $ n $ in $ \hat{V} $ is easy to write, it is the matrix element between $ \langle \mathrm{f} | $ and $ | \mathrm{i} \rangle $ of the product of $ n $ operators $ \hat{V} $ between which we insert $ n-1 $ operators $ G_0 (z) $ (two successive factors $ \hat{V} $ are always separated by a single factor $ G_0 $); here, $ G_0 (z) = (z-\hat{H}_0)^{-1} $ is the resolvent of the Hamiltonian without interaction and $ z = E_{\rm i}^{(0)}+\ii \eta $, $ \eta \to 0^+$. To understand what's going on, let's start with $ n = 3 $: the contribution $ \langle \mathrm{f} | \hat{V} G_0 (z) \hat{V} G_0 (z) \hat{V} | \mathrm{i} \rangle $ generates many diagrams after splitting $ \hat{V} $ into elementary processes as in equation (\ref{eq:033}), even taking into account the order of magnitude of the matrix elements (the largest are those $ \propto q^{1/2} $ of $ \hat{H}_3^{\phi \gamma} $) and parity changes in the number of phonons (on cannot take three times $ \hat{H}_3^{\phi \gamma} $ in the factors $ \hat{V} $) to eliminate the sub-leading or zero contributions. But in each diagram appears an arbitrary internal phonon wave vector $ \qq '' $, not constrained by the conservation of momentum, and on which it is therefore necessary to make a summation. Indeed, a factor $ \hat{V} $ annihilates the incident phonon $ \qq $, a factor $ \hat{V} $ (maybe the same) creates the emerging phonon $ \qq '$, therefore one of the three factors $ \hat{V} $ necessarily creates or annihilates a phonon of unconstrained wave vector $\qq '' $. We thus end up with the very simple upper bound 
\be
\label{eq:810}
\langle\mathrm{f}|V G_0(z) V G_0(z) V|\mathrm{i}\rangle 
=O\left(\sum_{\qq''}^{q''<\Lambda}\frac{(q q')^{1/2}q''}{\Delta E_1 \Delta E_2}
\right)= O(T^2 (q q')^{1/2})
\ee
since (i) each energy denominator is (at fixed $ \kk $) of order one in the wave vectors of the phonons therefore is $ \approx \Lambda = O (T) $, (ii) in the numerator, each absorption or emission of a phonon of wave number $ Q $ is accompanied by a factor $ \rho_Q \propto Q^{1/2} = O (T^{1/2}) $, and (iii) the summation on the wave vector of the internal phonon takes out as a factor the volume $ \Lambda^3 = O (T^3) $ below the cut-off in Fourier space. 
In the case of an arbitrary odd order $ n = 2s+1 $ in $ \hat{V} $, conservation of parity of the number of phonons in the transition $ | \mathrm{i} \rangle \to | \mathrm{f} \rangle $ imposes that there is not $ 2s+1 $ but at least $ 2s+2 $ absorptions or emissions of phonons, therefore at least $ 2s $ absorptions or emissions of phonon other than the incident phonon $ \qq $ and the emerging phonon $ \qq '$ (each factor $ \hat{V} $ absorbs or emits at least one phonon); it is then necessary to sum on at least $ s $ internal phonons of wave vectors $ \qq_1, \ldots, \qq_s $. In the case of an even order $ n = 2s $, there are for the same reasons at least $ 2s-2 $ absorptions or emissions of phonons other than $ \qq $ and $ \qq '$, and we must sum on at least $ s-1 $ internal phonons $ \qq_1, \ldots, \qq_{s-1} $. Hence the upper bound on the order $ n> 2 $ in $ \hat{V} $, \footnote{For $ n = 2 $, it would overestimate the real result $ O (q q ') $ by not taking account of the cancellation in expression (\ref{eq:038}) of the contributions $ \mathcal{T}_2 $ and $ \mathcal{T}_3 $ to their leading order $ T^0 $.} 
\be
\label{eq:811}
\langle\mathrm{f}| \hat{V}[G_0(z)\hat{V}]^{n-1}|\mathrm{i}\rangle
= \left\{
\begin{array}{ll}
O[(q q')^{1/2} T^{n-1}] & \mbox{ if } n \mbox{ odd} \\
O[(q q')^{1/2} T^{n-3}] & \mbox{ if } n \mbox{ even}
\end{array}
\right.
\ee
This upper bound is valid for a generic value $ \approx T $ of the complex energy $ z $. In the particular case $ z = E_{\rm i}^{(0)}+\ii \eta $ which interests us, certain diagrams of order $ n> 2 $ in $ \hat{V} $ like those of figure \ref{fig:diag}c are in fact infinite and must be removed from equation (\ref{eq:811}) for it to apply; this problem is discussed in the body of the article in the paragraph below equation (\ref{eq:342}). 

\paragraph{Analysis of order 2 in $ \hat{V} $} 
It remains to justify the fact that to order two in $ \hat{V} $, we can keep only the contributions (\ref{eq:038}) or, if you prefer, the diagrams in figure \ref{fig:diag}b. Each contribution quadratic in $ \hat{V} $ contains in the numerator the product of two matrix elements, chosen among those of $ \hat{H}_3^{\phi \gamma} \propto T^{1/2} $, $ \hat{H}_4^{\phi \gamma} \propto T $, $ \hat{H}_3^{\phi \phi} \sim \hat{H}_5^{\phi \gamma} \propto T^{3/2} $, $ \hat{H}_4^{\phi \phi} \sim \hat{H}_6^{\phi \gamma} \propto T^2 $, etc (even including the coupling of $ \gamma $ to the cube and the fourth power of the density fluctuations, and the four-phonon interaction). It contains an energy denominator that can be linearized in the phonon wave vectors and is therefore $ \propto T $. The contribution must not be a $ o (T) $. We can therefore {\sl a priori} take in the factors $ \hat{V} $: (i) twice $ \hat{H}_3^{\phi \gamma} $ (which gives the terms $ \mathcal{T}_2 $ and $ \mathcal{T}_3 $ in (\ref{eq:038})), (ii) twice $ \hat{H}_4^{\phi \gamma} $ (made negligible $ \approx (q q ')^{1/2} T^3 $ by inevitable summation on an internal wave vector), (iii) once $ \hat{H}_3^{\phi \gamma} $ and once $ \hat{H}_4^{\phi \gamma} $ (zero by non-conservation of the parity of the number of phonons), (iv) once $ \hat{H}_3^{\phi \gamma} $ and once $ \hat{H}_3^{\phi \phi} $ (which gives the terms $ \mathcal{T}_4 $ and $ \mathcal{T}_5 $ in (\ref{eq:038})), (v) once $ \hat{H}_3^{\phi \gamma} $ and once $ \hat{H}_5^{\phi \gamma} $ (negligible $ \approx (q q ')^{1/2} T^3 $ by inevitable summation on an internal wave vector). 

\paragraph{Matrix $ S $ between exact asymptotic states} 
We follow the approach of reference \cite{CCTbordeaux}, section B${}_{\rm III}$.3, by generalizing it to the case where the incident and emerging phonons, and not just the $\gamma$ quasiparticle, are dressed by virtual phonons. Recall the definition of the matrix elements of $S$ between the exact asymptotic states (\ref{eq:040}): 
\be
\label{eq:804}
\langle \mathrm{f}||S||\mathrm{i}\rangle = \lim_{t\to +\infty} \eee^{\ii E_{\rm f}t/2\hbar} \langle\mathrm{f}||U(t/2,-t/2)||\mathrm{i}\rangle \eee^{\ii E_{\rm i}t/2\hbar}=
\lim_{t\to +\infty} \eee^{\ii(E_{\rm i}+E_{\rm f})t/2\hbar} \int_{+\infty+\ii \eta}^{-\infty+\ii\eta} \frac{\dd z}{2\ii \pi} \eee^{-\ii z t/\hbar} \langle\mathrm{f}||G(z)||\mathrm{i}\rangle
\ee
where $ \eta \to 0^+$, $ E_{\rm i, f} $ are the exact energies (\ref{eq:039}) (for the moment not necessarily equal), $ U (t_{\rm f}, t_{\rm i}) $ is the evolution operator between the times $ t_{\rm i} $ and $ t_{\rm f} $ and $ G (z) = (z-\hat{H})^{-1} $ the resolvent of the full Hamiltonian $ \hat{H} $. Starting from the obvious relations (the purely $\gamma$-quasi-particulate component of $ \hat{H} $ commutes with the purely phononic operator $ \hat{B}_{\qq}^\dagger $), 
\be
\label{eq:805}
\hat{B}_{\qq}^\dagger (z-\hat{H})-(z-\hat{H})\hat{B}_{\qq}^\dagger= [\hat{H},\hat{B}_{\qq}^\dagger]=[\hat{V}_{\phi\gamma},\hat{B}_{\qq}^\dagger] + [\hat{H}_{\phi\phi},\hat{B}_{\qq}^\dagger]
\ee
we introduce the operator (\ref{eq:042}) with $ \QQ = \qq $ as follows to eliminate the phonon Hamiltonian $\hat{H}_{\phi\phi}$, 
\be
\label{eq:806}
\hat{B}_{\qq}^\dagger (z-\hat{H}-\hbar\omega_\qq) - (z-\hat{H}) \hat{B}_{\qq}^\dagger =[\hat{V}_{\phi\gamma},\hat{B}_{\qq}^\dagger] +\hat{\Delta}_{\qq}^\dagger
\ee
we divide by $ z-\hat{H} $ on the left and by $ z-\hat{H}-\hbar \omega_\qq $ on the right, and we make the result act on the stationary state $ || \gamma: \kk \rangle $ to obtain 
\be
\label{eq:807}
G(z)||\mathrm{i}\rangle = \frac{1}{z-E_{\rm i}} \left[||\mathrm{i}\rangle + G(z) \left( [\hat{V}_{\phi\gamma},\hat{B}_{\qq}^\dagger]+\hat{\Delta}_\qq^\dagger\right) ||\gamma:\kk\rangle\right] \quad \mbox{then}\quad \langle\mathrm{f}||G(z) = \frac{1}{z-E_{\rm f}} \left[ \langle\mathrm{f}|| + \langle\gamma:\kk'|| \left([\hat{B}_{\qq'}, \hat{V}_{\phi\gamma}]+\hat{\Delta}_{\qq'}\right) G(z)\right]
\ee
by doing the same, or by Hermitian conjugation and replacement of $ (z^*, \qq, \kk) $ by $ (z, \qq ', \kk') $. It remains to contract the first identity in (\ref{eq:807}) on the left by $ \langle \mathrm{f} || $ and to use the second identity in (\ref{eq:807}) to obtain 
\be
\label{eq:808}
\langle\mathrm{f}||G(z)||\mathrm{i}\rangle= \frac{\langle\mathrm{f}||\, ||\mathrm{i}\rangle}{z-E_{\rm i}} +\frac{\langle\gamma:\kk'|| \hat{B}_{\qq'} \left([\hat{V}_{\phi\gamma},\hat{B}_{\qq}^\dagger]+\hat{\Delta}_\qq^\dagger\right) ||\gamma:\kk\rangle+ \langle\gamma:\kk'||\left([\hat{B}_{\qq'},\hat{V}_{\phi\gamma}]+\hat{\Delta}_{\qq'}\right) G(z)  \left( [\hat{V}_{\phi\gamma},\hat{B}_{\qq}^\dagger]+\hat{\Delta}_\qq^\dagger\right) ||\gamma:\kk\rangle}{(z-E_{\rm i})(z-E_{\rm f})}
\ee
In free space $ L \to+\infty $ and long time limit, the curvilinear integral (\ref{eq:804}) is dominated by the residue of the integrand at the poles $ E_{\rm i} $ and $ E_{\rm f} $ of expression (\ref{eq:808}), knowing that the matrix element of $ G (z) $ in the numerator has no pole at these points \cite{CCTbordeaux}, and it reads according to the theory of distributions \cite{CCTbordeaux} $ \langle \mathrm{f} || S || \mathrm{i} \rangle = \langle \mathrm{f} || \, || \mathrm{i} \rangle-2 \ii \pi \delta (E_{\rm f}-E_{\rm i}) A_{\rm fi} $ with the sought transition amplitude (\ref{eq:041}) on the energy shell. 

\paragraph{Scattering of an unstable phonon} 
When the acoustic branch of the gas has a convex departure [$\gamma_\phi>0$ in equation (\ref{eq:001})], the incident phonon $ \qq $ is unstable and disintegrates with a rate $ \propto q^5 $ into a pair of phonons, themselves unstable. This Beliaev cascade makes the $ S $ matrix theory inapplicable. The idea that saves us is then to limit ourselves to a scattering time $ t $ long enough to ensure a quasi-conservation of undisturbed energy, but short enough to avoid Beliaev disintegration, by forming spatially localized incident wave packets by linear superposition of plane waves of wave vectors $ \qq $ for the phonon $ \phi $ and $ \kk $ for the $\gamma$ quasiparticle, with a Gaussian amplitude $ c (\qq, \kk) $ very peaked around the average values $ \qq_0 = q_0 \hat{\qq}_0 $ and $ \kk_0 = k_0 \hat{\kk}_0 $, i.e.\ with widths $ \Delta q \ll q_0 $ and $ \Delta k \ll k_0 $ (here we omit the problem of the dressing of asymptotic states and we assume, as we are suggested by (\ref{eq:003}), that $ \Delta q \ll \Delta k $). More precisely, we take as initial state in free space \footnote{In the absence of a quantization box, we 
use the condition of normalization of plane waves $ \langle \kk | \kk '\rangle = (2 \pi)^3 \delta (\kk-\kk ') $.} 
\be
\label{eq:820}
|\bar{\rm i}\rangle_t = \eee^{-\ii \hat{H}_0 (-t/2)/\hbar} 
\int \frac{\dd^3q\,\dd^3k}{(2\pi)^6} c(\qq,\kk) |\phi:\qq,\gamma:\kk\rangle
\ee
Since $ c (\qq, \kk) $ is real, the wave packets in $ | \bar{\rm i} \rangle $ would be localized in position space around the origin of coordinates (with widths $ 1/\Delta q $ and $ 1/\Delta k $) if the state vector had not been moved back in time by $ t/2 $ by a free evolution; they are then localized approximately in $-c \hat{\qq}_0 t/2 $ and $-v_{k_0} \hat{\kk}_0t/2 $, therefore well separated and \og asymptotically free \fg \, (without interaction) if 
\be
\label{eq:821}
v_{\rm rel}  t \gg \frac{1}{\Delta q}
\quad\mbox{with}\quad v_{\rm rel} = |c\hat{\qq}_0 - v_{k_0}\hat{\kk}_0|
\ee
the relative speed of the two incident packets. We evolve $ | \bar{\rm i} \rangle_t $ for a duration $ t $ (from time $-t/2 $ to $ t/2 $) with the full Hamiltonian $ \hat{H} $. The state becomes free again since the packets have separated, their overlap and their interaction lasting only a time $ \approx 1/(\Delta q v_{\rm rel}) $. It remains to project for analysis on the final plane waves $ | \mathrm{f} \rangle = | \phi: \qq ', \gamma: \kk' \rangle $ of energy $ E_{\rm f}^{(0)} $. 
After cancellation of the free evolution phase factor of $ \langle \mathrm{f} | $, we arrive at the effective scattering amplitude 
\begin{multline}
\label{eq:822}
A_{\mathrm{f}\bar{\mathrm{i}}}(t)\equiv \langle\mathrm{f}| \eee^{\ii\hat{H}_0t/2\hbar} \eee^{-\ii \hat{H} t/\hbar} |\bar{\rm i}\rangle_t
\simeq \int \frac{\dd^3q}{(2\pi)^3} c(\qq,\kk=\kk'+\qq'-\qq) 
\Bigg\{\frac{A_{\rm fi}(E_{\rm f}^{(0)})-A_{\rm fi}(E_{\rm i}^{(0)})}
{E_{\rm f}^{(0)}-E_{\rm i}^{(0)}} 
\cos\frac{(E_{\rm i}^{(0)}-E_{\rm f}^{(0)})t}{2\hbar} \\
-\ii \left[A_{\rm fi}(E_{\rm f}^{(0)})+A_{\rm fi}(E_{\rm i}^{(0)})\right]
\frac{\sin\frac{(E_{\rm i}^{(0)}-E_{\rm f}^{(0)})t}{2\hbar}}
{E_{\rm i}^{(0)}-E_{\rm f}^{(0)}}\Bigg\}
\end{multline}
To obtain the approximation in the third side of equation (\ref{eq:822}), we replaced the evolution operator by its expression in the form of a curvilinear integral (\ref{eq:804}), used the relation $ G (z) = G_0 (z)+G_0 (z) \hat{T} (z) G_0 (z) $ to make the matrix $ T $ appear with the complex energy $ z $, made the approximation as in reference \cite{CCTbordeaux} of keeping in the Cauchy formula only the residues of the integrand at the poles $ E_{\rm i}^{(0)} $ and $ E_{\rm f}^{(0)} $ originating from the factors $ G_0 (z) $ surrounding $ \hat{T} (z) $ \footnote{To second order in the interaction, $ \hat{T} (z) \simeq \hat{V}+\hat{V} G_0 (z) \hat{V} $, which suffices here, it is easy to include the contribution of the poles $ \epsilon_n $ de $ \langle \mathrm{f} | \hat{T} (z) | \mathrm{i} \rangle $, which adds between the braces of equation (\ref{eq:822}) the sum $ \sum_{n = 1}^{4} (\epsilon_n-E_{\rm i}^{(0)})^{-1} (\epsilon_n-E_{\rm f}^{(0)})^{-1} Z_n \exp [-\ii t \, (\epsilon_n-\frac{E_{\rm i}^{(0)}+E_{\rm f}^{(0)}}{2})/\hbar] $, where $ \epsilon_1 = \epsilon_{\kk+\qq}^{(0)} = \epsilon_{\kk '+\qq'}^{(0)} $, $ \epsilon_2 = \hbar \omega_\qq^{(0)}+\hbar \omega_{\qq '}^{(0)}+\epsilon_{\kk-\qq '}^{(0)} $, $ \epsilon_3 = \hbar \omega_{\qq'}^{(0)}+\hbar \omega_{\qq-\qq '}^{(0)}+\epsilon_\kk^{(0)} $, $ \epsilon_4 = \hbar \omega_\qq^{(0)}+\hbar \omega_{\qq '-\qq}^{(0)}+\epsilon_{\kk+\qq-\qq '}^{(0)} $ (i.e.\ the energies of the intermediate states of diagrams b2 to b5 in figure \ref{fig:diag}) and the associated residues of $ \langle \mathrm{f} | \hat{T} (z) | \mathrm{i} \rangle $ $ Z_n $ are regular functions of $ \qq, \qq ', \kk' $. If we take care to exclude in the phonon observation directions a cone around the direction $ \hat{\qq}_0 $ of the incident phonon, that is to say to impose $ | \hat{\qq} '\cdot \hat{\qq}_0 | <\cos\theta_c $ with $0<\cos\theta_c <1 $ fixed (to avoid that the Beliaev process in term $ \mathcal{T}_4 $ or that of Landau in $ \mathcal{T}_5 $ is resonant), we always find that the energy differences $ | \epsilon_n-E_{\rm i}^{(0)} | $ and $ | \epsilon_n-E_{\rm f}^{(0)} | $ are approximately greater than $ \hbar cq $ or $ \hbar c q '$, and we always have $ \hbar cq/2 \lesssim | \epsilon_n-(E_{\rm i}^{(0)}+E_{\rm f}^{(0)})/2 | $. The contribution of the poles of $ \langle \mathrm{f} | \hat{T} (z) | \mathrm{i} \rangle $ is therefore an oscillating function of time, which the average over $ \qq $ suppresses and makes negligible by a Gaussian-in-time factor $ \approx \exp [-(c \Delta q \, t)^2/2] $, as in equation (\ref{eq:824}).}, neglected the contribution of the free evolution term $ G_0 (z) $ under the condition $ | \qq '-\qq_0 | \gg \Delta q $, and finally posed $ A_{\rm fi} (E) = \langle \mathrm{f} | \hat{T} (E+\ii \eta) | \mathrm{i} \rangle $ with state $ | \mathrm{i} \rangle \equiv | \phi: \qq, \gamma: \kk = \kk '+\qq'-\qq \rangle $ of energy $ E_{\rm i}^{(0)} $ and $ \eta \to 0^+$. The cosine and cardinal sine prefactors are continuous and bounded functions of $ E_{\rm i}^{(0)} $; we can replace them by their value at the center of the wave packet $ (E_{\rm i}^{(0)} \simeq \bar{E}_{\rm i}^{(0)}) $ if their scale of variation $ E_{\rm pref} $ satisfies the condition (see footnote \ref{note:gauss}) 
\be
\label{eq:823}
\frac{\hbar}{t} \gg \frac{\mathrm{Var}\, E_{\rm i}^{(0)}}{E_{\rm pref}}
\quad \mbox{with}\quad E_{\rm pref} \approx \hbar c q_0\quad\mbox{and}\quad
\mathrm{Var}\, E_{\rm i}^{(0)} \approx (\hbar c \Delta q)^2
\ee
Here $ \bar{E}_{\rm i}^{(0)} $ and $ \mathrm{Var} \, E_{\rm i}^{(0)} $ are the mean and variance of $ E_{\rm i}^{(0)} \simeq \hbar cq+\epsilon_{\kk = \kk '+\qq'-\qq} $ for $ \qq $ distributed with the Gaussian law $ c (\qq, \kk '+\qq'-\qq) $. In calculating the mean of the cosine and the cardinal sine on the variable $ \qq $, the same condition (\ref{eq:823}) allows us to assume that $ E_{\rm i}^{(0)} $ also obeys a Gaussian law \footnote{\label{note:gauss} Suppose for simplicity that the $\gamma$ quasiparticle is strongly subsonic, $ | v_{k '}/c | \ll 1 $, and expand the phase shift during $ t $ in powers of fluctuations $ \delta \qq = \qq-\bar{\qq} $, $ [E_{\rm i}^{(0)}-\bar{E}_{\rm i}^{(0)}] t/\hbar \simeq ct \hat{\qq}_0 \cdot \delta \qq+O [ct (\delta q)^2/q_0] $. The linear term has a Gaussian distribution; the $ O $ is not Gaussian but negligible under the condition (\ref{eq:823}). Conversely, if $ E_{\rm i}^{(0)} $ is a Gaussian variable, it is easy to obtain the condition (\ref{eq:823}) of negligible variation of the prefactors in (\ref{eq:822}), assuming that these vary linearly with $ E_{\rm i}^{(0)} $.}, in which case 
\be
\label{eq:824}
\langle\cos \frac{(E_{\rm i}^{(0)}-E_{\rm f}^{(0)})t}{2\hbar}\rangle_\qq
=\exp\left[-t^2 \mathrm{Var}\, E_{\rm i}^{(0)}/8 \hbar^2\right]
\cos \frac{(\bar{E}_{\rm i}^{(0)}-E_{\rm f}^{(0)})t}{2\hbar}
\ee
\bea
\label{eq:825a}
\left\langle\frac{\sin\frac{(E_{\rm i}^{(0)}-E_{\rm f}^{(0)})t}{2\hbar}}
{E_{\rm i}^{(0)}-E_{\rm f}^{(0)}}\right\rangle_\qq &=&
\frac{(\pi/2)^{1/2}}{\Delta E_{\rm i}^{(0)}}
\exp\left[-\frac{(E_{\rm f}^{(0)}-\bar{E}_{\rm i}^{(0)})^2}{2\,\mathrm{Var}\, E_{\rm i}^{(0)}}\right]
\re\Phi\left(\frac{\Delta E_{\rm i}^{(0)}t}{2\hbar\sqrt{2}} +
\ii\frac{|E_{\rm f}^{(0)}-\bar{E}_{\rm i}^{(0)}|}{\sqrt{2}\Delta E_{\rm i}^{(0)}}
\right) \\
\label{eq:825b}
&\simeq&
\left\{
\begin{array}{l}
\displaystyle\frac{(\pi/2)^{1/2}}{\Delta E_{\rm i}^{(0)}}
\exp\left[-(E_{\rm f}^{(0)}-\bar{E}_{\rm i}^{(0)})^2/2\,\mathrm{Var}\, E_{\rm i}^{(0)}\right] \ \ \mbox{if}\ \ |E_{\rm f}^{(0)}-\bar{E}_{\rm i}^{(0)}|< t\,\mathrm{Var}\, E_{\rm i}^{(0)}/2\hbar \\
\displaystyle\frac{\sin\frac{(\bar{E}_{\rm i}^{(0)}-E_{\rm f}^{(0)})t}{2\hbar}}
{\bar{E}_{\rm i}^{(0)}-E_{\rm f}^{(0)}} \exp\left[-t^2 \mathrm{Var}\, E_{\rm i}^{(0)}/8 \hbar^2\right]\ \ \mbox{otherwise} 
\end{array}
\right.
\eea
where $ \Delta E_{\rm i}^{(0)} $ is the standard deviation (square root of the variance) and $ \Phi (z) = (2/\sqrt{\pi}) \int_0^z \dd u \exp (-u^2) $ is the error function \footnote{We get the error function using $ \sin x/x = \int_{-1}^{1} \dd u \, \eee^{\ii ux}/2 $ and exchanging the integrations on $ u $ and on energy. The approximate forms in (\ref{eq:825b}) follow from the asymptotic behavior of the error function in the first quadrant: if $ z = \rho \exp (\ii \theta) $ and $ \rho $ tends to $+\infty $ at fixed $ \theta $, then $ \Phi (z) \to 1 $ if $ \theta \in [0, \pi/4 [$ and $ \Phi (z) \sim-\exp (-z^2)/(z \sqrt{\pi}) $ if $ \theta \in] \pi/4, \pi/2 [$.}. Let us choose for example $ \Delta q \propto \Delta k \propto q_0^3 \approx T^3 $ (which gives sufficient precision on $ q $, see footnote \ref{note:lourdeur}) and a scattering time $ t \propto q_0^{-4} $, and let $ q_0 $ and the gas temperature tend to zero at fixed $ k_0 $. Then conditions (\ref{eq:821}) and (\ref{eq:823}) are asymptotically satisfied, and everything happens as if the $S$ matrix was given by the usual expression $ \langle \mathrm{f} | S | \mathrm{i} \rangle = \langle \mathrm{f} | \mathrm{i} \rangle-2 \ii \pi \langle \mathrm{f} | \hat{T} (E_{\rm i}^{(0)}+\ii \eta) | \mathrm{i} \rangle \delta (E_{\rm f}^{(0)}-E_{\rm i}^{(0)}) $: this is obvious for the first case of equation (\ref{eq:825b}), which can be found by the usual substitution $ \sin (\epsilon t/2 \hbar)/\epsilon \to \pi \delta (\epsilon) $; the second case of (\ref{eq:825b}) and equation (\ref{eq:824}) do not correspond to a energy-conserving Dirac distribution but this does not matter because they are exponentially suppressed. As the scattering time $ t $ is infinitely shorter than the lifetime $ \approx q_0^{-5} $ of the incident and emerging phonons, we can rightly neglect their Beliaev instability, that is to say their real rather than virtual emission of phonons, in the calculation of the scattering amplitude on a $\gamma$ quasiparticle to leading order in $ q $. 

\paragraph{Scattering in matter} 
As scattering of phonon $ \qq $ on the $\gamma$ quasiparticle actually occurs in a thermal gas of phonons, it seems incorrect, for the applications of section \ref{sec:carac}, to calculate the scattering amplitude in vacuum and not to take into account stimulated emission of internal phonons in the diagrams b4 and b5 of figure \ref{fig:diag} in the already populated modes of wave vectors $ \qq-\qq' $ and $ \qq '-\qq $. At first glance, it would therefore be necessary to correct the last two terms of result (\ref{eq:038}) as follows: 
\be
\label{eq:830}
\mathcal{T}_4 \stackrel{?}{\longrightarrow}(1+\bar{n}_{\qq-\qq'})\,\mathcal{T}_4
\quad\mbox{and}\quad 
\mathcal{T}_5 \stackrel{?}{\longrightarrow}(1+\bar{n}_{\qq'-\qq})\,\mathcal{T}_5
\ee
where the $ \bar{n} $ are the average occupation numbers of the modes. But, if the modes $ \qq-\qq '$ and $ \qq'-\qq $ are already populated, it is necessary to take into account the fact that phonons of these modes can participate temporarily in the scattering (they must remain in equal numbers in the initial state and in the final state). This introduces two new diagrams. The first one, of amplitude $ \mathcal{T}_4 '\bar{n}_{\qq-\qq'} $, describes the absorption of a phonon $ \qq-\qq '$ by the $\gamma$ quasiparticle of wave vector $ \kk $ (which brings it directly into its final state $ \kk '$), then the Beliaev decay of the incident phonon $ \qq $ into two phonons $ \qq '$ and $ \qq-\qq' $, which provides the expected final phonon and restores the mode $ \qq-\qq '$ to its original occupation number. The second one, of amplitude $ \mathcal{T}_5 '\bar{n}_{\qq'-\qq} $, describes Beliaev's merging of the incident phonon $ \qq $ with a phonon $ \qq'-\qq $ (which provides the final phonon $ \qq '$), followed by the emission of a phonon $ \qq'-\qq $ by the $\gamma$ quasiparticle of wave vector $ \kk $, which puts it directly in its final state $ \kk '$ and identically repopulates the mode $ \qq'-\qq $. None of these new diagrams are related, and they are the only possible ones to leading order. The right correction to apply to (\ref{eq:038}) is therefore 
\be
\label{eq:831}
\mathcal{T}_4 \longrightarrow (1+\bar{n}_{\qq-\qq'})\,\mathcal{T}_4 +
\bar{n}_{\qq-\qq'} \mathcal{T}_4'
\quad\mbox{and}\quad 
\mathcal{T}_5 \longrightarrow(1+\bar{n}_{\qq'-\qq})\,\mathcal{T}_5
+ \bar{n}_{\qq'-\qq} \mathcal{T}_5'
\ee
In equation (\ref{eq:341}) [(\ref{eq:342})], we see that $ \mathcal{T}_4 '$ ($ \mathcal{T}_5 '$) has the same numerator as $ \mathcal{T}_4 $ ($ \mathcal{T}_5 $), that its energy denominator $ \epsilon_\kk^{(0)}+\hbar \omega_{\qq-\qq '}^{(0)}-\epsilon_{\kk'}^{(0)} $ ($ \hbar \omega_\qq^{(0)}+\hbar \omega_{\qq '-\qq}^{(0)}-\hbar \omega_{\qq'}^{(0)} $) is generally unrelated to that of $ \mathcal{T}_4 $ ($ \mathcal{T}_5 $), except on the energy shell (\ref{eq:035}) where it is the exact opposite and where $ \mathcal{T}_4 '=-\mathcal{T}_4 $ ($ \mathcal{T}_5 '=-\mathcal{T}_5 $). The occupation numbers in the corrected form (\ref{eq:831}) then cancel exactly, \footnote{A similar cancellation appears in reference \cite{Annalen}.} and result (\ref{eq:038}) established in vacuum also applies in matter. 

\section{Calculation of spatial diffusion at low temperature by the regression theorem} 
\label{app:Fokkergen} 

Equation (\ref{eq:146}) gives a low temperature equivalent of the spatial diffusion coefficient of the $\gamma$ quasiparticle in a phonon gas. In the case where the dispersion relation $ \epsilon_k $ of the quasiparticle reaches its minimum in a wave number $ k_0> 0 $, the justification provided by the main text, using a minimal Langevin stochastic model, is not fully convincing. We present here a demonstration based on the regression theorem and on a systematic low temperature expansion of the solution of the corresponding Fokker-Planck equation. We also justify expression (\ref{eq:146}) in the transition zone between the case $ k_0 \equiv $ and the case $ k_0> 0 $, which must be treated separately. 

\paragraph{Regression theorem} 
Knowledge of the spatial diffusion coefficient requires that of the correlation function $ C_{ij} (\tau) $ of the coordinates of the velocity vector in the stationary state, see equations (\ref{eq:140}, \ref{eq:141}, \ref{eq:142}, \ref{eq:143}). By isotropy, we limit ourselves to the calculation of $ C_{zz} (\tau) $, $ \tau \geq 0 $. In the Fokker-Planck framework, we then have the following regression theorem: 
\be
\label{eq:701}
C_{zz}(\tau)=\int \frac{\dd^3k}{(2\pi)^3} \frac{1}{\hbar}
\frac{\dd\epsilon_k}{\dd k} \hat{k}_z \Pi(\kk,\tau)
\ee
where the distribution $ \Pi (\kk, \tau) $ has evolved for a time $ \tau $ according to the Fokker-Planck equation (\ref{eq:103}) from the initial condition 
\be
\label{eq:702}
\Pi(\kk,0)= \frac{1}{\hbar} \frac{\dd\epsilon_k}{\dd k} \hat{k}_z \Pi_0(k)
\ee
with $ \Pi_0 (k) $ the normalized rotationally invariant stationary solution of equation (\ref{eq:103}), and $ \hat{k}_z = k_z/k $. Since $ \Pi (\kk, \tau) $ is of angular momentum one along $Oz$, we can set $ \Pi (\kk, \tau) = \Pi_1 (k, \tau) \hat{k}_z $ and obtain the equation of motion 
\begin{multline}
\label{eq:703}
\frac{\partial}{\partial \tau}\Pi_1(k,\tau) = -\frac{2 F(k) \Pi_1(k,\tau)}{\hbar k}
-\frac{1}{\hbar} \frac{\partial}{\partial k} [F(k) \Pi_1(k,\tau)]
+\frac{1}{\hbar^2} \frac{\partial^2}{\partial k^2} [D_\sslash(k)\Pi_1(k,\tau)] \\
+\frac{1}{\hbar^2 k} \frac{\partial}{\partial k} \{ [4 D_\sslash(k)-2 D_\perp(k)]
\Pi_1(k,\tau)\} 
+\frac{1}{\hbar^2 k^2}[2 D_\sslash(k)-4 D_\perp(k)] \Pi_1(k,\tau)
\end{multline}
Equation (\ref{eq:300}) would allow to express $ \Pi_0 (k) $ in terms of the mean force $ F (k) $ and momentum diffusion coefficients $ D_{\sslash, \perp} (k) $. We find it more enlightening to write it in the form $ \Pi_0 (k) = \exp (-\beta \tilde{\epsilon}_k)/Z $, where $ Z $ is a normalization factor, and to eliminate the force $ F (k) $ in terms of the effective dispersion relation $ \tilde{\epsilon}_k $. As we learn from the comparison of equations (\ref{eq:300}, \ref{eq:301}), this is submitted to the constraint $\tilde{\epsilon}_k-\epsilon_k=O(T^2)$ when $T\to 0$ with $k-k_0=O(T^{1/2})$, so it can, as the true dispersion relation $ \epsilon_k $, be expanded around the location $\tilde{k}_0$ of its minimum (equal to $\Delta_*$ by convention); $\tilde{k}_0$ is zero if $k_0\equiv 0$, and depends on temperature but remains very close to $k_0$ if $k_0>0$, and the expansion coefficients depend weakly on temperature too: \footnote{\label{note:faible} If we accept that the $\phi-\gamma$ scattering amplitude (\ref{eq:041}) has an expansion in integer powers of $q$ at fixed $\kk$, then $F(k)$, $D_{\perp,\sslash}(k)$ and $\tilde{\epsilon}_k$ have an expansion in integer powers of $T$ at fixed $k$, and one has $\tilde{k}_0/k_0=1+O(T^2)$, $\tilde{m}_*/m_*=1+O(T)$, $\tilde{b}/b=1+O(T)$, $\tilde{\ell}/\ell=O(1)$.}
\be
\label{eq:704}
\tilde{\epsilon}_k \underset{k\to{\tilde{k}}_0}{=} \Delta_* + \frac{\hbar^2(k-{\tilde{k}}_0)^2}{2{\tilde{m}}_*}
+ \frac{\hbar^2(k-{\tilde{k}}_0)^3{\tilde{b}}}{3{\tilde{m}}_*} + \frac{\hbar^2 (k-{\tilde{k}}_0)^4{\tilde{\ell}}^2}{4{\tilde{m}}_*}
+O(k-{\tilde{k}}_0)^5
\ee
In other words, one should not believe in the predictions taken from the Fokker-Planck equation (an approximation of the master equation (\ref{eq:100}) in the limit (\ref{eq:003})) if they involve the small deviation between the effective masses $ m_* $ and $\tilde{m}_*$, between the lengths $ b $ and $\tilde{b}$, between the minimizing wave numbers $k_0$ and $\tilde{k}_0$, or if they depend on the length $ \tilde{\ell} $ or on higher order coefficients in expansion (\ref{eq:704}). Let us carry out a last transformation: the operator on the functions of $ k $ defined by the right-hand side of equation (\ref{eq:703}) is not self-adjoint, but can be made so by the change of function $ \Pi_1 (k) = k^{-1} \exp (-\beta \tilde{\epsilon}_k/2) \psi (k) $. After formal temporal integration, we end up for $ \tau \geq 0 $ with a beautiful expression (the factor $ 3 $ in the denominator comes from angular integration): 
\be
\label{eq:705}
C_{zz}(\tau)= 
\frac{\langle\psi_{\rm S}|\exp(-\mathcal{H}_1\tau)|\psi_{\rm S}\rangle}
{3\hbar^2 \int_0^{+\infty}\dd k \, k^2 \eee^{-\beta\tilde{\epsilon}_k}}
=\sum_{n\in\mathbb{N}} 
\frac{|\langle\psi_{\rm S}|\phi_n\rangle|^2 \eee^{-\omega_n t}}
{3\hbar^2 \int_0^{+\infty}\dd k \, k^2 \eee^{-\beta\tilde{\epsilon}_k}}
\quad\mbox{with}\quad
\left\{
\begin{array}{l}
\displaystyle
\mathcal{H}_1 = -\frac{\dd}{\dd k} \frac{D_\sslash(k)}{\hbar^2}
\frac{\dd}{\dd k} + U_1(k) \\
\\
\displaystyle
\langle k|\psi_{\rm S}\rangle = k\frac{\dd{\epsilon}_k}{\dd k} 
\eee^{-\beta \tilde{\epsilon}_k/2}
\end{array}
\right.
\ee
We have introduced the spectral decomposition of the fictitious Hamiltonian $ \mathcal{H}_1 $, which is selfadjoint for the one-dimensional scalar product $ \langle \phi_a | \phi_b \rangle = \int_0^{+\infty} \dd k \, \phi_a^* (k) \phi_b (k) $, the eigenstate $ | \phi_n \rangle $ of eigenvalue $ \omega_n $ being normalized. The potential is equal to
\be
\label{eq:706}
U_1(k)=\frac{1}{\hbar^2}\left[\frac{1}{4} D_\sslash(k)
\left(\beta\frac{\dd\tilde{\epsilon}_k}{\dd k}\right)^2
-\frac{1}{2} D_\sslash(k)\,\beta \frac{\dd^2\tilde{\epsilon}_k}{\dd k^2}
-\frac{1}{2}\frac{\dd D_\sslash(k)}{\dd k} 
\beta\frac{\dd\tilde{\epsilon}_k}{\dd k}
-\frac{D_\sslash(k)}{k} \beta \frac{\dd\tilde{\epsilon}_k}{\dd k}
+\frac{1}{k} \frac{\dd D_\sslash(k)}{\dd k} + \frac{2 D_\perp(k)}{k^2}
\right]
\ee

\paragraph{In the case $ k_0> 0 $} 
To take the low temperature limit in the case $ k_0> 0 $, we carry out a suitable change of scale on wave vector and time by setting $ k-\tilde{k}_0 = x k_* $, $ \tau = \bar{\tau}/\tilde{\Gamma} $, and therefore $ \mathcal{H}_1 = \bar{\mathcal{H}}_1 \tilde{\Gamma} $, $ \omega_n = \bar{\omega}_n \tilde{\Gamma} $ and $ \phi_n (k) = k_*^{-1/2} \bar{\phi}_n (x) $, with thermal wave number $ k_* = (\tilde{m}_* k_B T)^{1/2}/\hbar $ and rate $\tilde{\Gamma}$ build on the model of rate $\Gamma$ in equation (\ref{eq:149}). It remains to expand $ \bar{\mathcal{H}}_1 $ in powers of $ k_* $ at fixed $ x $ ($ x $ spans $ \mathbb{R} $ in this limit), then to process successive orders by perturbation theory: 
\be
\label{eq:707}
\bar{\mathcal{H}}_1= -\frac{\dd}{\dd x} \frac{D_\sslash(\tilde{k}_0+x k_*)}
{D_\sslash(\tilde{k}_0)} \frac{\dd}{\dd x} + \frac{\hbar^2 k_*^2}
{D_\sslash(\tilde{k}_0)} U(\tilde{k}_0 + x k_*)
\stackrel{x\, \mbox{\scriptsize fixed}}{\underset{k_*\to 0}{=}}
\bar{\mathcal{H}}_1^{(0)}+\bar{\mathcal{H}}_1^{(1)}+\bar{\mathcal{H}}_1^{(2)}
+\ldots
\ee
We will also need to expand the source wave function to sub-leading order and the denominator of $ C_{zz} (\tau) $ in equation (\ref{eq:705}) to leading order: 
\be
\label{eq:708}
\langle k|\psi_{\rm S}\rangle \stackrel{x\,\mbox{\scriptsize fixed}}
{\underset{k_*\to 0}{=}} \frac{\hbar^2 \tilde{k}_0 k_*}{\tilde{m}_*} \eee^{-\beta\Delta_*/2}
x e^{-x^2/4}\left[1+x\left(\frac{k_*}{\tilde{k}_0}+k_*\tilde{b}\right)
-\frac{1}{6} k_*\tilde{b}x^3 + O(k_*^2)\right],
\int_0^{+\infty}\!\!\!\!\dd k\, k^2 \eee^{-\beta\tilde{\epsilon}_k}
\underset{k_*\to 0}{\sim}
\tilde{k}_0^2k_*\eee^{-\beta\Delta_*} \int_{-\infty}^{+\infty}\!\!\!\!
\dd x\,\eee^{-x^2/2}
\ee
At order zero in $ k_* $, we come across a reduced one-dimensional harmonic oscillator Hamiltonian (mass equal to $ 1/2 $, reduced Planck constant equal to $ 1 $, angular oscillation frequency equal to $ 1 $) and spectrum $ \bar{\omega}_n^{(0)} = n $: 
\be
\label{eq:709}
\bar{\mathcal{H}}_1^{(0)} = - \frac{\dd^2}{\dd x^2} + \frac{1}{4} x^2
-\frac{1}{2} = \hat{a}^\dagger \hat{a} \quad\mbox{with}\quad\hat{a}=\frac{1}{2}
x + \frac{\dd}{\dd x}\ ,\ \hat{a}^\dagger=\frac{1}{2} x 
-\frac{\dd}{\dd x} \ , \ [\hat{a},\hat{a}^\dagger]=1
\ee
To leading order, the source wave function is therefore proportional to the wave function $ \bar{\phi}_1^{(0)} (x) $ of the harmonic oscillator first excited state. The contribution of excited states $ n> 0 $ in equation (\ref{eq:705}) is thus dominated by $ n = 1 $ and we keep 
\be
\label{eq:710}
C_{zz}^{n\neq 0}(\tau)\stackrel{\bar{\tau}\,\mbox{\scriptsize fixed}}
{\underset{k_*\to 0}{\sim}}
\frac{k_B T}{3 \tilde{m}_*}\eee^{-\bar{\tau}}
\ee
We thus justify the first term in square brackets in equation (\ref{eq:145}), considering footnote \ref{note:faible}. The contribution of $ n = 0 $ is more difficult to obtain. Indeed, the overlap of $ \psi_{\rm S} $ with the ground state $ \bar{\phi}_0 $ is zero at the order of $ \bar{\mathcal{H}}_1^{(0)} $, and so is the ground state angular eigenfrequency $ \omega_0 $, which leads to an indeterminate form $ 0/0 $ after integration of $ C_{zz} (\tau) $ over time as in equation (\ref{eq:143}). It is necessary to go to order one in $ k_* $ to obtain a nonzero overlap, 
\begin{multline}
\label{eq:711}
\bar{\mathcal{H}}_1^{(1)}= k_* \frac{\frac{\dd D_\sslash}{\dd k}({\tilde{k}}_0)}
{D_\sslash({e\tilde{k}}_0)}
\left(-\frac{\dd}{\dd x} x \frac{\dd}{\dd x} + \frac{1}{4}x^3-x\right)
+ k_* {\tilde{b}} \left(\frac{1}{2}x^3-x\right)
-\frac{k_*}{{\tilde{k}}_0} x = 
\left(\frac{k_*\frac{\dd D_\sslash}{\dd k}({\tilde{k}}_0)}{D_\sslash({\tilde{k}}_0)}
+\frac{3}{2} k_* {\tilde{b}}\right) 
(\hat{a}^{\dagger 2}\hat{a}+\hat{a}^\dagger\hat{a}^2) \\
+\frac{1}{2} k_* {\tilde{b}} (\hat{a}^{\dagger 3}+\hat{a}^3)
+\left(\frac{1}{2} k_* {\tilde{b}} -\frac{k_*}{{\tilde{k}}_0}\right) (\hat{a}+\hat{a}^\dagger)
\Longrightarrow |\bar{\phi}_0\rangle
\underset{k_*\to 0}{=}|\bar{\phi}_0^{(0)}\rangle
-\left(\frac{1}{2} k_* {\tilde{b}} -\frac{k_*}{{\tilde{k}}_0}\right) |\bar{\phi}_1^{(0)}\rangle
-\frac{k_*{\tilde{b}}}{\sqrt{6}}|\bar{\phi}_3^{(0)}\rangle
+ O(k_*^2) \\
\Longrightarrow \frac{|\langle\psi_{\rm S}|\phi_0\rangle|^2}{3\hbar^2
\int_0^{+\infty} \dd k\, k^2 \eee^{-\beta\tilde{\epsilon}_k}}
\underset{k_*\to 0}{\sim} \frac{k_B T}{3 {\tilde{m}}_*} \left(\frac{2 k_*}{{\tilde{k}}_0}\right)^2
\end{multline}
and to order two in $ k_* $ to obtain a nonzero ground state angular eigenfrequency: 
\begin{multline}
\label{eq:712}
\bar{\mathcal{H}}_1^{(2)}=
\frac{k_*^2\frac{\dd^2 D_\sslash}{\dd k^2}({\tilde{k}}_0)}{2 D_\sslash({\tilde{k}}_0)}
\left[-\frac{\dd}{\dd x}x^2\frac{\dd}{\dd x}
+\frac{1}{4}x^4-\frac{3}{2}x^2\right]
+\frac{k_*^2\frac{\dd D_\sslash}{\dd k}({\tilde{k}}_0)}{D_\sslash({\tilde{k}}_0)}
\left[\frac{{\tilde{b}}}{2}(x^4-3 x^2)+\frac{1-x^2}{{\tilde{k}}_0}\right]
+\frac{1}{4} k_*^2 {\tilde{b}}^2 x^4 + \frac{1}{2} k_*^2 {\tilde{\ell}}^2 (x^4-3 x^2) \\
+\frac{k_*^2}{{\tilde{k}}_0^2}(1-{\tilde{k}}_0{\tilde{b}}) x^2+
\frac{2 k_*^2 D_\perp({\tilde{k}}_0)}{{\tilde{k}}_0^2 D_\sslash({\tilde{k}}_0)}
\Longrightarrow \bar{\omega}_0 \underset{k_*\to 0}{\sim}
\langle\bar{\phi}_0^{(0)}|\bar{\mathcal{H}}_1^{(2)}|\bar{\phi}_0^{(0)}\rangle
+\sum_{n>0}\frac{|\langle\bar{\phi}_n^{(0)}|\bar{\mathcal{H}}_1^{(1)}|
\bar{\phi}_0^{(0)}\rangle|^2}{-n}=\frac{2 k_*^2}{{\tilde{k}}_0^2}\frac{D_\perp({\tilde{k}}_0)}
{D_\sslash({\tilde{k}}_0)}
\end{multline}
knowing that $ \langle x^2 \rangle = 1 $, $ \langle x^4 \rangle = 3 $ and $ \langle \frac{\dd}{\dd x} x^2 \frac{\dd}{\dd x} \rangle =-3/4 $ where the expectation value is taken in the state $ | \bar{\phi}_0^{(0)} \rangle $. We justify the second term in square brackets in equation (\ref{eq:145}) and therefore, in total, the equivalent (\ref{eq:146}) of the spatial diffusion coefficient \footnote{There is a a quick and elegant way to find these results. By restricting the time-dependent Fokker-Planck equation to rotationally invariant distributions, then applying the same function change as in $ \Pi_1 (k) $, we get the fictitious self-adjoint Hamiltonian $ \mathcal{H}_0 $. In this isotropic sector, the Fokker-Planck equation has a nonzero stationary solution $ \Pi_0 $. The ground state $ | \psi_0 \rangle $ of $ \mathcal{H}_0 $ is therefore known exactly, $ \psi_0 (k) = \mathcal{N}_0 k \exp (-\beta \tilde{\epsilon}_k/2) $ with $ \mathcal{N}_0 = [\int_0^{+\infty} \dd k \, k^2 \exp (-\beta \tilde{\epsilon}_k)]^{1/2} $, and its eigenvalue is exactly zero. Now, as the calculation shows, we simply have $ \mathcal{H}_1 = \mathcal{H}_0+2 D_\perp (k)/(\hbar k)^2 $, so that $ \bar{\mathcal{H}}_1 $ differs from $ \bar{\mathcal{H}}_0 $ by a second order term $ 2 k_*^2 D_\perp (\tilde{k}_0+x k_*)/[D_\sslash (\tilde{k}_0) (\tilde{k}_0+x k_*)^2] $. The ground state of $ \mathcal{H}_1 $ therefore coincides with $ | \psi_0 \rangle $ to sub-leading order, and we have $ \langle \psi_{\rm S} | \phi_0 \rangle \sim \langle \psi_{\rm S} | \psi_0 \rangle $; to the same order, we can replace $ \dd \epsilon_k/\dd k $ by $ \dd \tilde{\epsilon}_k/\dd k $ in $ \langle k | \psi_{\rm S} \rangle $. A simple integration by parts (you have to integrate $ \frac{\dd \tilde{\epsilon}_k}{\dd k} \exp (-\beta \tilde{\epsilon}_k) $ and take the derivative of $ k^2 $), then the replacement of the slowly varying factors $ k $ in the numerator and $ k^2 $ in the denominator by $ \tilde{k}_0 $ and $ \tilde{k}_0^2 $, give the overlap equivalent as in equation (\ref{eq:711}). Similarly, to leading order, the ground state angular eigenfrequency $ \omega_0 $ of $ \mathcal{H}_1 $ is obtained by treating the difference between $ \mathcal{H}_1 $ and $ \mathcal{H}_0 $ to first order of perturbation theory, $ \omega_0 \sim \langle \psi_0 | 2 D_\perp (k)/(\hbar k)^2 | \psi_0 \rangle = \int_0^{+\infty} \dd k \, 2 D_\perp (k) \exp (-\beta \tilde{\epsilon}_k)/\int_0^{+\infty} \dd k \, (\hbar k)^2 \exp (-\beta \tilde{\epsilon}_k) $. By approximating the slowly varying factors $ 2 D_\perp (k) $ and $ (\hbar k)^2 $ by their value in $ \tilde{k}_0 $, we find the equivalent of $ \omega_0 $ given in equation (\ref{eq:712}).} \footnote{We have also successfully verified the validity at low temperature of the spatial diffusion coefficient predicted by the Fokker-Planck approximation by numerically solving the master equation (\ref{eq:100}) with the microreversible $\phi-\gamma$ scattering amplitude (\ref{eq:303}) for the simple model $ \epsilon_k = \Delta_*+(\hbar^2/2 m_*) [(k^2-k_0^2)/2 k_0]^2 $ ($ m_*> 0, k_0> $ 0) and $ R_k (u, u ', w) = (1-\eta u^2) (1-\eta u'^2) $, leading to $ \alpha = \frac{32}{1575} (3 \eta^2-10 \eta+15) (15 \eta^2-42 \eta+35)$, $ \mathcal{D}_\perp (k_0) = \frac{32}{1575} (3 \eta^2-10 \eta+15) (3 \eta^2-14 \eta+35)\nequiv\alpha$ and $\frac{\dd}{\dd k}\mathcal{D}_{\sslash}(k_0)=0$ thus to a nonzero mean force $ F (k_0) $ at order $ T^9 $ (except if $ \eta = 0 $). The error made by Fokker-Planck on $\mathcal{D}^{\rm spa}$ is $<10\%$ for all $\eta\in[0,3]$ as soon as $k_*/k_0<1/20$.  The calculation uses the quantum regression theorem to access the velocity correlation function, which is then inserted in expression (\ref{eq:143}) of $ \mathcal{D}^{\rm spa} $. In the zero angular momentum sector, the master equation writes $ \partial_t \Pi_0 (k, t) =-\gamma (k) \Pi_0 (k, t)+\int_0^{+\infty} \dd k '\, \gamma_0 (k' \to k) \Pi_0 (k ', t) $; in practice, we ensure probability conservation by numerically calculating all the feeding rates $ \gamma_0 (k '\to k) $ then deducing the departure rates from the relation $ \gamma (k) = \int_0^{+\infty} \dd k '\, (k'/k)^2 \gamma_0 (k \to k ') $. In the unit angular momentum sector, the master equation writes $ \partial_t \Pi_1 (k, t) =-\gamma (k) \Pi_1 (k, t)+\int_0^{+\infty} \dd k '\, \gamma_1 (k' \to k) \Pi_1 (k ', t) $ (the departure rates are unchanged). Microreversibility requires $ k^2 \exp (\beta \epsilon_k) \gamma_n (k '\to k) = k'^2 \exp (\beta \epsilon_{k '}) \gamma_n (k \to k') $, $ n = 0 $ or $ 1 $, which brings us back to integral operators with symmetric kernels (this is numerically advantageous) by the change of function $ \Pi_n (k) = k^{-1} \exp (-\beta \epsilon_k/2) \psi_n (k) $. Expression (\ref{eq:705}) then applies, provided that we replace $ \tilde{\epsilon}_k $ by $ \epsilon_k $ and $-\mathcal{H}_1 $ by the integral operator relating $ \partial_t \psi_1 $ to $ \psi_1 $.}.

\paragraph{Between cases $ k_0 \equiv 0 $ and $ k_0> 0 $} 
At the transition point between these two cases, the effective mass $ m_* $ is infinite, $ 1/m_* = 0 $, the dispersion relation $ \epsilon_k $ varies quartically with the wave number near its minimum, and none of the calculations of the spatial diffusion coefficient presented so far in this work apply. To connect the two limiting cases (for example when density $\rho$ varies), it is necessary to use an approximation of degree 4 for the dispersion relation, 
\be
\label{eq:720}
\epsilon_k \simeq \Delta_0 + \frac{\hbar^2 k^2}{2 m_{\rm eff}}+\frac{\hbar^2 k^2}{4m} 
(k l)^2
\ee
The length $l$ is essentially constant but the variation of the effective mass $ m_{\rm eff} $ in $ k = 0 $ (not to be confused with that $ m_* $ at the minimum) describes the passage of a dispersion relation reaching its minimum in $ k = 0 $ (case $ 1/m_{\rm eff} > 0 $) to a dispersion relation reaching its minimum in a nonzero wave number (case $ 1/m_{\rm eff} <0 $). We then make a magnification around the transition point, looking in $1/m_{\rm eff}$-space and $k$-space at scales $ T^\nu $ such that the last two terms of equation (\ref{eq:720}) are of the same order of magnitude $ \approx k_B T $. We therefore introduce the dimensionless wave number $ \kappa $ and the dimensionless parameter $ \delta $ controlling the transition: 
\be
\label{eq:721}
\kappa = k \left(\frac{\hbar^2 l^2}{m k_B T}\right)^{1/4} \quad
\mbox{and}\quad \delta = \frac{m}{m_{\rm eff}} 
\left(\frac{\hbar^2}{m k_B T l^2}\right)^{1/2}
\ee
It remains to take the limit $ T \to 0 $ at fixed $ \kappa $ and $ \delta $. We approximate the mean force and the momentum diffusion coefficients by their leading order at the transition point, i.e.\ in the reduced form (\ref{eq:106}) by $ \mathcal{F} (k) \simeq-\alpha_0 e_k $ and $ \mathcal{D}_\perp (k) \simeq \mathcal{D}_\sslash (k) \simeq \alpha_0 $, where $ \alpha_0 $ is the reduced friction coefficient at the transition \footnote{We check indeed in equation (\ref{eq:052}) that, for the dispersion relation (\ref{eq:720}) and for a finite limit of $ \dd (m/m_{\rm eff})/\dd \rho $ at the transition $ \delta = 0 $, $ R_k (u, u ', w) \to (1/2) (\check{e}_{x, 0}+w \check{e}_{\rho, 0}) $ where $ \check{e}_{x, 0} $ and $ \check{e}_{\rho, 0} $ are the $\delta=0$ values of $ \check{e}_x $ and $ \check{e}_\rho $ defined by equation (\ref{eq:050}). This is what we get by taking $ m/m_* = 0 $ in equation (\ref{eq:051}).}. It remains to introduce the reduced time 
\be
\label{eq:722}
\theta = t/t_0 \quad \mbox{with}\quad t_0\equiv \frac{\hbar}{m c^2}
\frac{15}{\pi^5} \left(\frac{m c^2}{k_B T}\right)^{17/2} 
\frac{(\hbar\rho^{1/3}/mc)^6}{\alpha_0 m c l/\hbar} 
\ee
so that the random walk of the $\gamma$ quasiparticle in Fourier space is described by a universal dimensionless stochastic process with a parameter $ \delta $ and with the same Gaussian noise as equation (\ref{eq:110}), 
\be
\label{eq:723}
\dd\mbox{\kapb} = -\dd\theta\, \mathbf{grad}_{\mbox{\kapb}} \veps(\kappa)
+ \sqrt{2\dd\theta}\,\mbox{\etab}
\quad \mbox{with}\quad \veps(\kappa)=\frac{1}{2}\kappa^2\delta+\frac{1}{4}\kappa^4
\ee
To extract the spatial diffusion coefficient as defined by equation (\ref{eq:143}), we go through the regression theorem and the Fokker-Planck equation as we did at the beginning of this \ref{app:Fokkergen} and we get (setting $ \veps' = \dd \veps (\kappa)/\dd \kappa $) 
\be
\label{eq:724}
\mathcal{D}^{\rm spa}\stackrel{\delta\,\mbox{\scriptsize fixed}}
{\underset{T\to 0}{\sim}} 
\frac{\hbar}{m}\frac{\alpha_0^{-1}\Delta^{\rm spa}(\delta)}
{\frac{\pi^5}{15}\left(\frac{mc}{\hbar\rho^{1/3}}\right)^6
\left(\frac{k_B T}{mc^2}\right)^7}
\quad\mbox{with}\quad \Delta^{\rm spa}(\delta)=
\frac{\langle\mathrm{s}|\hat{h}^{-1}|\mathrm{s}\rangle}
{3\int_0^{+\infty} \dd\kappa\, \kappa^2\eee^{-\veps(\kappa)}}
\quad\mbox{and}\quad
\left\{
\begin{array}{l}
\displaystyle\hat{h}=-\frac{\dd^2}{\dd\kappa^2} +\frac{2}{\kappa^2}+
\frac{1}{4}\veps'^2-\frac{1}{\kappa}\veps'-\frac{1}{2}\veps'' \\
\\
\displaystyle
\langle\kappa|\mathrm{s}\rangle = \kappa \veps' \eee^{-\veps(\kappa)/2}
\end{array}
\right.
\ee
Here we formally integrated over time the velocity correlation function written as in equation (\ref{eq:705}), which introduced the inverse of the fictitious Hamiltonian operator $ \hat{h}$. Amazingly we know how to analytically determine the action of $ \hat{h}^{-1} $ on the source vector $ | \mathrm{s} \rangle $ (it is easy to verify by action of $ \hat{h} $ on the result): 
\be
\label{eq:725}
\langle\kappa|\hat{h}^{-1}|\mathrm{s}\rangle = \kappa^2 \eee^{-\veps(\kappa)/2}
\Longrightarrow \Delta^{\rm spa}(\delta)=
\frac{\int_0^{+\infty} \dd\kappa \, \kappa^3 \veps' \eee^{-\veps(\kappa)}}
{3\int_0^{+\infty} \dd\kappa\, \kappa^2\eee^{-\veps(\kappa)}}
\ee
A simple integration by parts in the numerator of (\ref{eq:725}) then leads to the remarkable result $ \Delta^{\rm spa} (\delta) \equiv 1 $, \footnote{As the reader will have understood, this conclusion is independent of the particular form of the reduced dispersion relation $ \veps (\kappa) $.} in perfect agreement with the prediction (\ref{eq:146}), which therefore applies everywhere in the transition zone between the cases $ k_0 \equiv 0 $ and $ k_0> 0 $. 


\end{document}